%% file: cpc-hlt.tex
\documentclass[ALICE,manyauthors]{cernphprep}

\usepackage[comma,square,numbers,sort&compress]{natbib}
\usepackage{hyperref}
\usepackage{lineno}

\newcommand{\fig}[1]{Fig.~\ref{fig:#1}}
\newcommand{\figur}[1]{Figure~\ref{fig:#1}}
\newcommand{\tab}[1]{Tab.~\ref{tab:#1}}
\newcommand{\tabl}[1]{Table~\ref{tab:#1}}

\newcommand*{\eg}{e.\,g.\@\xspace}
\newcommand*{\ie}{i.\,e.\@\xspace}

\newcommand{\run}[1]{Run~#1}

\newcommand{\pt}{$p_{\mathrm{T}}$}

\newcommand{\comment}[1]{} 

\begin{document}%

\begin{titlepage}
\PHyear{2018}
\PHnumber{337}      
\PHdate{17 December}  
%

\title{Real-time data processing in the ALICE High Level Trigger at the LHC}
\ShortTitle{The ALICE High Level Trigger}   

\Collaboration{ALICE Collaboration\thanks{See Appendix~\ref{app:collab} for the list of collaboration members}}
\ShortAuthor{ALICE Collaboration} 

\begin{abstract}
At the Large Hadron Collider at CERN in Geneva, Switzerland, atomic nuclei are collided at ultra-relativistic energies.
Many final-state particles are produced in each collision and their properties are measured by the \mbox{ALICE} detector.
The detector signals induced by the produced particles are digitized leading to data rates that are in excess of~$48$\,GB/s.
The \mbox{ALICE} High Level Trigger (HLT) system pioneered the use of FPGA- and GPU-based algorithms to reconstruct charged-particle trajectories and reduce the data size in real time.
The results of the reconstruction of the collision events, available online, are used for high level data quality and detector-performance monitoring and real-time time-dependent detector calibration.
The online data compression techniques developed and used in the \mbox{ALICE} HLT have more than quadrupled the amount of data that can be stored for offline event processing.
\end{abstract}

\end{titlepage}
\setcounter{page}{2}

\section*{Outline of this article}

In the following, after introducing the \mbox{ALICE} (A Large Ion Collider Experiment) apparatus and highlighting specific detector subsystems relevant to this article, the \mbox{ALICE} High Level Trigger (HLT) architecture and the system software that operates the compute cluster are presented.
Thereafter, the custom Field Programmable Gate Array (FPGA) based readout card, which is employed to receive data from the detectors, is described.
An overview of the most important processing components employed in the HLT follows.
The updates made to the HLT for LHC \run{2}, that provided the capability to operate at twice the event rate compared to LHC \run{1}, are discussed.
The track and event reconstruction methods used, along with the quality of their performance are highlighted.
The presentation of the \mbox{ALICE} HLT is concluded with an analysis of the maximum feasible data and event rates, along with an outlook in particular to LHC \run{3}.

\section{The ALICE detector}

\begin{figure}[!ht]
\begin{centering}
\includegraphics[width=0.47\textwidth]{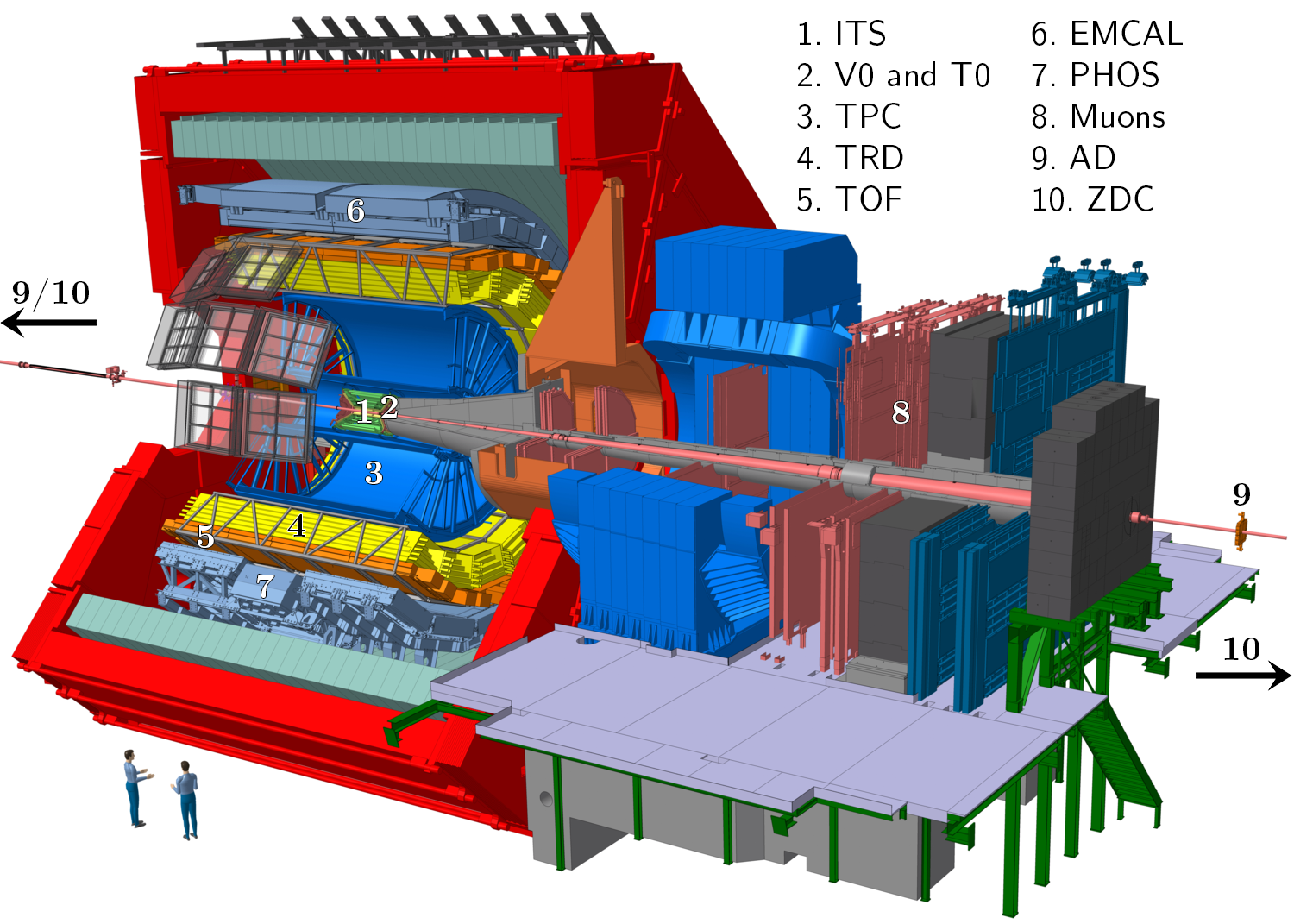}
\par\end{centering}
\caption{The ALICE detector system at the LHC.
}
\label{fig:alice}
\end{figure}

The \mbox{ALICE} apparatus \cite{bib:alice_citation} comprises various detector systems (\fig{alice}), each with its own specific technology choice and design, driven by the physics requirements and the experimental conditions at the LHC \cite{bib:lhc}.
The most stringent design constraint is the extreme charged particle multiplicity density (d$N_{\rm{ch}}$/d$\eta$) in heavy-ion collisions, which was measured at midrapidity to be 1943 $\pm$ 54 in the 5\% most central (head-on) Pb--Pb events at $\sqrt{s_{\rm{NN}}}$~ = $5.02$ TeV \cite{MultPbPb5}.
The main part of the apparatus is housed in a solenoidal magnet, which generates a field of~$0.5$\,T within a volume of~$1600$\,m$^3$.
The central barrel of \mbox{ALICE} is composed of various detectors for tracking and particle identification at midrapidity.
The main tracking device is the Time Projection Chamber (TPC)~\cite{tpc}.
In addition to tracking, it provides particle identification information via the measurement of the specific ionization energy loss (d$E$/d$x$).
The momentum and angular resolution provided by the TPC is further enhanced by using the information from the six layer high-precision silicon Inner Tracking System (ITS) \cite{aliceits}, which surrounds the beam pipe.
Outside the TPC there are two large particle identification detectors: the Transition Radiation Detector (TRD) \cite{Acharya:2017lco} and the Time-Of-Flight (TOF) \cite{Cortese:2002kf}.
The central barrel of \mbox{ALICE} is augmented by dedicated detectors that are used to measure the energy of photons and electrons, the Photon Spectrometer (PHOS) \cite{phos} and ElectroMagnetic Calorimeter (EMCal) \cite{emcal}.
In the forward direction of one of the particle beams is the muon spectrometer \cite{bib:muon}, with its own large dipole magnet.
In addition, there are other fast-interaction detectors including the V0, T0 \cite{forward}, and Zero Degree Calorimeter (ZDC) \cite{zdc}.
As the TPC is the most relevant for the performance of the HLT a more detailed description of it follows.

The TPC is a large cylindrical, gas-filled drift detector with two readout planes at its end-caps.
A central high voltage membrane provides the electric drift field and divides the total active volume of~$85$\,m$^3$ into two halves.
Each charged particle traversing the gas in the detector volume produces a trace of ionization along its own trajectory.
The ionization electrons drift towards the readout planes, which are subdivided into~$18$ trapezoidal readout sectors.
The readout sectors are segmented into~$15488$ readout pads each, arranged in~$159$ consecutive rows in radial direction.
Upon their arrival at the readout planes, ionization electrons induce electric signals on the readout pads.
For an issued readout trigger, the signals are digitized by a~$10$\,bit ADC at a frequency of~$10$\,MHz, sampling the maximum drift time of about~$100$\,$\mu$s into~$1000$ time bins.
This results in a total of~$5.5 \cdot 10^8$ ADC samples containing the full digitized TPC pulse height information.
The size of data corresponding to a single collision event is about $700$\,MB.
A zero-suppression algorithm implemented in an ASIC reduces the proton-proton TPC event size to typically~$100$\,kB.
The exact event size depends on the background, trigger setting, and interaction rate.
Central Pb--Pb collisions produce up to~$100$\,MB of TPC data, which can grow up to around~$200$\,MB with pile-up.
The TPC is responsible for the bulk of the data rate in \mbox{ALICE}.
In \run{2}, when operated at event rates of up to~$2$\,kHz (pp and p--Pb) and~$1$\,kHz (Pb--Pb), it reads out up to~$40$\,GB/s.
In addition, the total readout rate has a contributixon of a few GB/s from other ALICE detectors, some of them operating at trigger rates up to~$3.5$\,kHz.
The volume of data taken at these rates exceeds the capacity for permanent storage considerably.

The amount of data that is stored can be reduced in a number of ways.
The most widely used methods are compression of raw data (using either lossless or lossy schemes) and online selection of a subset of physically interesting events (triggering), which discards a certain fraction of the data read out by the detector~\cite{trigger1,trigger2,trigger3}.
A hierarchical trigger system performs this type of selection by having the lower hardware levels base their decision only on a subset of the data recorded by trigger detectors.
The highest trigger level is the software-based High Level Trigger (HLT), which has access to the entire detector data set.

\section{The High Level Trigger (HLT)}

\label{sec:hlt}

\subsection{From LHC \run{1} commissioning to LHC \run{2} upgrades}

A first step in transforming raw data to fully reconstructed physics information in real time was achieved with the beginning of LHC \run{1} on November~23$^\mathrm{rd}$, 2009, when protons collided in the center of the \mbox{ALICE} detector for the first time.
On the morning of December~6$^\mathrm{th}$, stable beams at an energy of~$450$\,GeV per beam were delivered by the LHC for the first time, and the HLT reconstructed the first charged-particle tracks from pp collisions by processing data from all available \mbox{ALICE} detectors.
Though the HLT was designed as trigger and was operated as such at the start of \run{1}, the collaboration found that by using it for data compression one could record all data to storage, thus optimizing the use of beam time.
This was possible due to the the quality of the online reconstruction and the increased bandwidth to storage.
Throughout \run{1} the HLT was successful as an online reconstruction and data compression facility.

\begin{figure}[htb]
\begin{centering}
\includegraphics[width=0.45\textwidth]{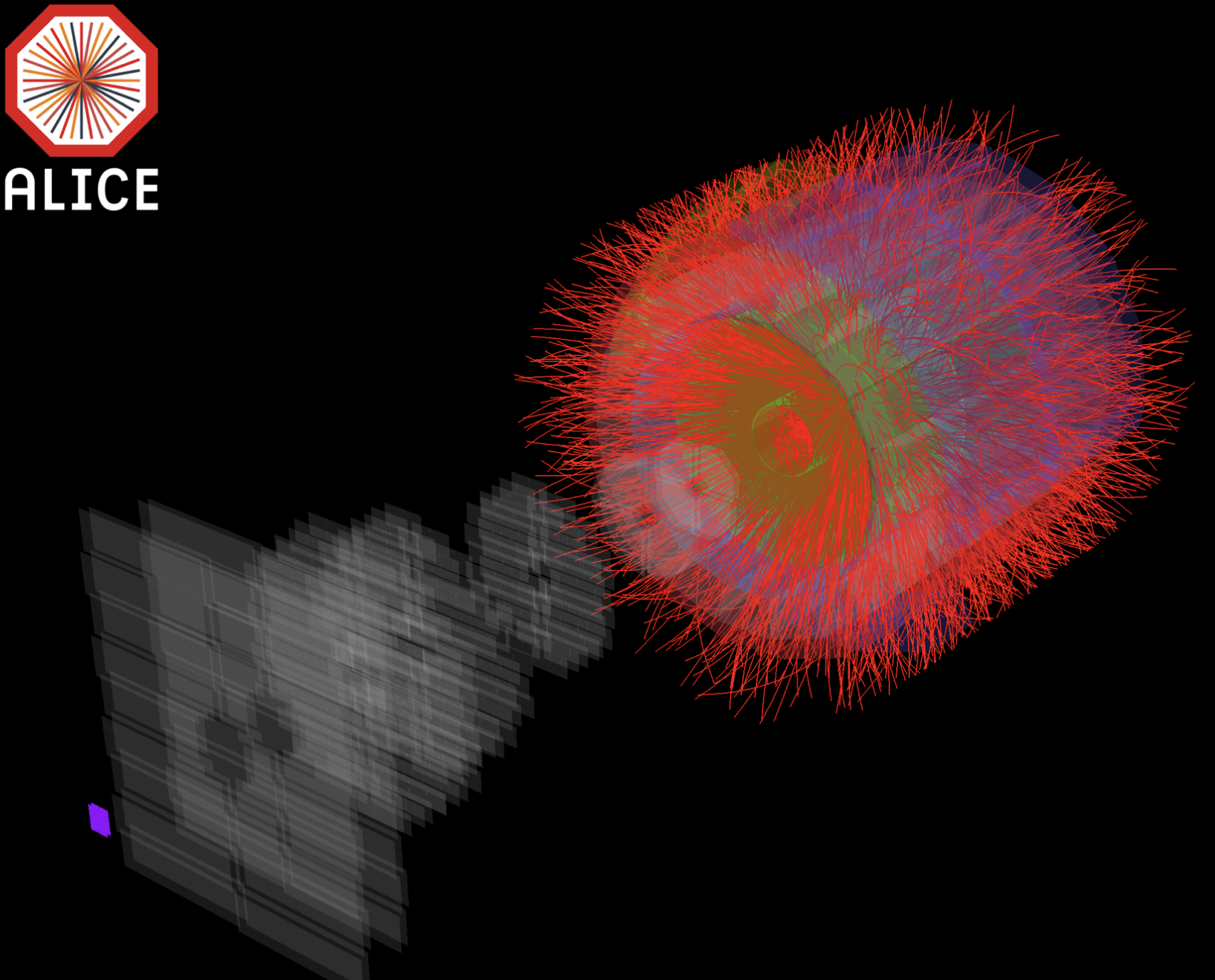}
\par\end{centering}
\caption{Visualization of a heavy-ion collision recorded in ALICE with tracks reconstructed in real time on the GPUs of the HLT.}
\label{fig:first_gpu_pbpb}
\end{figure}

After the LHC \run{1}, that lasted to the beginning of 2013, parts of the \mbox{ALICE} detector were upgraded for LHC \run{2}, which started in~2015.
The most important change was the upgrade of the TPC readout electronics, employing a new version of the Readout Control Unit (RCU2) \cite{bib:rcu2} which uses the updated optical link speed of~$3.125$\,Gbps instead of the previous readout rate of~$2.125$\,Gbps.
The upgrades, along with an improved TPC readout scheme, doubled the theoretical maximum TPC readout data rate to~$48$\,GB/s, thus allowing \mbox{ALICE} to record twice as many events.
In addition, the HLT farm underwent a consolidation phase during that period in order to be able to cope with the increased data rate of \run{2}.
This update improved several parts of the HLT based on the experience from \run{1}.
While the HLT processed up to~$13$\,GB/s of TPC data in \run{1}~\cite{hlt-run1-2011}, the new HLT infrastructure allows for the processing of the full~$48$\,GB/s (see Section \ref{sec:maxrate}).
\figur{first_gpu_pbpb} shows a screenshot of the online event display during a \run{2} heavy-ion run\footnote{A run is defined as a limited period of data taking with similar detector and data-taking conditions.} with active GPU-accelerated online tracking in the HLT, of which will be described in the following.

\subsection{General description}
\label{sec:general}
The main objective of the \mbox{ALICE} HLT is to reduce the data volume that is stored permanently to a reasonable size, so to fit in the allocated tape space.
The baseline for the entire HLT operation is full real-time event reconstruction.
This is required for more elaborate compression algorithms that use reconstructed event properties.
In addition, the HLT enables a direct high-level online Quality Assurance (QA) of the data received from the detectors, which can immediately reveal problems that arise during data taking.
Several of the \mbox{ALICE} sub-detectors (like the TPC) are so called drift-detectors that are sensitive to environmental conditions like ambient temperature and pressure.
Thus a precise event reconstruction requires detector calibration, which in turn requires results from a first reconstruction as input.
It is natural then to perform as much calibration as possible online in the HLT, which is also immediately available for offline event reconstruction, and thus reduces the required offline compute resources.
In summary, the HLT tasks are online reconstruction, calibration, quality monitoring and data compression.

\begin{figure*}[t]
\begin{centering}
\includegraphics[width=0.95\textwidth]{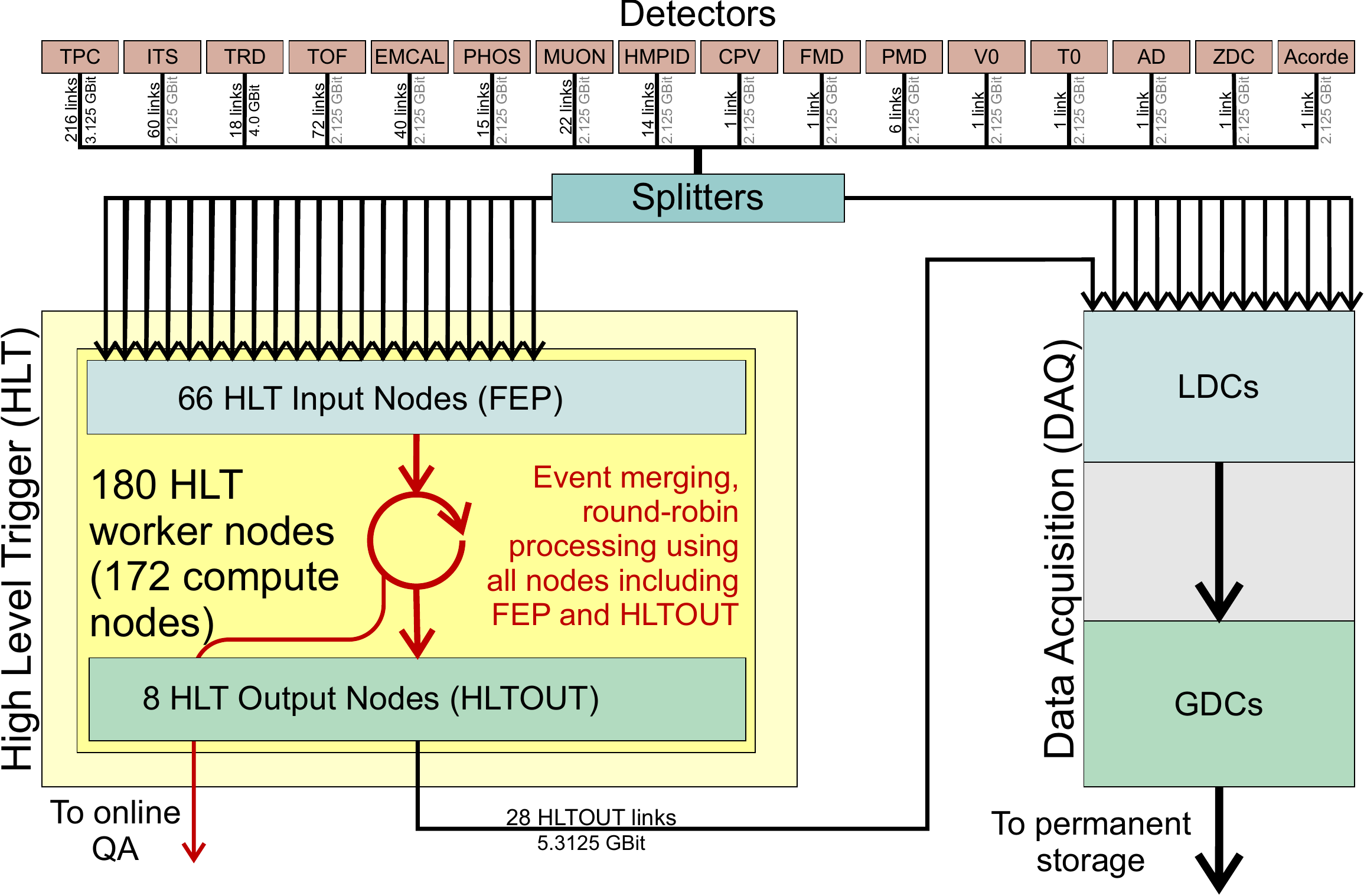}
\par\end{centering}
\caption{
The \mbox{ALICE} HLT in the data readout scheme during \run{2}.
 In the DAQ system the data flow through the local and the global data concentrators, LDC and GDC, respectively.
 In parallel, HLT ships QA and calibration data via dedicated interfaces.
}
\label{fig:overview}
\end{figure*}

The HLT is a compute farm composed of~$180$~worker nodes and~$8$~infrastructure nodes.
It receives an exact copy of all the data from the detector links.
After processing the data, the HLT sends its reconstruction output to the Data Acquisition (DAQ) via dedicated optical output links.
Output channels to other systems for QA histograms, calibration objects, etc., are described later in this paper.
In addition the HLT sends a trigger decision.
The decision contains a readout list, which specifies the output links that are to be stored and are to be discarded by DAQ.
A collision event is fully accepted if all detector links are allowed to store data and rejected if the decision is negative for all links.
Data on some links may be replaced by issuing a negative decision for those links and injecting (reconstructed) HLT data instead.
DAQ buffers all the event fragments locally and waits for the readout decision from the HLT, which has an average delay of 2--4 seconds for Pb--Pb data, while in rare cases the maximum delay reaches 10 seconds.
Then, DAQ builds the events using only the fraction of the links accepted by the HLT plus the HLT payloads and moves the events first to temporary storage and later to permanent storage.
\figur{overview} illustrates how the HLT is integrated in the \mbox{ALICE} data readout scheme.

The compute nodes use off-the-shelf components except for the Read Out Receiver Card (RORC - outlined in Section~\ref{sec:rorc}), which is a custom FPGA-based card developed for \run{1} and \run{2}.
During LHC \run{1} the HLT farm consisted of~$248$ servers including $117$ dedicated Front-End Processor (FEP) nodes equipped with RORCs for receiving data from the detectors and sending data to DAQ.
The remaining servers were standard compute nodes with two processors each, employing AMD Magny-Cours twelve-core CPUs and Intel Nehalem Quad-core CPUs.
A subset of~$64$ compute nodes was equipped with NVIDIA Fermi GPUs as hardware accelerators for track reconstruction, described in Section~\ref{sec:tracking}.
In addition, there were around~$20$ infrastructure nodes for provisioning, storage, database service and monitoring.
Two independent networks connected the cluster: a gigabit Ethernet network for management and a fast fat-tree InfiniBand QDR~$40$\,GBit network for data processing.
Remote management of the compute nodes was realized via the custom developed FPGA-based CHARM card \cite{charmcard} that emulates and forwards a VGA interface, as well as the BMC (Board Management Controller) iKVM (Keyboard, Video, Mouse over IP) available as IPMI (Intelligent Platform Management Interface) standard in new compute nodes \cite{ipmi}.

\begin{table}[htb]
\begin{center}
\caption{Overview of the HLT \run{1} and \run{2} production clusters.}
\scalebox{0.85}{
\begin{tabular}{lrr}
\hline
 & \run{1} farm & \run{2} farm \\
\hline
CPU cores & Opteron / Xeon & Xeon E5-2697 \\
& \llap{2784 }cores, up to 2.27\,GHz & 4480 cores, 2.7\,GHz \\
GPUs & \llap{64 }$\times$ GeForce GTX480 & 180 $\times$ FirePro S9000 \\
Total memory & 6.1\,TB & 23.1\,TB \\
Total nodes & 248 & 188 \\
Infrastructure nodes & 22 & 8 \\
Worker nodes & 226 & 180 \\
Compute nodes (CN) & 95 & 172 \\
Input nodes & 117 & \textit{(subset of CNs)} 66 \\
Output nodes & 14 & 8 \\
Bandwidth to DAQ & 5 GB/s & 12 GB/s \\
Max. input bandwidth & 25 GB/s & 48 GB/s \\
Detector links & 452 & 473 \\
Output links & 28 & 28 \\
\hline
RORC type & H-RORC & C-RORC \\
Host interface & PCI-X & PCI-Express \\
Max. PCI bandwidth & 940\,MB/s & 3.6\,GB/s \\
Optical links & 2 & 12 \\
Max. link bandwidth & $2.125$\,Gbps & $5.3125$\,Gbps \\
Clock frequency & 133.3\,MHz & 312.5\,MHz \\
On-board memory & 128\,MB & up to 16\,GB \\
\hline
\end{tabular}}
\label{tab:hltfarm}
\end{center}
\end{table}

In 2014, a new HLT cluster was installed for \run{2} replacing the older servers, in particular the \run{1} FEP nodes, which were operational since~2008, during system commissioning.
The availability of modern hardware, specifically the faster PCI Express interface and network interconnect, allowed for a consolidation of the different server types.
The \run{2} HLT employs~$188$~ASUS ESC4000 G2S servers with two twelve-core Intel Xeon IvyBridge E5-2697 CPUs running at 2.7\,GHz and one AMD S9000 GPU each.
In order to exclude possible compatibility problems before purchase, a full HLT processing chain was stress tested on the SANAM ~\cite{bib:sanampaper} compute cluster at the GSI Helmholtz Centre for Heavy-Ion Research using almost identical hardware.
The front-end and output functionality was integrated into $66$ input nodes and $8$ output nodes, where the input nodes serve also as compute nodes.
They were equipped with RORCs for input and output allowing for a better overall resource utilization of the processors, while the infrastructure nodes of the same server type were kept separate.
This reduction in the total number of servers also reduced the required rack-space and number of network switches and cables.
Furthermore, the fast network was upgraded to~$56$\,GBit FDR InfiniBand.
Table~\ref{tab:hltfarm} gives an overview of the \run{1} and \run{2} computing farms.

Considering the requirement of high reliability, which is driven among other things by the operating cost of the LHC, a fundamental design criterion is the robustness of the overall system with regard to component failure.
Therefore, all the infrastructure nodes are duplicated in a cold-failover configuration.
The workload is distributed in a round-robin fashion among all compute nodes, so that if one pure compute node fails it can easily be excluded from the data-taking period.
Potentially the failover requires a reboot and a restart of the \mbox{ALICE} data taking.
This scenario only takes a few minutes, which is acceptable given the low failure rate of the system; for instance, there were only~9 node failures in 1409 hours of operation during~2016.
A more severe problem would be the failure of an input node, because in that case the HLT is unable to receive data from several optical links.
Even though there are spare servers and spare RORCs, manual intervention is needed to reconnect the fibers if the FEP node cannot be switched on remotely.
However, this scenario occurred only twice in all the years of HLT operation (from 2009 to~2017).
Since the start of \run{2}, the entire production cluster is connected to an online uninterruptible power supply.

Since the installation of the \run{2} compute farm, parts of the former compute infrastructure are reused as a development cluster, to allow for software development and realistic scale testing without disrupting the data taking activities.
Additionally, the development cluster is used as an opportunistic GRID compute resource (see Section~\ref{sec:griduse}) and an integration cluster for the \mbox{ALICE} Online-Offline ($\text{O}^2$) computing upgrade foreseen for \run{3} ~\cite{bib:o2}.
The $\text{O}^2$ project includes upgrades to the \mbox{ALICE} computing model, a software framework that integrates the online and offline data processing, and the construction of a new computing facility.

\subsection{The Common Read-Out Receiver Card}
\label{sec:rorc}

\begin{figure}[ht]
  \centering
  \includegraphics[width=0.8\textwidth]{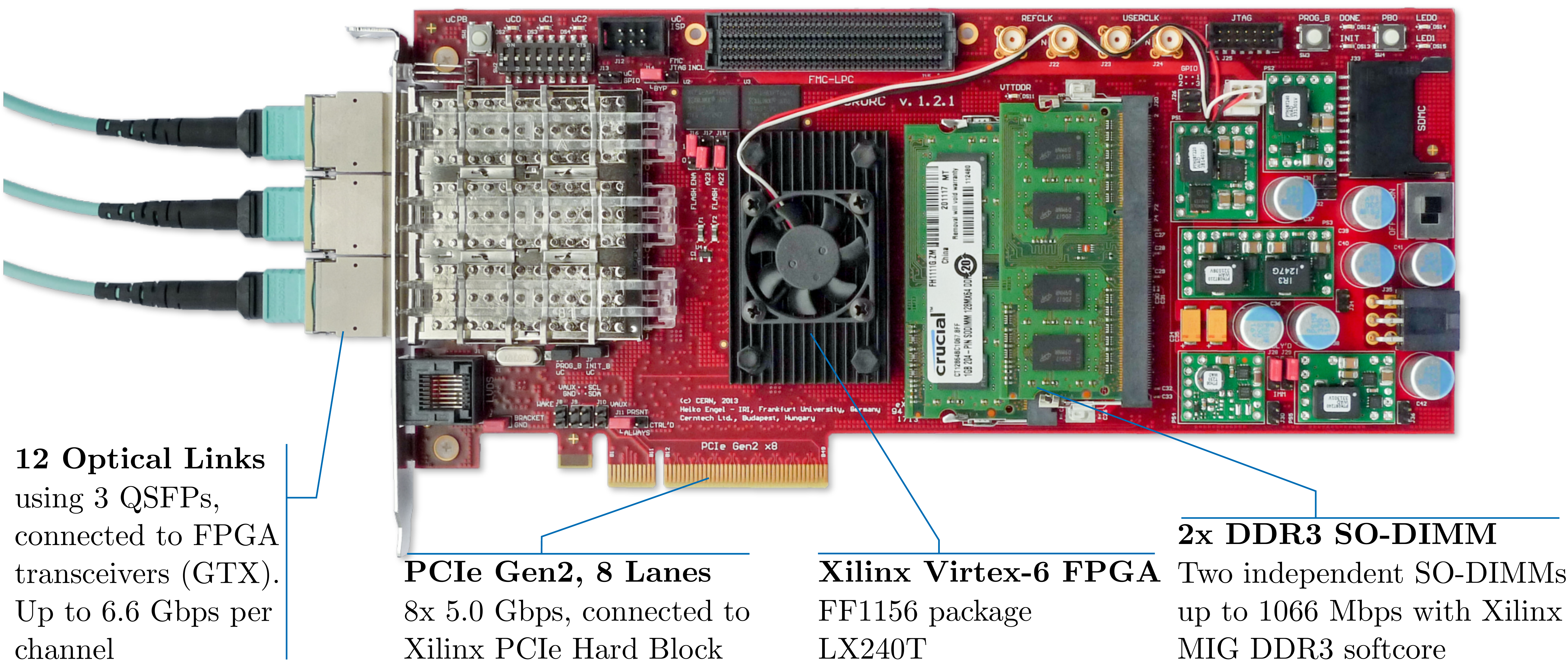}
  \caption{The Common Read-Out Receiver Card.}
  \label{fig:rorc}
\end{figure}

The Read-Out Receiver Card (\mbox{RORC}) is the main input and output interface of the HLT for detector data.
It is an FPGA-based server plug-in board that connects the optical detector links to the HLT cluster and serves as the first data processing stage.
During \run{1} this functionality was provided by the HLT-dedicated RORC (\mbox{H-RORC}) \cite{bib:torsten_diss}, a PCI-X based FPGA board that connects to up to two optical detector links at 2.125\,Gbps.
The need for higher link rates, the lack of the PCI-X interface on recent server PCs, as well as the limited processing capabilities of the H-RORC with respect to the \run{2} data rates required a new RORC for \run{2}.
None of the commercially available boards were able to provide the required functionality, which led to the development of the Common Read-Out Receiver Card (\mbox{C-RORC}) as a custom readout board for \run{2}.
The hardware was developed in order to enable the readout of detectors at higher link speeds, extend the hardware-based online processing of detector data, and provide state-of-the-art interfaces with a common hardware platform.
Additionally, technological advancements enabled a factor six higher link density per board and therefore reduced the number of boards required for the same amount of optical links compared to the previous generation of RORCs.
One HLT \mbox{C-RORC} receives up to~12 links.
A photograph of the board is shown in \fig{rorc}.
The \mbox{C-RORC} has been part of the production systems of \mbox{ALICE} DAQ, \mbox{ALICE} HLT and ATLAS trigger and data acquisition since the start of \run{2} \cite{c-rorc}.
The FPGA handles the data stream from the links and directly writes the data into the RAM of the host machine using Direct Memory Access (DMA).
A minimal kernel adapter in combination with a user space device driver based on the Portable Driver Architecture (PDA) \cite{eschweiler:2014:pda} provides buffer management, flow control, and user-space access to the data on the host side.
A custom DMA engine in the firmware enables a throughput of 3.6\,GB/s from device to host.
This is enough to handle the maximum input bandwidth of the TPC as the biggest data contributor (1.9 GB/s per \mbox{C-RORC}), the TRD as the detector with the fastest link speed (6 links at 2.3\,GB/s per \mbox{C-RORC}), and a fully equipped \mbox{C-RORC} with 12 links at 2.125\,Gbps (2.5\,GB/s).
The \mbox{C-RORC} FPGA implements a cluster finding algorithm to process the TPC raw data at an early stage.
This algorithm is further described in Section~\ref{sec:hwcf}.
The \mbox{C-RORC} can be equipped with several GB of on-board memory, used for data replay purposes.
Generated, simulated, previously recorded, or even faulty detector data can be loaded into this on-board RAM and played back as if it were coming via the optical links.
The HLT output \mbox{FPGAs} can be configured in a way to discard data right before it would be sent back to the DAQ system.
The data replay can be operated independently from any other \mbox{ALICE} online system, detector, or LHC operational state.
In combination with a configurable replay event rate, the data replay functionality provides a powerful tool to verify, scale, and benchmark the full HLT system.
This feature is essential for the optimizations presented in Section~\ref{sec:maxrate}.
The \mbox{C-RORC}s are integrated into the HLT data transport framework as data-source components for detector data input via optical links and as sink components to provide the HLT results to the DAQ system.
The \mbox{C-RORC} FPGA firmware and its integration into the HLT is further described in \cite{bib:twepp2015}.
The data from approximately 500 links, at link rates between 2.125\,Gbps and 5.3125\,Gbps, is handled via 74 \mbox{C-RORC}s that are installed in the HLT.
\\

\subsection{Cluster commissioning, software deployment, and monitoring}

The central goal for managing the HLT cluster is automation that minimizes the need for manual interventions and guarantees that the whole cluster is in a consistent state that can be easily controlled and modified if needed.
Foreman \cite{Foreman} is used to automatize the basic installation of the servers via PXE-boot.
The operating system (OS) that is currently used on all of the servers is CERN CentOS 7.
Once the OS is installed on these servers, Puppet \cite{Puppet} controls and applies the desired configuration to each server.
Puppet efficiently integrates into Foreman and allows for servers to be organized into groups according to different roles and apply changes to multiple servers instantaneously.
With this automatized setup the complete cluster can be rebuilt, including the final configuration, in roughly three hours.
For both the production and development clusters several infrastructure servers are in place, providing different services like DNS, DHCP, NFS, databases, or private network monitoring.
Critical services are redundant to reduce the risk of cluster failure in case there is a problem with a single infrastructure server.

The monitoring of the HLT computing infrastructure is done using the open source tool Zabbix \cite{zabbix}.
It allows administrators to gather metrics, be aware of the nodes health status, and react to undesired states.
More than 100 metrics per node are being monitored, such as temperature, CPU load, network traffic, free disk space, disk-health status, and failure rate on the network fabric.
The monitoring system automatizes many tasks that would require administrators' intervention.
These preemptive measures offer the possibility to replace hardware beforehand, \ie during technical shutdowns, and to avoid failures during data taking.
HLT administrators receive a daily report of the system status and, in addition, e-mail notifications when certain metrics exceed warning thresholds.
For risky events there are automated actions in place.
For instance, several shutdown procedures are performed when the node temperature reaches critical values, in order to prevent damage to the servers.

In addition to Zabbix, \mbox{ALICE} has developed a custom distributed log collector called InfoLogger.
A parser script is employed that scans all error messages stored to the logs to find important problems in real time.
These alerts can also help the detector experts with the monitoring of their systems, including automated alarms sent via e-mail or SMS.

This configuration lowers the complexity of managing a heterogeneous system with around~$200$ nodes for a period of at least~$10$ years, reducing the number of trained on-site engineers required for operation.

\subsection{Alternative use cases of the HLT farm}
\label{sec:griduse}
\begin{figure}[htb]
	\begin{centering}
		\includegraphics[width=0.65\textwidth]{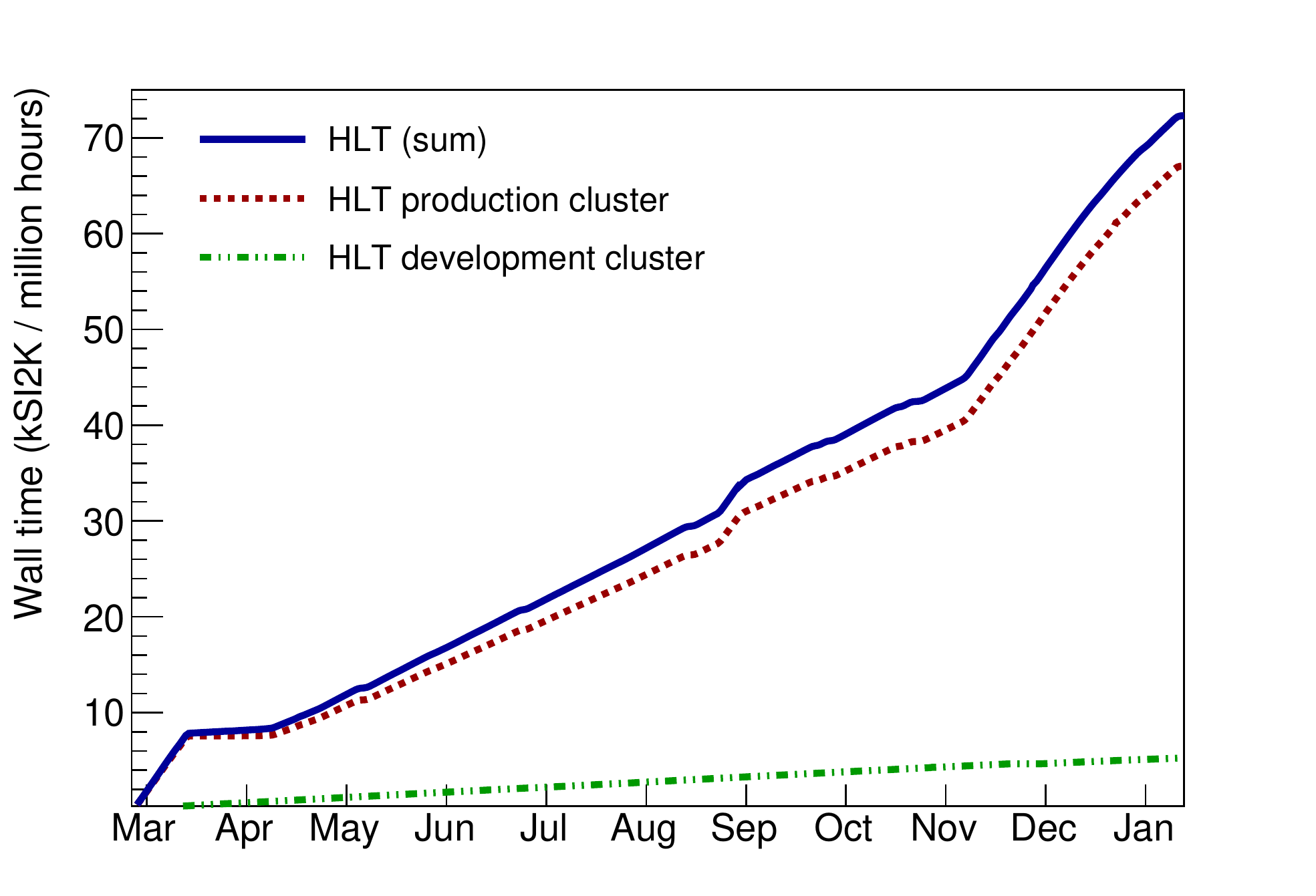}
		\par\end{centering}
	\caption{Contribution of the HLT production (dotted line) and development (double-dotted dashed line) clusters to the WLCG between March 2017 and January 2018, with the sum of both contributions shown as the solid line.}
	\label{fig:grid}
\end{figure}

In order to maximize the usage of the servers during times when there are no collisions, a Worldwide LHC Computing Grid (WLCG)~\cite{Bird:2014ctt} configuration was developed for the cluster in cooperation with the \mbox{ALICE} offline team.
The first WLCG setup used OpenStack~\cite{openstack} Virtual Machines (VM) to produce \mbox{ALICE} Monte Carlo (MC) simulations of particle collision events.
In 2017, the WLCG setup was improved to use Docker~\cite{docker} containers instead of OpenStack VMs, which allows for more flexibility and therefore improves efficiency with the available resources.
The containers are spawned for just one job and destroyed after the job finishes.
During pp data taking a part of the production cluster is contributed to the WLCG setup.
During phases without data taking, like LHC year-end shutdowns and technical stops, the whole HLT production cluster is operated as a WLCG site as long as it is not needed for tests of the HLT system.
\figur{grid} shows the aggregated wall time of the new Docker setup from March 2017 onward.
The steeper slope represents periods when the complete cluster is assigned to WLCG operation, while the plateau indicates a phase of full scale framework testing.
The HLT production cluster provides a contribution to the \mbox{ALICE} MC simulation compute time with this opportunistic use on a best-effort basis.
The WLCG setup of the HLT focuses on MC simulations because these require less storage and network resources than general \mbox{ALICE} Grid jobs and are thus ideally suited for opportunistic operation without side effects.

The HLT development cluster, introduced in Section~\ref{sec:general}, is composed of approximately 80 older servers.
Not only does it allow for ongoing development of the current framework, of which runs on the production cluster, but it can also used for tests of the future framework for \run{3}.
During periods when no development is taking place, 60 of the nodes act as second WLCG site, in addition to the opportunistic use of the production cluster, donating the compute resources to \mbox{ALICE} MC jobs.
To guarantee that there is no interference with data taking, the HLT development cluster is completely separated from the production environment.
The development cluster is installed in different racks and also uses a different private network, which has no direct connection to the production cluster.
For WLCG operation, the HLT internal networks and the network used for WLCG communication were completely separated via VLANs configured at switch level.

\subsection{HLT architecture and data transport software framework}
\label{sec:framework}

In order to transform the raw detector signals into physical properties all ALICE detectors have developed reconstruction software, like TPC cluster finding (Section~\ref{sec:hwcf}) and track finding (Section~\ref{sec:tracking}) algorithms.
In the HLT the data processing is arranged in a pipelined data-push architecture.
The reconstruction process starts with local clusterization of the digitized data, continues with track finding for individual detectors, and ends with the creation of the Event Summary Data (ESD).
The ESD is a complex ROOT~\cite{bib:root} data structure that holds all of the reconstruction information for each event.

In addition to the core framework described in the this section, a variety of interfaces exist to other \mbox{ALICE} subsystems~\cite{bib:hlt1}.
These include the command and control interface to the Experiment Control System (ECS), the Shuttle system used for storing calibration objects for offline use, the optical links to DAQ, the online event display, and Data Quality Monitoring (DQM) for online visualization of QA histograms.

\begin{figure*}[htb]
\begin{centering}
\includegraphics[width=0.95\textwidth]{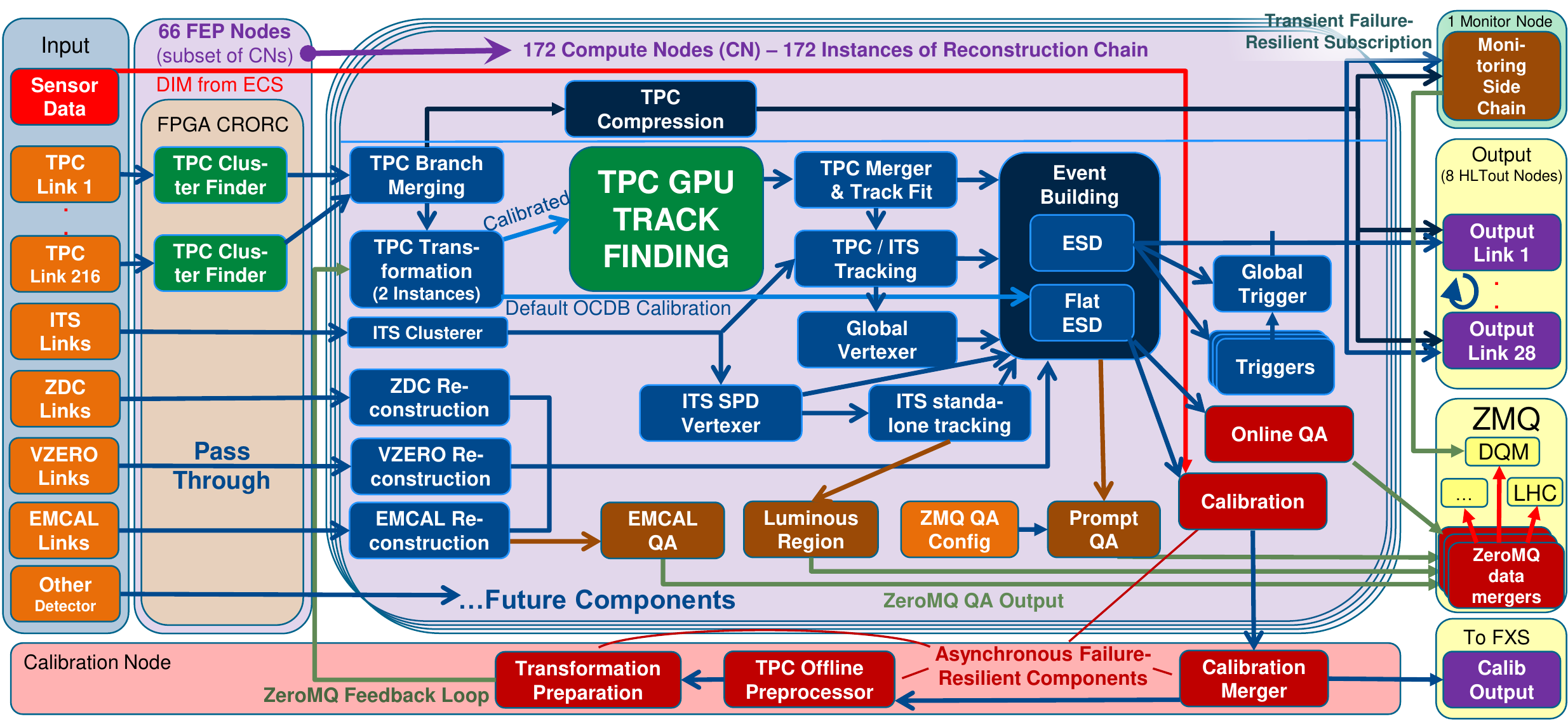}
\par\end{centering}
\caption{Schema of the HLT components.
  The colored boxes represent processes accelerated by GPU/FPGA (green), normal processes (blue), processes that produced HLT output that is stored (dark blue), entities that store data (purple), asynchronous failure-resilient processes (dark red), classical QA components that use the original HLT data flow (brown), input (orange), and sensor data (red).
  Incoming data are passed through by the \mbox{C-RORC} FPGA cards or processed internally.
  The input nodes locally merge data from all links belonging to one event.
  The compute nodes then merge all fragments belonging to one event and run the reconstruction.
  The bottom of the diagram shows the asynchronous online calibration chain with a feedback loop as described in Section~\ref{sec:calibration}.
 }
\label{fig:components}
\end{figure*}

The \mbox{ALICE} HLT uses a modular software framework consisting of separate components, which communicate via a standardized publisher-subscriber interface designed to cause minimal overhead for data transport~\cite{framework1,framework2}.
Such components can be data sources that feed into the HLT processing chain, either from the detector link or from other sources like TPC temperature and pressure sensors.
Data sinks extract data from the processing chain and send the reconstructed event and trigger decision to DAQ via the output links.
Other sinks ship calibration objects or QA histograms, which are stored or visualized.
In addition to source and sink components, analysis or worker components perform the main computational tasks in such a processing chain and are arranged in a pipelined hierarchy.
\figur{components} gives an overview of the data flow of the most relevant components currently running in the HLT.
A component reads a data set (if it is not a source), processes it, creates the output and proceeds to the next data set.
Although each component processes only one event at a time, the framework pipelines the events such that thousands of events can be either in-chain in the cluster or also on a single server.
Merging of event fragments, scattering of events among multiple compute nodes for load balancing, and network transfer are all handled via special processing components provided by the framework and are transparent to the worker processes.
Components situated on the same compute node pass data via a shared-memory based zero-copy scheme.
With respect to \run{1} the framework underwent a revision of the interprocess-scheduling approach.
The old approach, using POSIX pipes, began to cause a significant CPU load through many system calls and was consequently replaced by a shared-memory based communication.

Presently, the user simply defines the processing chain with reconstruction, monitoring, calibration, and other processing components.
The user also defines the inputs for all components as well as the output at the end of the processing chain.
The full chain is started automatically and distributed in the cluster.
The processing configuration can be annotated with hints to guide the scheduling.
In order to minimize the data transfer, the chain usually starts with local processing components on the front-end nodes (like the TPC cluster finder presented in Section~\ref{sec:hwcf}).
In the end, after the local steps have reduced the data volume, all required event fragments are merged on one compute node for the global event reconstruction.

The data transport framework is based on three pillars.
There is a primary reconstruction chain which processes all the recorded events in an event-synchronous fashion.
It performs the main reconstruction and data compression tasks and is responsible for receiving and sending data.
This main chain is the backbone of the HLT event reconstruction and its stability is paramount for the data taking efficiency of \mbox{ALICE}.

The second pillar is the data monitoring side chains, which run in parallel at low rates on the compute nodes.
These subscribe transiently to the output of a component of the main chain.
In this way, the side chains cannot break or stall the HLT main chain.

For \run{2} a third pillar was added, based on Zero-MQ (Zero Message Queue) message transfer \cite{bib:chep2016hlt}, which provides similar features compared to the main chain but runs asynchronously.
Currently, it is used for the monitoring and calibration tasks and does not merge fragments of one event but instead it is fed with fully reconstructed events from the main chain.
It processes as many events as possible on a best-effort basis, skipping events when necessary.
Results of the distributed components are merged periodically to combine statistics processed by each instance.
The same Zero-MQ transport is also used as an interface to DQM and as external interface which allows detector experts to query merged results of QA components running in the HLT.

The transport framework is not restricted to closed networks or computing clusters.
A proof-of-principle test of the framework used locally in the HLT cluster deploys a global processing chain for a Grid-like real-time data processing.
This framework was distributed on a North-South axis between Cape Town in South Africa and Troms\o{} in northern Norway, with Bergen (Norway), Heidelberg (Germany), and Dubna (Russia) as additional participating sites~\cite{hltglobal}.
The concepts developed for the HLT are the basis for the new framework of the \mbox{ALICE} $\text{O}^2$ computing upgrade.

\subsection{Fault tolerance and dynamic reconfiguration}

Robustness of the main reconstruction chain is the most important aspect from the point of view of data taking efficiency.
Therefore, the HLT was designed with several failure resiliency features.
All infrastructure services run on two redundant servers and compute node failures can be easily compensated for.
Experimental and non-critical components can run in a side-chain or asynchronously via Zero-MQ, separate from the main chain.

Also the main chain itself has several fault tolerance features.
Some components use code from offline reconstruction, or code written by the teams responsible for certain detector development, and hence they are not developed considering the high-reliability requirements of the HLT.
Nevertheless, the HLT must still ensure stable operation in case of critical errors like segmentation faults.
Thus, all components run in different processes, which are isolated from each other by the operating system.
In case one component fails, the HLT framework can transparently cease the processing of that component for a short time, and then later restart the component.
Although the event is still processed, the result of that particular component for this event and possibly several following events are lost.
This loss of a single instance causes only a marginal loss of information.

\section{Fast algorithms for fast computers}

Since the TPC produces~$91.1$\% (Pb--Pb) and $95.3$\% (pp) of the data volume\footnote{Values from the 8 kHz Pb--Pb and 200 kHz pp data taking runs of 2015} and, also because of the sheer data volume, event reconstruction of the TPC data including clusterizing and tracking is the most compute intensive task of the HLT.
This makes the TPC the central detector for the HLT.
Its raw data are the most worthwhile target for data compression algorithms.
Since a majority of the compute cycles are spent processing TPC data, it is mandatory that the TPC reconstruction code is highly efficient.
It is the TPC reconstruction that leverages the compute potential of both the FPGA and GPU hardware accelerators in the HLT.
Furthermore, since it is an ionization detector, TPC calibration is both challenging and essential.

Here, a selection of important HLT components, following the processing of the TPC data in the chain is described.
The processing of the TPC data starts with the clusterization of the raw data, which happens in a streaming fashion in the FPGA while the data are received at the full optical speed.
Two independent branches follow, where one component compresses the TPC clusters and replaces the TPC raw data with compressed HLT data.
The second branch starts with the TPC track reconstruction using GPUs, continues with the creation of the ESD, and runs the TPC calibration and QA components.

\subsection{Driving forces of information science}

The design of the \mbox{ALICE} detector dates back two decades.
At that time, the LHC computing needs could not be fulfilled based on existing technology but relied on extrapolations according to Moore's Law~\cite{moore}.
Indeed the performance of computers has improved by more than three orders of magnitude since then, but the development of microelectronics has reached physical limits in recent years.
For example, processor clock rates have not increased significantly since~2004.
To increase computing power various levels of parallelization are implemented, such as the use of multi- or many-core processors, or by supporting SIMD (Single Instruction, Multiple Data) vector-instructions.
At this point in time computers do not become faster for single threads but they can become more powerful if parallelism is exploited.
Although these developments were only partially foreseeable at the beginning of the \mbox{ALICE} construction phase, they have been taken into account for the realization of the HLT.

\subsection{Fast FPGA cluster finder for the TPC}

\label{sec:hwcf}

\begin{figure}[htb]
\begin{centering}
\includegraphics[width=0.4\textwidth]{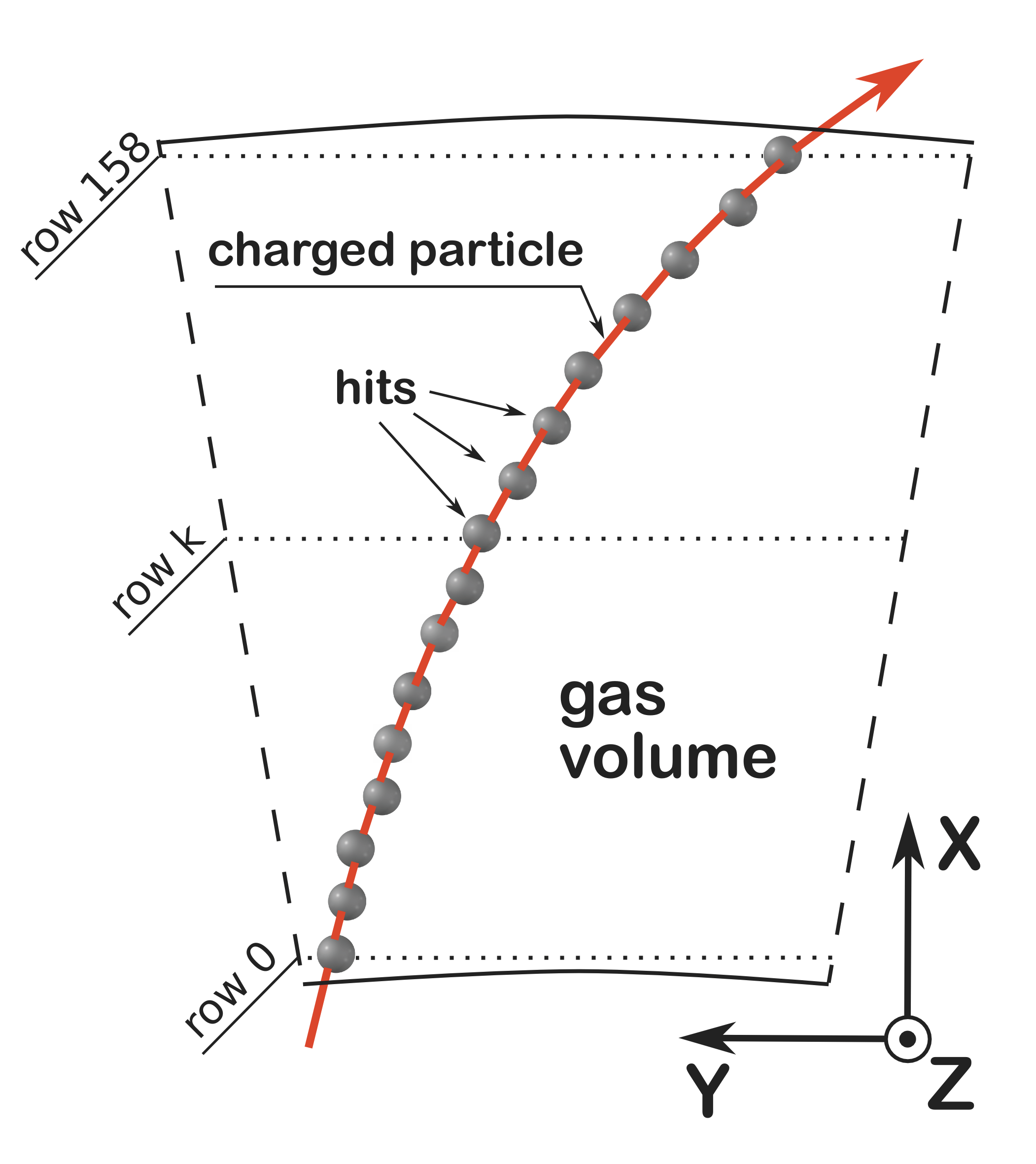}
\par\end{centering}
\caption{Schematic representation of the geometry of a TPC sector.
  Local $y$ and $z$ coordinates of a charged-particle trajectory are measured at certain $x$ positions of 159 readout rows, providing a chain of spatial points (hits) along its trajectory.
}
\label{fig:slice}
\end{figure}

At the beginning of the reconstruction process the so-called clusters of locally adjacent signals in the TPC have to be found.
\figur{slice} shows a schematic representation of a cross-section of a trapezoidal TPC sector, where the local coordinate system is such that in the middle of the sector the $x$-axis points away from the interaction point.
One can imagine a stack of~2D pad-time planes ($y$-$z$ plane in \fig{slice}) in which a charged particle traversing the detector creates several neighboring signals in each~2D plane.
The exact position of the intersection between the charged-particle trajectory and the~2D plane can be calculated by using the weighted mean of the signals in the plane, \ie by determining their center of gravity.
The HLT cluster-finder algorithm can be broken down into three separate steps.
Firstly, the relevant signals have to be extracted from raw data and the calibration factors are applied.
Next, neighboring signals and charge peaks in time-direction are identified and the center of gravity is calculated.
Finally, neighboring signals in the TPC pad-row direction ($x$-$y$ plane) are merged to form a cluster.
These reconstructed clusters are then passed on to the subsequent reconstruction steps, such as the track finding described in Section~\ref{sec:tracking}.

By design, the TPC cluster-finder algorithm is ideally suited for the implementation inside an FPGA~\cite{fpga2}, which supports small, independent and fast local memories and massively parallel computing elements.
The three processing steps are mutually independent and are correspondingly implemented as a pipeline, using fast local memories as de-randomizing interfaces between these stages.
In order to achieve the necessary pipeline throughput, each pipeline stage implements multiple custom designed arithmetic cores.
The FPGA based RORCs are required as an interface of the HLT farm to the optical links.
By placing the online processing of the TPC data in the FPGA, the data can be processed on-the-fly.
The hardware cluster finder is designed to handle the data bandwidth of the optical link.
Finally, a compute node receives the TPC clusters, computed in the FPGA, directly into its main memory.

An offline reference implementation of the cluster finding exists but is far too slow to be implemented online.
Rather, the offline cluster finder is used as a reference for both the physics performance and the processing speed.
In comparison to the hardware cluster finder executed on the FPGA, it performs additional and more complex tasks.
These include checking TPC readout pads for baseline shifts and, if present, applying corrections and deconvoluting overlapping clusters using a Gaussian fit to the cluster shapes, which are simply split in the hardware version.
Additional effects such as missing charge in the gaps between TPC sectors and malfunctioning TPC channels are considered.
Finally, after the application of the drift-velocity calibration, cluster positions are transformed into the spatial $x$, $y$, and $z$ coordinate system.
In the HLT, a separate transformation component performs this spatial transformation as a later step.
The evaluation in Section~\ref{sec:hwcfperf} demonstrates that the HLT hardware cluster finder delivers a performance comparable to the offline cluster finder.

\begin{figure}[ht]
  \centering
  \includegraphics[width=0.8\textwidth]{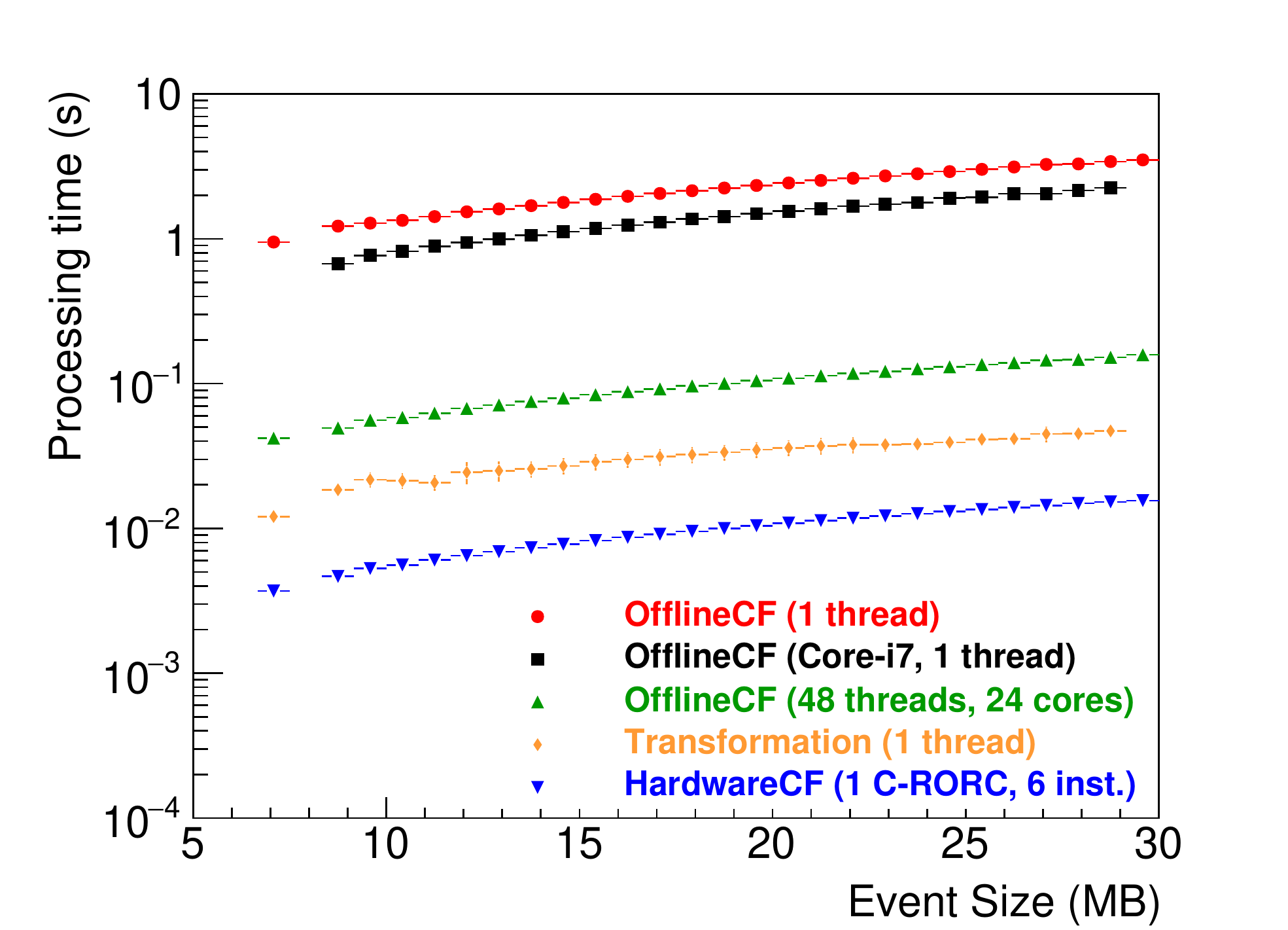}
  \caption{Processing time of the hardware cluster
    finder and the offline cluster finder.
    The measurements were performed on an HLT node (circles, triangles, diamonds), a newer Core-i7 6700K CPU (squares), and on the \mbox{C-RORC} (inverted triangles).
  }
  \label{fig:hwcf_vs_offline}
\end{figure}

Benchmarks have shown that one \mbox{C-RORC} with six hardware cluster finder (HardwareCF) instances is about a factor~$10$ faster than the offline cluster finder (OfflineCF) using~48 threads on an HLT node, as shown in \fig{hwcf_vs_offline}.
The software processing time measurements were done on a HLT node with dual Xeon E5-2697 CPUs for the single-threaded variant, the multi-threaded variant as well as the cluster transformation component.
The single-threaded variant was also evaluated on a Core-i7 6700k CPU to show the performance improvements of using the same implementation on a newer CPU architecture.
The measurements were also performed on the \mbox{C-RORC}.

Several factors increase the load on the hardware cluster finder in \run{2}.
The \mbox{C-RORC} receives more links than the former \mbox{H-RORC} of \run{1}, with the FPGA implementing six instead of the previously two instances of the cluster finder.
The TPC RCU2 sends the data at a higher rate, up to~$3.125$\,Gbps.
In addition, during 2015 and 2016, the TPC was operated with argon gas instead of neon yielding a higher gain factor, which resulted in a higher probability of noise over the zero-suppression threshold.
In this situation, the cluster finder detects a larger number of clusters, though a significantly large fraction of these are fake.
In addition, the readout scheme of the RCU2 was improved, disproportionately increasing the data rate sent to the HLT compared to the link speed, yielding a net increase of a factor of 2.
These modifications also required the clock frequency of the hardware cluster finder to be disproportionately scaled up compared to the link rate in order to cope with the input data rates.
Major portions of the online cluster finder were adjusted, further pipelined, and partly rewritten to achieve the required clock frequency and throughput.
The peak-finding step of the algorithm was replaced with an improved version more resilient to noise.
This filtering reduces the number of noise induced clusters found, relaxes the load on the merging stage, and thus reduces the cluster finder output data size.
The reduced output size, in combination with improvements to the software based data compression scheme, increases the overall data compression factor of the HLT (see Section~\ref{sec:compression}).

\subsubsection{Physics performance of the HLT cluster finder for the TPC}

\label{sec:hwcfperf}

In order to reduce the amount of data stored on tape, the TPC raw data are replaced by clusters reconstructed in the HLT.
The cluster-finder algorithm must be proven not to cause any significant degradation to the physical accuracy of the data.
The offline track reconstruction algorithm was improved by better taking into account the slightly different behavior of the HLT cluster finder and its center of gravity approach compared to the offline cluster finder.
The performance of the algorithm has been evaluated by looking at the charged-particle tracks reconstructed with the improved version of the offline track-reconstruction algorithm, described in Section~\ref{sec:tracking}.

\begin{figure}[htb]
\begin{centering}
\includegraphics[width=0.8\textwidth]{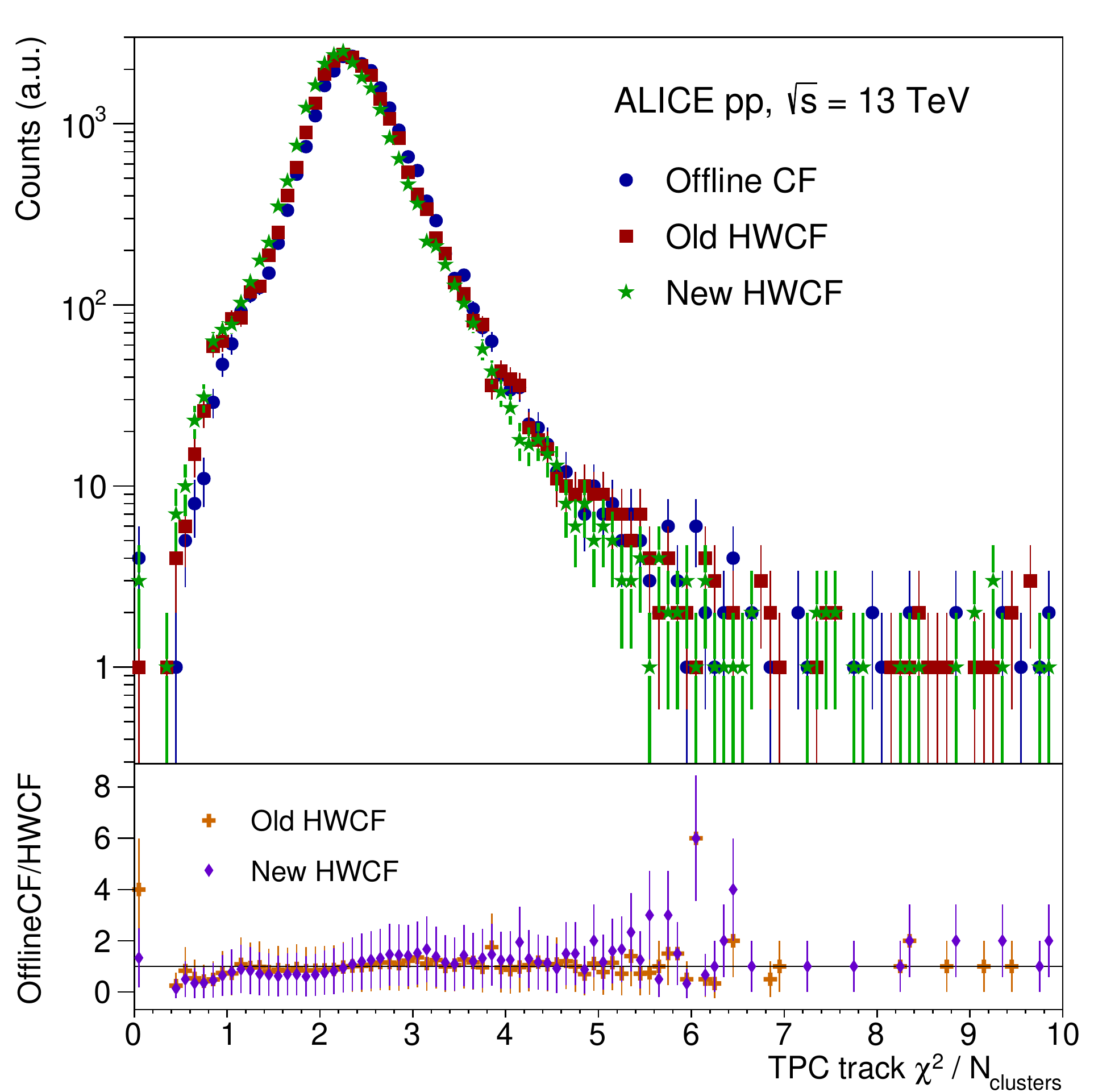}
\par\end{centering}
\caption{The upper panel shows the distribution of TPC track~$\chi^2$ residuals from offline track reconstruction obtained using total cluster charges from offline cluster finder (Offline CF) and different versions of the HLT hardware cluster finder (HWCF).
 Tracks, reconstructed using the TPC and ITS points, satisfy the following selection criteria: pseudorapidity $|\eta| < 0.8$ and $N_{\text{TPC clusters}} \geq 70$.
 The ratios of the distributions obtained using the offline cluster finder and the HLT cluster finder are shown in the lower panel.}
\label{fig:tpc-track-chi2}
\end{figure}

The important properties of the clusters are the spatial position, the width, and the charge deposited by the traversing particle.
\figur{tpc-track-chi2} compares the~$\chi^2$ distribution of TPC tracks reconstructed by the offline tracking algorithm using TPC clusters produced using either the HLT hardware cluster finder or the offline version.
Since the cluster errors coming from a fit to the track are parameterized and not derived from the width of the cluster, the~$\chi^2$ distribution is proportional to the average cluster-to-track residual.
On a more global level, the cluster positions in the ITS are used to evaluate the track resolution of the TPC.
The TPC track is propagated through the ITS volume and the probability of finding matching ITS spatial points is analyzed.
Since the ITS cluster position is very precise it is a good metric for TPC track quality.
However, because the occupancy for heavy-ion collisions is high, the matching requires an accurate position of the TPC track with a good transverse momentum (\pt{}) fit for precise extrapolation.
It was found that there are no significant differences in track resolution and $\chi^2$ between the offline cluster finder and the new HLT cluster cluster finder, with the old HLT hardware cluster finder yielding a slightly worse result.

\begin{figure}[!t]
\begin{centering}
\includegraphics[width=0.8\textwidth]{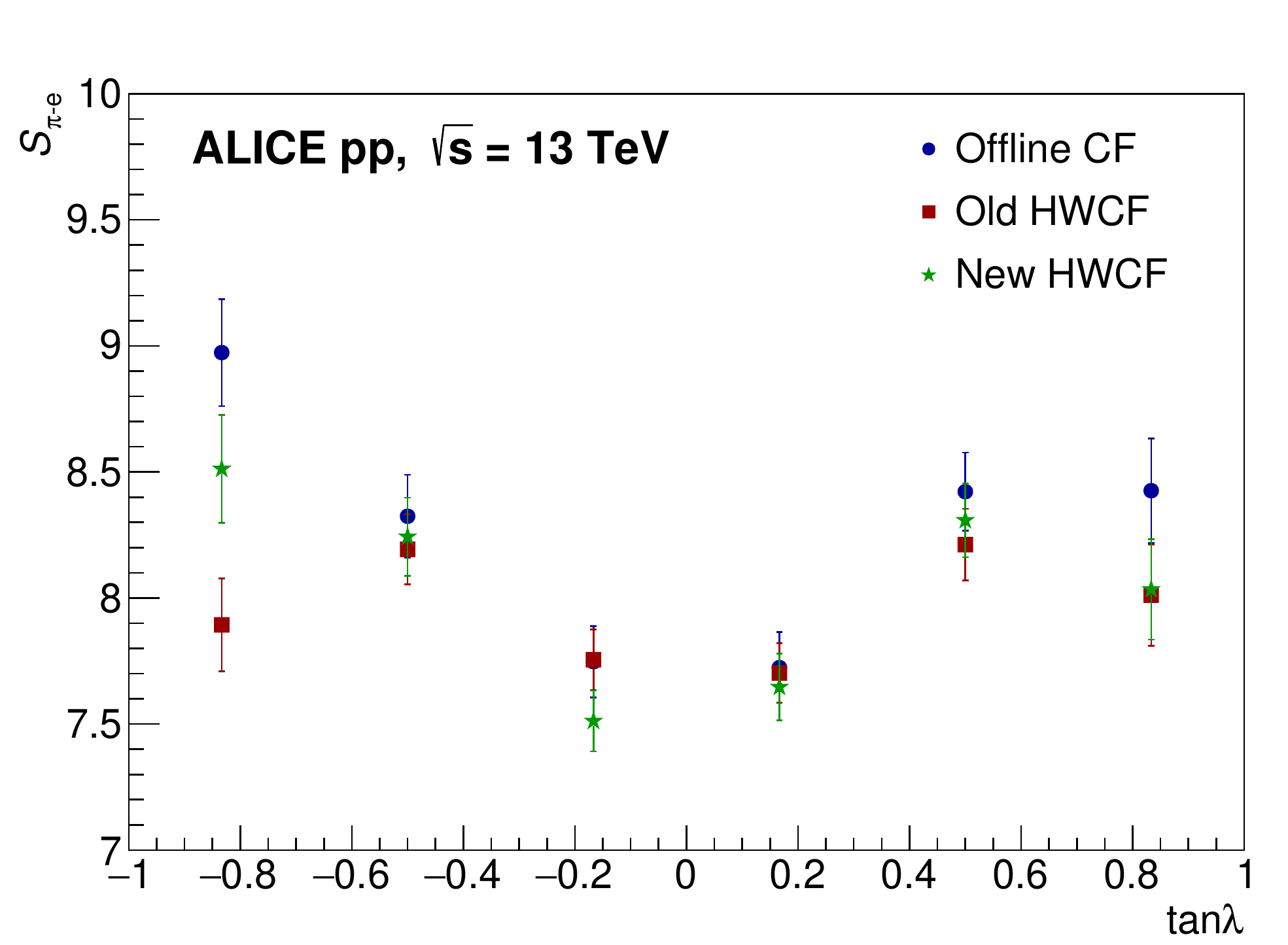}
\par\end{centering}
  \caption{Separation power ($S_{\pi-e}$) of pions and electrons (minimum ionizing particles, \ie pions at 0.3 to 0.6 GeV/$c$ versus electrons from gamma conversions at 0.35 to 0.5 GeV/$c$) as a function of the track momentum dip angle, where $\tan\mathrm{\lambda} = p_{\mathrm{z}}/p_{\mathrm{T}}$.
  Comparison of d$E$/d$x$ separation power using total cluster charges from Offline CF and different versions of the HWCF.
}
\label{fig:tpc-dedx-separation-power}
\end{figure}

\figur{tpc-dedx-separation-power} shows the d$E$/d$x$ separation power as a measure of the quality of the HLT cluster charge reconstruction.
Here, the separation power is defined as the d$E$/d$x$ separation between the pions and electrons scaled by the resolution.
Since the d$E$/d$x$ is calculated from the cluster charge, an imprecise charge information would deteriorate the d$E$/d$x$ resolution and  consequently separation power.
Within the statistical uncertainty no substantial difference is observed between the offline and hardware cluster-finder algorithms.

\subsection{Track reconstruction in the TPC}

\label{sec:tracking}

In \mbox{ALICE} there are two different TPC-track reconstruction algorithms. One is employed for offline track reconstruction and the other is the HLT track reconstruction algorithm. In this section, the HLT algorithm is described and its performance compared to that of the offline algorithm.

In the HLT, following the cluster finder step, the reconstruction of the trajectories of the charged particles traversing the TPC is performed in real time.
The \mbox{ALICE} HLT is able to process pp collisions at a rate of~$4.5$\,kHz and central heavy-ion collisions at~$950$\,Hz (see Sec.~\ref{sec:maxrate}), corresponding to a data rate of~$48$\,GB/s, which is above the maximum deliverable rate from the TPC.

The TPC track reconstruction algorithm has two steps, namely the in-sector track-segment finding within individual TPC sectors and the segment merger, which concludes with a full track refit.
The in-sector tracking is the most compute intense step of online event reconstruction, therefore it is described in more detail in the following subsection.

\subsubsection{Cellular automaton tracker}

\label{sec:ca}

Based on the cluster-finder information, clusters belonging to the same initial particle trajectory are combined to form tracks.
This combinatorial pattern recognition problem is solved by a track finder algorithm.
Since the potential number of cluster combinations is quite substantial, it is not feasible to calculate an exact solution of the problem in real time.
Therefore, heuristic methods are applied.
One key issue is the dependence of reconstruction time on the number of clusters.
Due to the large combinatorial background, \ie the large number of incorrectly combined clusters from different tracks, it is critical that the dependence is linear in order to perform online event processing.
This was achieved by developing a fast algorithm for track reconstruction based on the cellular automaton principle~\cite{bib:simdkalman,catracking} and the Kalman filter~\cite{bib:kalman} for modern processors~\cite{chep2016gpu}.
The processing time per track is~$5.4$\,$\mu$s on an AMD S9000 GPU.
The tracking time per track increases linearly with the number of tracks, and is thus independent of the detector occupancy, as shown in Sec.~\ref{sec:gpuperf}.

\begin{figure}[t]
\begin{centering}
\includegraphics[width=0.55\textwidth]{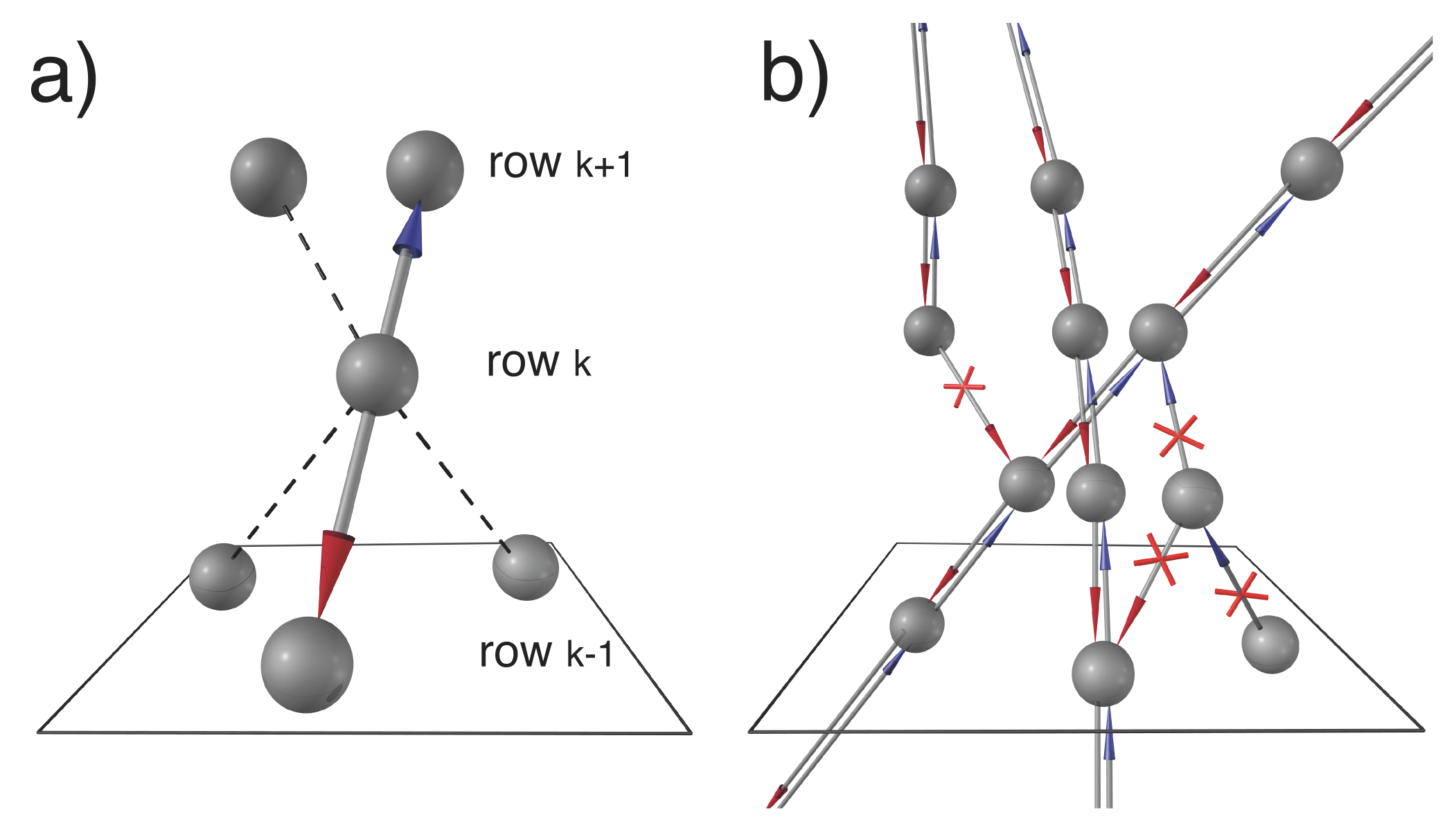}
\par\end{centering}
\caption{Cellular automaton track seeding steps.
  a) Neighbor finder. Each cluster at a row $k$ is linked to the best pair of its neighbors from the next and the previous row.
  b) Evolution step. Non-reciprocal links are removed, chains of reciprocal links define the tracklets.
}
\label{fig:tracking_ca_steps}
\end{figure}

The track finder algorithm starts with a combinatorial search of track candidates (tracklets), which is based on the cellular automaton method.
Local track segments are created from spatially adjacent clusters, eliminating non-physical cluster combinations.
In the two-stage combinatorial processing, the neighbor finder matches, for each cluster at a row~$k$, the best pair of neighboring clusters from rows~$k + 1$ and~$k - 1$, as shown in \fig{tracking_ca_steps} (left).
The neighbor selection criterion requires the cluster and its two best neighbors to form the best straight line, in addition to having a loose vertex constraint.
The links to the best two neighbors are stored.
Once the best pair of neighbors is found for each cluster, a consequent evolution step determines reciprocal links and removes all non-reciprocal links (see \fig{tracking_ca_steps}) (right).

A chain of at least two consecutive links defines a tracklet, which in turn defines the particle trajectory.
The geometrical trajectories of the tracklets are fitted with a Kalman filter.
Then, track candidates are constructed by extending the tracklets to contain clusters close to the trajectory.
A cluster may be shared among track candidates; in this case it is assigned to the candidate that best satisfies track quality criteria like the track length and $\chi^2$ of the fit.

This algorithm does not employ decision trees or multiple track hypotheses.
This simple approach is possible due to the abundance of clusters for each TPC track and it results in a linear dependence of the processing time on the number of clusters.

Following the in-sector tracking the segments found in the individual TPC sectors are merged and the final track fit is performed.
A flaw in this approach is that if an in-sector track segment is too short, \eg having on the order of 10 clusters, it might not be found by the in-sector tracking algorithm.
This is compensated for by a posterior step, that treats tracks ending at sector boundaries close to the inner or outer end of the TPC specially, by extrapolating the track through the adjacent sector, and picking up possibly missed clusters~\cite{bib:chep}.
The time overhead of this additional step is less than~$5$\% of the in-sector tracking time.

The HLT track finder demonstrates an excellent tracking efficiency, while running an order of magnitude faster than the offline finder, while also achieving comparable resolution.
Corresponding efficiency and resolution distributions extracted from Pb--Pb events are shown in Section~\ref{sec:gpuperf}.
The advantages of the HLT algorithm are a high degree of locality and the allowance of a massively parallel implementation, which is outlined in the following sections.

\subsubsection{Track reconstruction on CPUs}
\label{sec:trackingvc}

Modern CPUs provide SIMD instructions allowing for operation on vector data with a potential to speed up corresponding to the vector width (to-date a factor up to 16 is achievable with the AVX512 instruction set).
Alternatively, hardware accelerators like GPUs offer vast parallelization opportunities.
In order to leverage this potential in the track finder, all the computations are implemented as a simple succession of arithmetic operations on single precision floats.
An appropriate vector class and corresponding data structures were developed, yielding a vectorized version of the tracker that can run on both the Xeon Phi and standard CPUs using their vector instructions, or additionally in a scalar way.
Data access is the most challenging part.
The main difficulty is the fact that all tracklets have different starting rows, lengths, and number of clusters requiring random access into memory instead of vector loads.
While the optimized and vectorized version of the Kalman filter itself yielded a speedup of around~$3$ over the initial scalar version, the overall speedup was however smaller.
Therefore, the track reconstruction is performed on GPUs.
Due to the random memory access during the search phase, it is impossible to create a memory layout optimized for SIMD.
This poses a bottleneck for the GPU as well, but it is less severe due to the higher memory bandwidth and better latency hiding of the GPU.
The vector library developed in the scope of this evaluation is available as the open source Vc library~\cite{vc}.
It was integrated into ROOT and is part of the C++ Parallelism technical specification~\cite{bib:parallelism2ts}.
The optimized data layout originally developed for fast SIMD access has also proven very efficient for parallelization on GPUs.

\subsubsection{Track reconstruction on GPUs}

The alternative many-core approach using GPUs as general purpose processors is currently employed in the HLT.
All steps of the cellular automaton tracker and the Kalman filter can be distributed on many independent processors.
In order to be independent from any GPU vendor, the HLT code must not rely exclusively on a proprietary GPU programming framework.
The fact that the reconstruction code is used in the \mbox{ALICE} offline framework, AliRoot, and that it is written in C++ poses several requirements on the GPU API.
Currently, the HLT tracking can optionally use both the CUDA framework for NVIDIA GPUs or the OpenCL framework with C++ extensions for AMD GPUs.
Even though OpenCL is an open, vendor-independent framework, the current HLT code is limited to AMD because other vendors do not yet support the C++ kernel language.
C++ templates avoid code duplication for class instances residing in the different OpenCL memory scopes.
The new OpenCL 2.2 standard specifies a C++ kernel language very similar to the extension currently used, which will allow for an easy migration.
The tracking algorithm is written such that a common source file in generic C++ contains the entire algorithm representing more than~$90$\% of the code.
Small wrappers allow the execution of the code on different GPU models and also on standard processors, optionally parallelized via OpenMP.
This aids in avoiding division between GPU and CPU code bases and thus reduces the maintenance effort~\cite{bib:chep2015} since improvements to the tracking algorithm are developed only once.
All optimizations are parameterized and switchable, such that each architecture (CPU, NVIDIA GPU, AMD GPU) can use its own settings for optimum performance.

\begin{figure}[htb]
\begin{centering}
\includegraphics[width=0.8\textwidth]{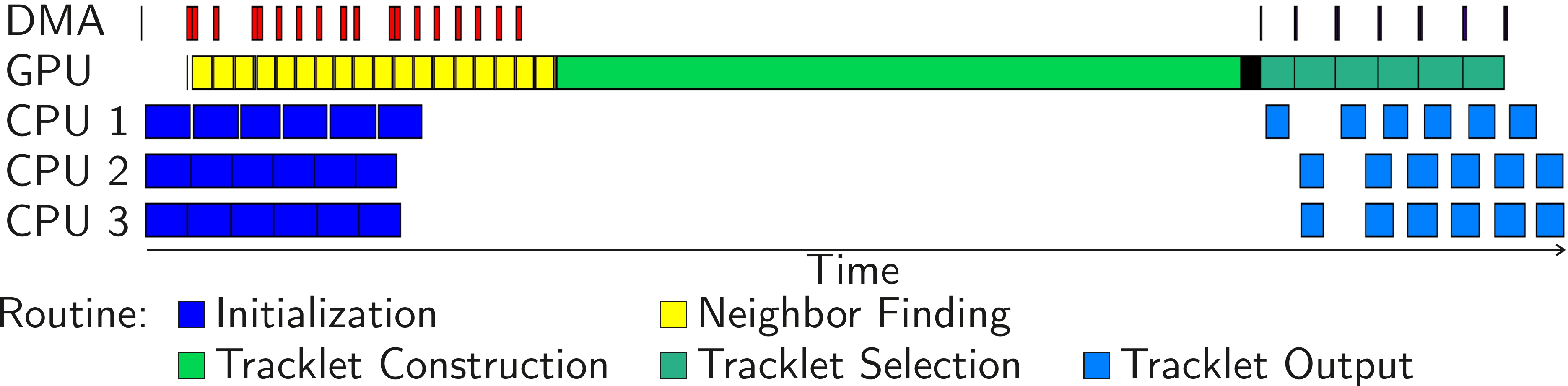}
\par\end{centering}
\caption{Visualization of the pipelined GPU processing of the track reconstruction using multiple CPU cores to feed data to the GPU.}
\label{fig:tracking_pipeline}
\end{figure}

One such optimization for GPUs is pipelined processing: the execution of the track reconstruction on the GPU, the initialization and output merging on the CPU, as well as the DMA transfer, all happen simultaneously (\fig{tracking_pipeline}).
The pipeline hides the DMA transfer time and the CPU tasks and keeps the GPU executing kernels more than~$95$\% of the time.
On top of that, multiple events are processed concurrently to make sure all GPU compute units are always fully used~\cite{chep2016gpu}.
One obstacle already mentioned in Section~\ref{sec:trackingvc} is the different starting rows and lengths of tracks, which prevent optimum utilization of the GPU's single instruction, multiple thread units.
A dynamic scheduling which, after processing a couple of rows, redistributes the remaining workload among the GPU threads was implemented.
This reduces the fraction of wasted GPU resources due to warp-serialization due to a track that has ended while another track is still being followed.

\subsubsection{Performance of the track reconstruction algorithm}

\label{sec:gpuperf}

\begin{figure}[htb]
\begin{centering}
\includegraphics[width=0.8\textwidth]{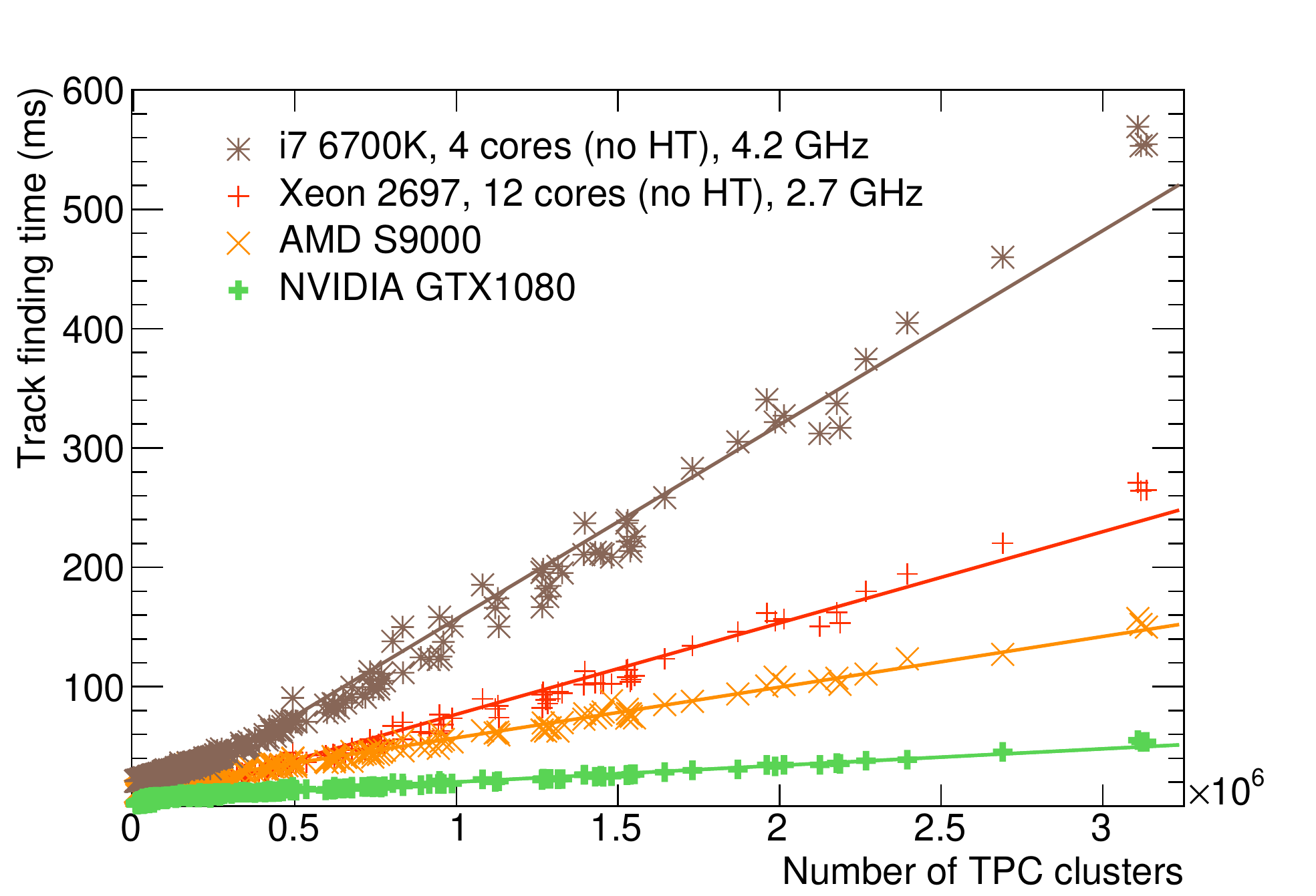}
\par\end{centering}
  \caption{Time required for execution of the tracking algorithm on CPUs and on GPUs as function of the input data size expressed in terms of the number of TPC clusters.
  The lines represent linear fits to the distributions.
  The merging and refitting times are not included in the track finding time.}
\label{fig:tracking_time_size}
\end{figure}

The dependence of the tracking time on input data size expressed in terms of the number of TPC clusters is shown in \fig{tracking_time_size}.
The hardware used for the HLT performance evaluation is the hardware of the HLT \run{2} farm, which consists of the already several years old Intel Xeon 2697 CPU and AMD FirePro S9000 GPU.
The compute time using a modern system, \ie an Intel Skylake CPU (i7 6700K) or NVIDIA GTX1080 GPU, is also shown and demonstrates that newer GPU generations yield the expected speedup.
On both CPU and GPU architectures, the compute time grows linearly with the input data size.
For small events, the GPU cannot be fully utilized and the pipeline-initialization time becomes significant, yielding a small offset for empty events.
With no dominant quadratic complexity in the tracking algorithm an excellent scaling to large events is achieved.
The CPU performance is scaled to the number of physical CPU cores via parallel processing of independent events, which scales linearly, while the tracking on GPUs processes a single event in one go.
Only one CPU socket of the HLT \run{2} farm's server is used to avoid NUMA (Non Uniform Memory Architecture).

\begin{figure}[htb]
\begin{centering}
\includegraphics[width=0.8\textwidth]{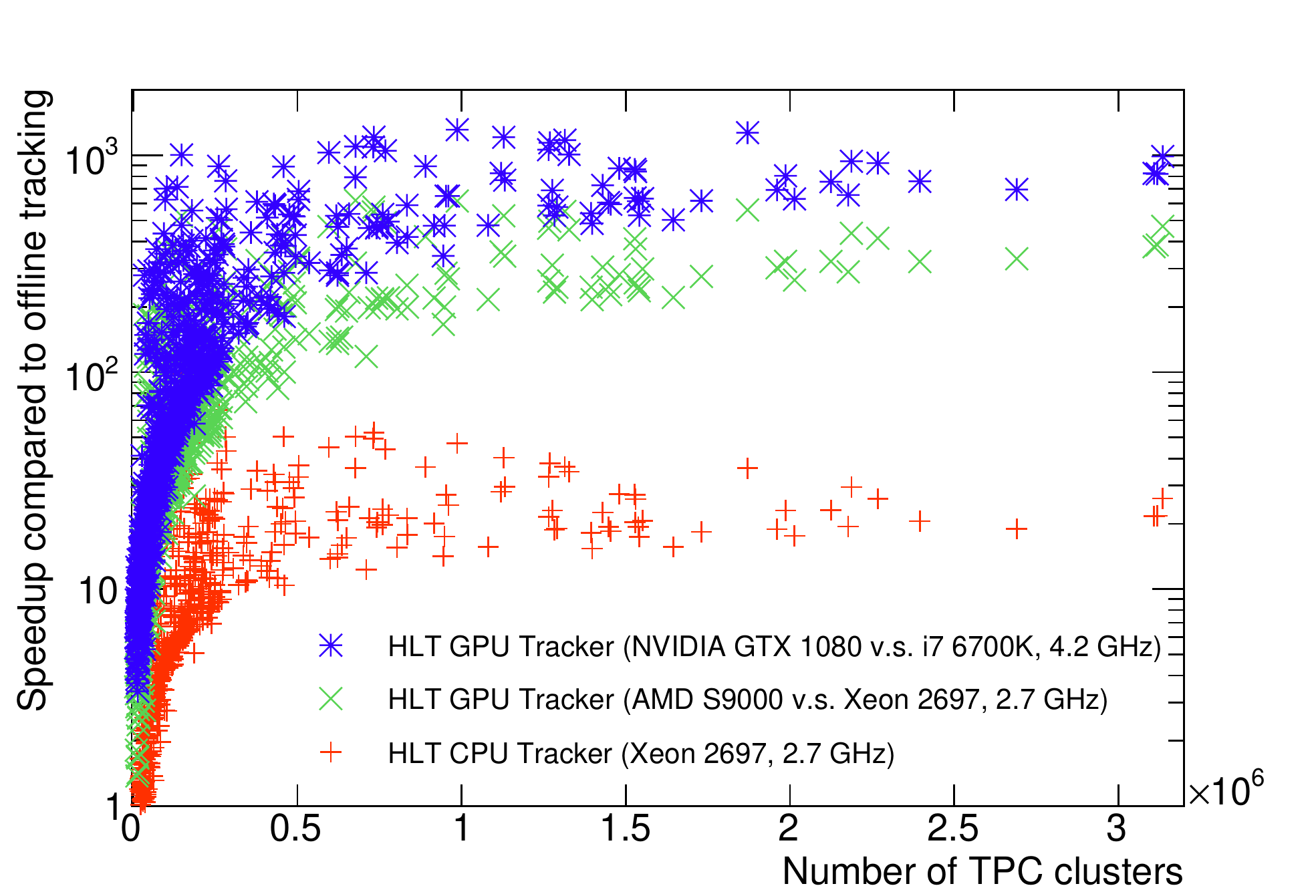}
\par\end{centering}
\caption{Speedup of HLT tracking algorithm executed on GPUs and CPUs compared to the offline tracker normalized to a single core and corrected for the serial processing part that the CPU contributes to GPU tracking as a function of the input data size expressed in terms of the number of TPC clusters. The plus markers show the speedup as a function of the number of TPC clusters with the HLT tracking executed on the CPU. The cross(astrisk) markers show the speedup obtained with the tracking executed on a older(newer) GPU.}
\label{fig:tracking_speedup}
\end{figure}

The overall speedup achieved by the HLT GPU tracking is shown in \fig{tracking_speedup}.
It is computed as the ratio of the processing time of offline (CPU) tracking and the single-core processing time of GPU tracking.
Here the CPU usage-time for pre- and post-processing of GPU tracking scaled by the average number of CPU cores used during the steps of GPU tracking is folded out of the total CPU-tracking time.
For the CPU version of the HLT tracking algorithm, this is exactly the speedup.
For the GPU version, this is the number of CPU cores equivalent, tracking-performance-wise, to one GPU.
In this case, the full track reconstruction duration includes the merging and refitting time, whereas for \fig{tracking_time_size} the non tracking-related steps of the offline tracking, \eg d$E$/d$x$ calculation, are disabled.
Overall, the HLT tracking algorithm executed on the CPU is 15--20 times faster than the offline tracking algorithm.
One GPU of the HLT \run{2} farm replaces more than 15 CPU cores in the server, for a total speedup factor of up to 300, with respect to offline tracking.
The CPU demands for pre- and post-processing of the old AMD GPUs in the HLT server are significantly greater than for newer GPUs since the AMD GPUs lack the support for the OpenCL generic address space required by several processing steps.
The newer NVIDIA GTX1080 GPU model supports offloading of a larger fraction of the workload and is faster in general, replacing up to 40 CPU cores of the Intel Skylake (i7 6700K) CPU, or up to 800 Xeon 2697 CPU cores when compared to offline tracking.
Overall, in terms of execution time, a comparable performance is observed for the currently available AMD and NVIDIA GPUs.
It has to be noted that HyperThreading was disabled for the measurements of \fig{tracking_time_size} and \fig{tracking_speedup}.
With HyperThreading, the Intel Core i7 CPU's total event throughput was 18\% higher.
The GPU throughput can also be increased by processing multiple independent events in parallel.
A throughput increase of 32\% is measured, at the expense of some latency on the AMD S9000~\cite{chep2016gpu}.
For \fig{tracking_speedup}, the better GPU performance would also require more CPU cores for pre- and post-processing, such that these speedups basically cancel each other out after the normalization to a CPU core.
The tracking algorithm has proven to be fast enough for the LHC \run{3}, in which \mbox{ALICE} will process time frames of up to 5 overlapping heavy-ion events in one TPC drift time.

GPU models used in the HLT farms of both \run{1} and \run{2} offered a tracking performance equivalent to a large fraction of the CPU cores on an HLT node.
Thus, by equipping the servers with GPUs the required size of the farm was nearly reduced by a half.
The cost savings compared to tracking on the processors in a traditional farm was around half a million CHF for \run{1} and is above one million CHF for \run{2}, not including the savings accrued by having a smaller network, less infrastructure, and lower power consumption.
If the HLT only used CPUs, online track reconstruction of all events, using the HLT algorithm, would be prohibitively expensive.
Running the offline track reconstruction online would accordingly be even more expensive.
This shows that fast tracking algorithms that exploit the capabilities of hardware accelerators are mandatory for future high luminosity heavy-ion experiments like \mbox{ALICE} in the LHC \run{3} or at the experiments that will be setup at the Facility for Antiproton and Ion Research (FAIR) at GSI~\cite{bib:fair}.

\begin{figure*}[t]
\begin{centering}
\includegraphics[width=0.95\textwidth]{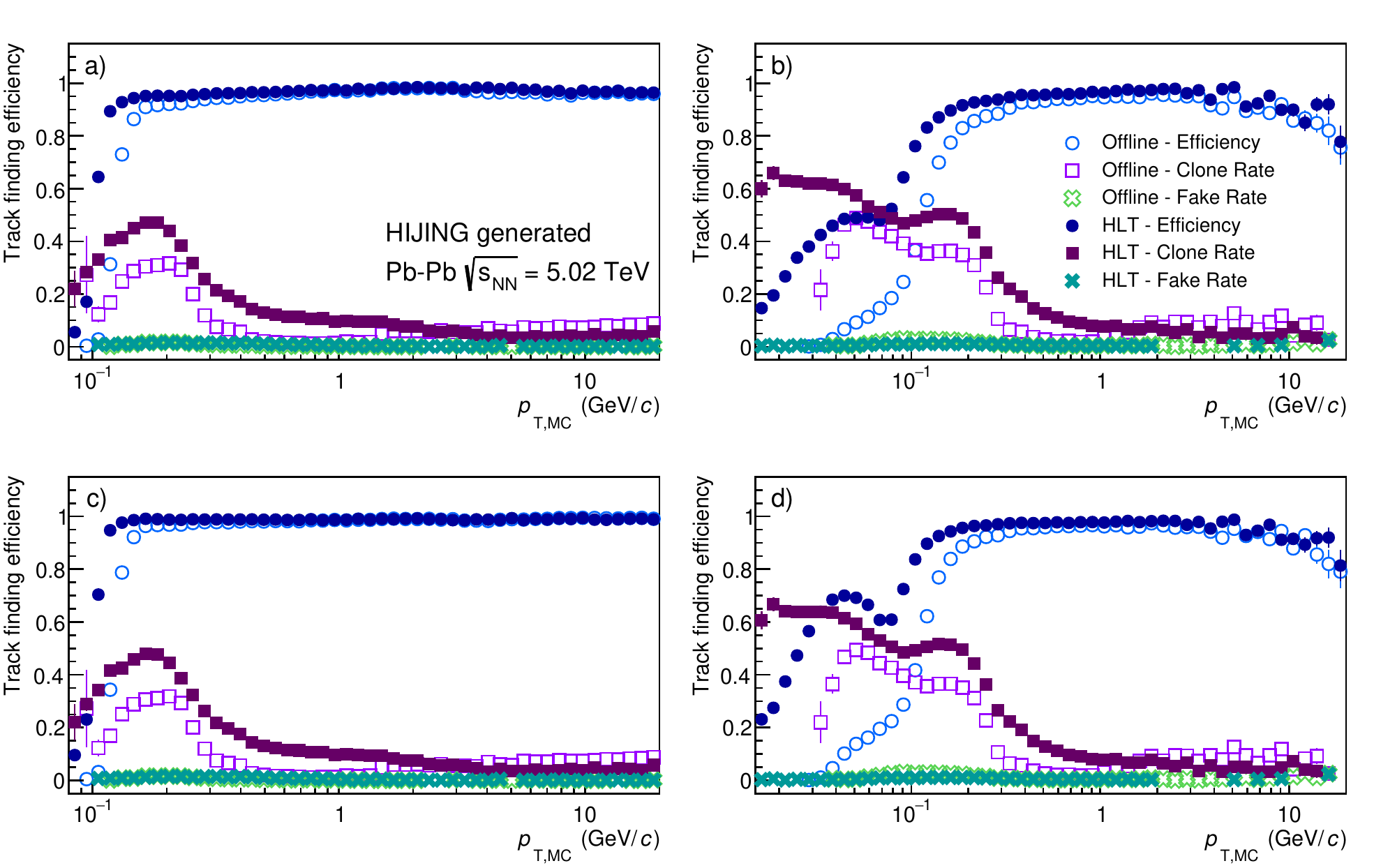}
\par\end{centering}
  \caption{Tracking efficiency of the HLT and offline trackers as function of the transverse momentum calculated as the ratio of reconstructed tracks and simulated tracks in HIJING generated Pb--Pb events at $\sqrt{s_{\rm{NN}}}$ = $5.02$\,TeV, shown for tracks that are a) primary, b) secondary, c) findable primary, and d) findable secondary.
Findable tracks are defined as reconstructed tracks that have at least~$70$~clusters in the TPC.
}
\label{fig:tracking_efficiency}
\end{figure*}

The tracking efficiencies, in terms of the fraction of simulated tracks reconstructed by offline and HLT algorithms, are shown in \fig{tracking_efficiency}.
These efficiencies calculated using a HIJING \cite{Gyulassy:1994ew} simulation of Pb--Pb collision events at $\sqrt{s_{\rm{NN}}}$ = $5.02$.
The figure distinguishes between primary and secondary tracks as well findable tracks.
Findable tracks are reconstructed tracks that have at least 70 clusters in the TPC, and both offline and HLT algorithms achieve close to 100\% efficiency for findable primaries.
In comparison, when the track sample includes tracks which are not physically in the detector acceptance or tracks with very few TPC hits the efficiency is lower.
The minimum transverse momentum measurable for primaries reaches down to~$90$\,MeV/$c$, as tracks with lower \pt{} do not reach the TPC.
The HLT tracker achieves a slightly higher efficiency for secondary tracks because of the usage of the cellular automaton seeding without vertex constraint.
In preparation for \run{3}, the HLT tracking has also been tuned for the low-\pt{} finding efficiency in order to improve looper-track identification required for the $\text{O}^2$ compression~\cite{bib:o2}.
Both offline and HLT trackers have negligible fake rates, while HLT shows a slightly lower clone rate at high-\pt{}, which is due to the approach used for sector tracking and merging.
The clone rate increases significantly for low-\pt{} secondaries, in particular for the HLT.
This is not a deficit of the tracker but rather is caused by looping tracks inside the TPC for which the merging of the multiple legs of the loop is not yet implemented.

\begin{figure*}[t]
\begin{centering}
\includegraphics[width=0.95\textwidth]{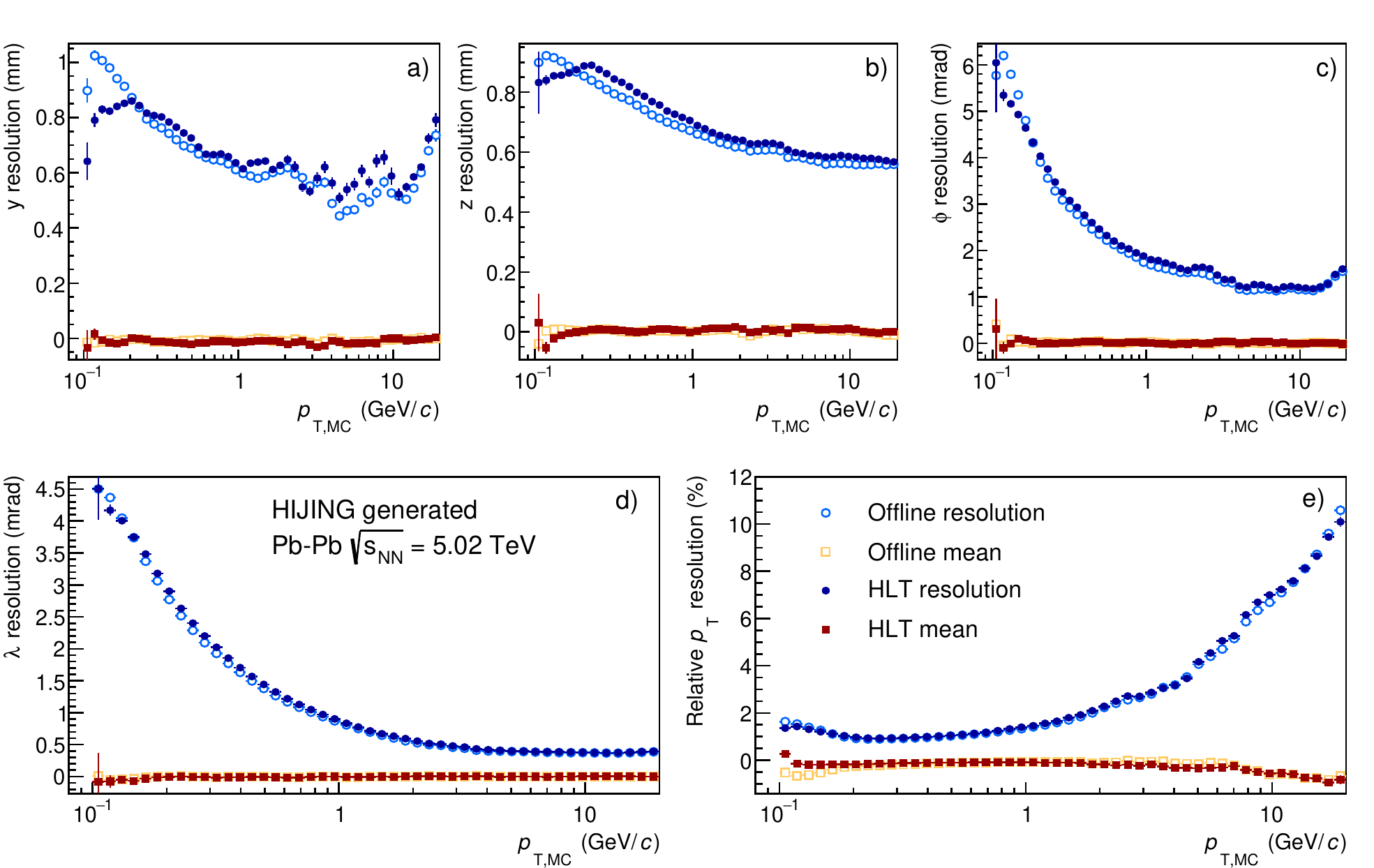}
\par\end{centering}
  \caption{Mean value and track parameter resolutions of the HLT and offline trackers as function of the transverse momentum measured in HIJING generated Pb--Pb events at $\sqrt{s_{\rm{NN}}}$ = $5.02$\,TeV.
  The resolution of a) $y$ and b) $z$ spatial positions, c) azimuthal angle ($\phi$), d) lambda ($\lambda$), and e) relative transverse momentum are shown.}
\label{fig:tracking_resolution}
\end{figure*}

The track resolution with respect to the track parameters of the MC track taken at the entrance of the TPC is shown in \fig{tracking_resolution}.
These track parameters include the $y$ and $z$ spatial positions in the local coordinate system (see \fig{slice}), the transverse momentum (\pt{}), the azimuthal ($\phi$) and dip ($\lambda$) angles.
The HLT tracker shows only a nearly negligible degradation compared to the offline algorithm.
In order to provide a fair comparison of the tracking algorithms independent from calibration, the offline calibration was used in both cases.
This guarantees the exact same transformation of TPC clusters from pad, row, and time to spatial coordinates and the same parameterization of systematic cluster errors due to distortions in the TPC that result from an accumulation of space charge at high interaction rates.
Even though the calibration is the same, offline performs some additional corrections to account for the space-charge distortions, e.\,g.~a correction of the covariance matrix that takes the correlation of systematic measurement errors in locally distorted regions into account.
The mean values of the distributions obtained from the HLT and offline trackers are identical and the trackers do not show a significant bias for either of the track parameters.
The remaining differences in the resolution originate from TPC space-charge distortions, since this correction is not yet implemented in the HLT tracker.
This was verified by using MC simulations without the space-charge distortions, where differences in the resolution distribution mostly disappeared.

Overall, the HLT track reconstruction performance is comparable with offline track reconstruction.
Speeding up the computation by an order of magnitude introduces only a minor degradation of the track resolution compared to offline.
A comparison of efficiency and resolution of GPU and CPU version of the HLT tracking yields identical results.
However, the bit-level CPU and GPU results are not 100\% comparable because of different floating point rounding and concurrent processing.

\subsection{TPC online calibration}

\label{sec:calibration}

High quality online tracking demands proper calibration objects.
Drift detectors, like the TPC, are sensitive to changes in the environmental conditions such as the ambient pressure and/or temperature.
Therefore, precise calibration of the electron drift velocity is crucial in order to properly relate the measured arrival time to the TPC end-caps spatial positions along the $z$ axis.
Spatial and temporal variations of the properties of the gas inside the TPC as well as the geometrical misalignment of the TPC and ITS contribute to misalignment of individual track segments belonging to a single particle.
Corrections for these effects are found by comparing independently fitted TPC track parameters with those found in the ITS \cite{Krzewicki:2013dkt}.
For the online calibration, the cycle starts by collecting data from processing components, which run in parallel on all the HLT nodes.
When the desired amount of events (roughly~3000~Pb--Pb events) is obtained, the resulting calibration parameters are merged and processed.
To account for their time dependence, the procedure is repeated periodically.
At the beginning of the run, no valid online calibration exists.
Therefore, the HLT starts the track reconstruction with a default calibration until the online calibration becomes available after the first cycle.

\begin{figure}[htb]
 \begin{centering}
 \includegraphics[width=0.8\textwidth]{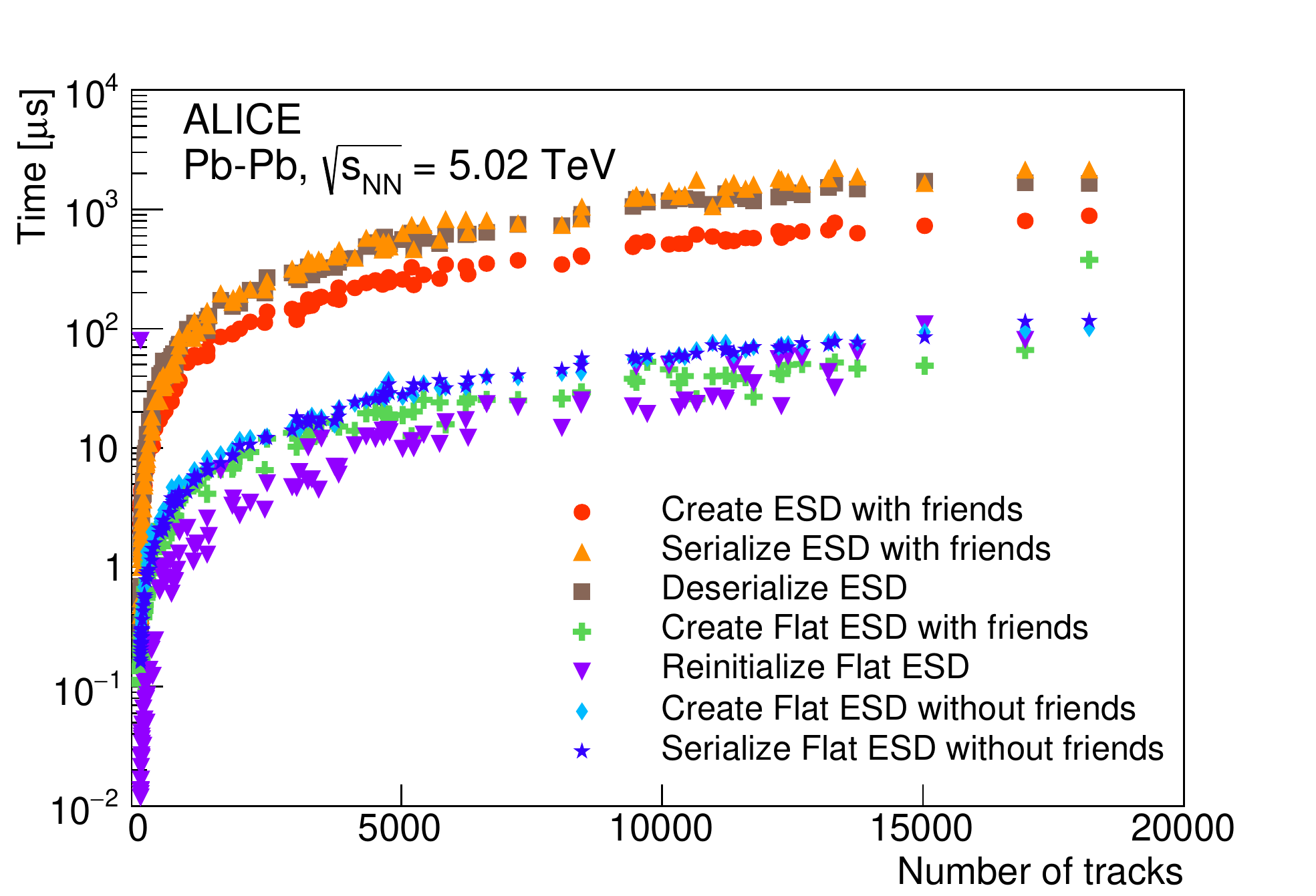}
\par\end{centering}
\caption{Time required for the creation, serialization, and deserialization of the Flat ESD vs.~the standard ESD for offline analysis as a function of the number of TPC tracks.}
\label{fig:esdcost}
\end{figure}

The offline TPC drift-velocity calibration is implemented within the \mbox{ALICE} analysis framework, which is optimized for the processing of ESDs.
In addition, the calibration algorithm produces a ROOT object called ESD friend, which contains additional track information and cluster data.
Since it is relatively large, the ESD friend is not created for each event, rather it is stored for the events that are used for the calibration.
Within the HLT framework the data are transferred between components via contiguous buffers.
Hence these ESD objects must be serialized before sending and deserialized after receiving a buffer.
Since this flow, comparable to online reconstruction, is resource-hungry a custom data representation was developed, called Flat ESD.
Although the Flat ESD shares the same virtual interface with the ESD, the underlying data store of the flat structure is a single contiguous buffer.
By design it has zero serialization/deserialization overhead.
There is only a negligible overhead related to the virtual function table pointer restoration.
Overall, creation, serialization, and deserialization of the Flat ESD is more than~10 times faster compared to the standard ESD used in offline analysis, as demonstrated in \fig{esdcost}.

The HLT provides a wrapper to execute offline code inside the HLT online processing framework using offline configuration macros.
The calibration components on each compute node process the calibration tasks asynchronously with respect to the main in-chain data flow.
Once sufficient calibration data are collected, the components send their output to an asynchronous data merger.
The merged calibration objects are then sent to a single asynchronous process which calculates the cluster transformation maps.
These maps are used to correct the cluster position before the track finder algorithm is executed in order to avoid to interfere with the main HLT chain.
Finally, finishing the cycle, these maps are distributed back to the beginning of the chain where they are used in the online reconstruction.
The cycle is illustrated in \fig{components}.
The calibration objects from a cycle are used until the following cycle finished and the output is available.
The asynchronous transport uses ZeroMQ.

Depending on the availability of computing resources, which rely on beam and trigger conditions, the HLT runs up to 3 calibration worker processes per node on its 172 compute nodes.
Events for the calibration are processed distributedly by the $3 \times 172$ instances of the calibration task.
This number is a parameter: more instances would need more compute resources but in turn would yield more data for calibration in a shorter amount of time.
A sufficient calibration precision requires approximately 3000 Pb--Pb event, which are collected in roughly 2 minutes with the number of instances mentioned above.
The subsequent merging of the data, transformation map calculation and distribution to all reconstruction processes takes about another 30 seconds.
While the TPC drift time calibration is stable within a 15 minute time window, the total calibration cycle time never exceeds this stable calibration time window.

\begin{figure}[tp]
\centering
\begin{subfigure}{0.6\textwidth}
\includegraphics[width=0.95\textwidth]{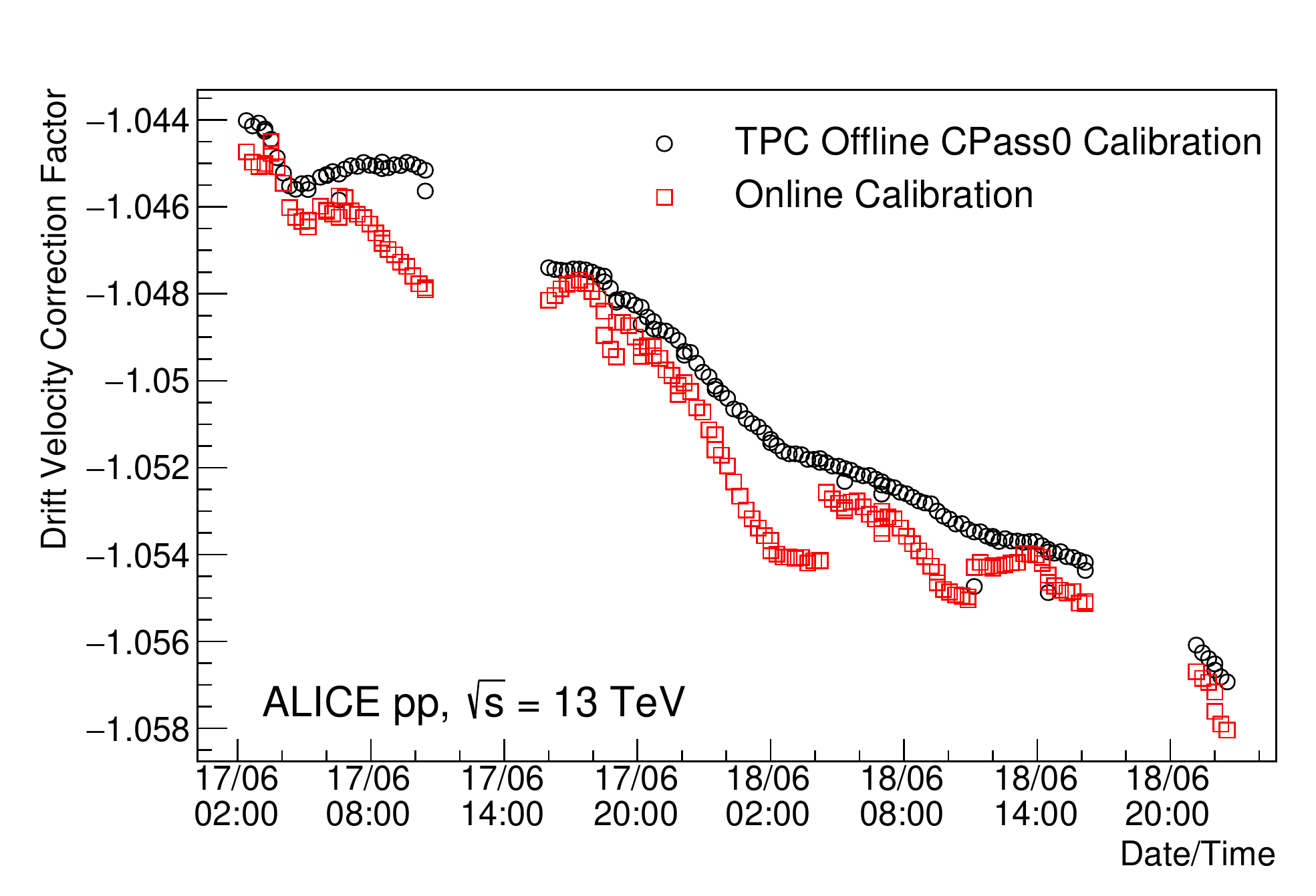}
 \subcaption{TPC drift-time correction factor for a period of $12$ hours measured during pp data taking at $\sqrt{s}$ = $13$\,TeV.}
 \label{fig:drifttime}
 \end{subfigure}

 \begin{subfigure}{0.6\textwidth}
 \includegraphics[width=0.95\textwidth]{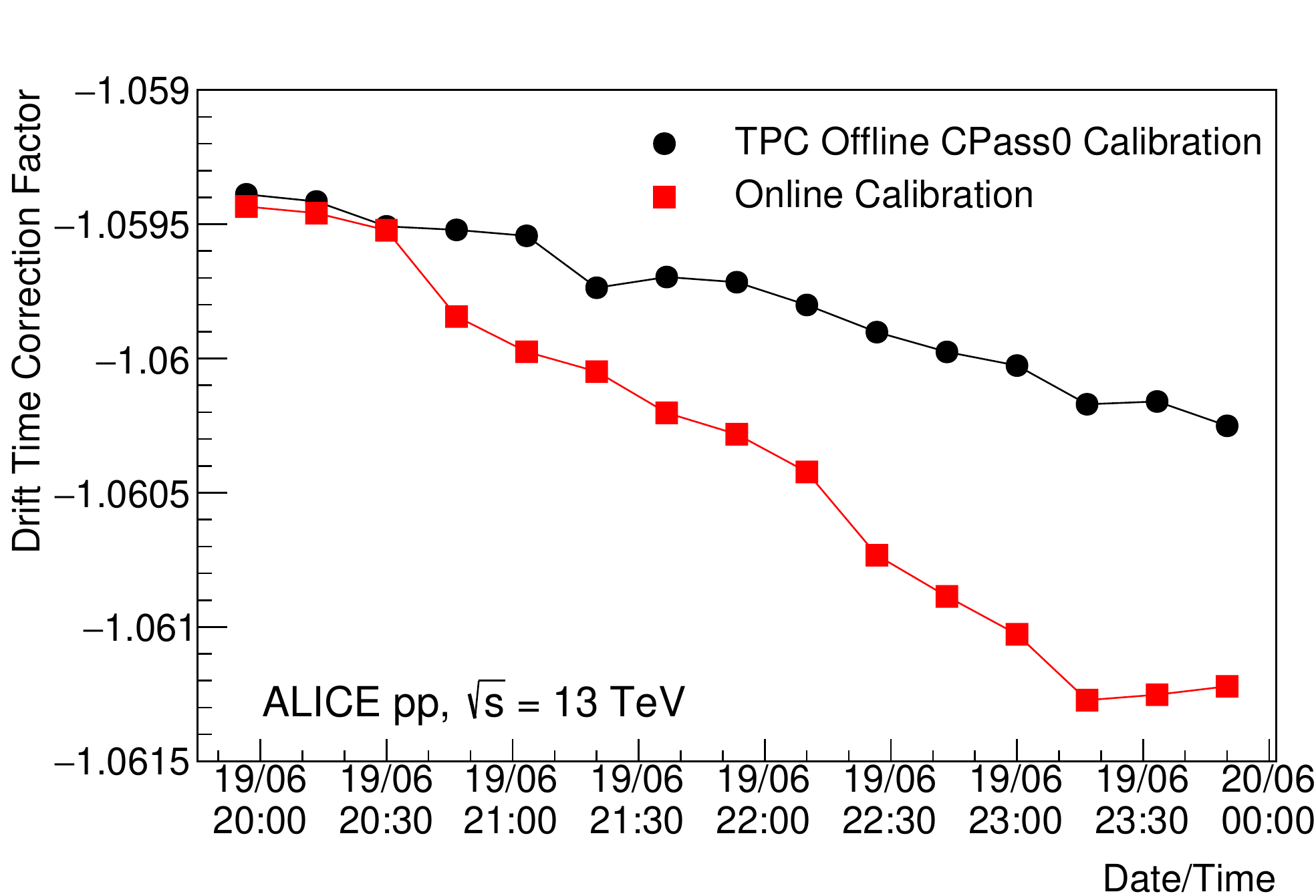}
 \subcaption{TPC drift-time correction factor for a duration of one run (4 hours) measured during pp data taking at $\sqrt{s}$ = $13$\,TeV. The correct temperature stored at the beginning of the run was included manually.}
 \label{fig:drifttime-correct-temperature}
 \end{subfigure}

 \begin{subfigure}{0.6\textwidth}
 \includegraphics[width=0.95\textwidth]{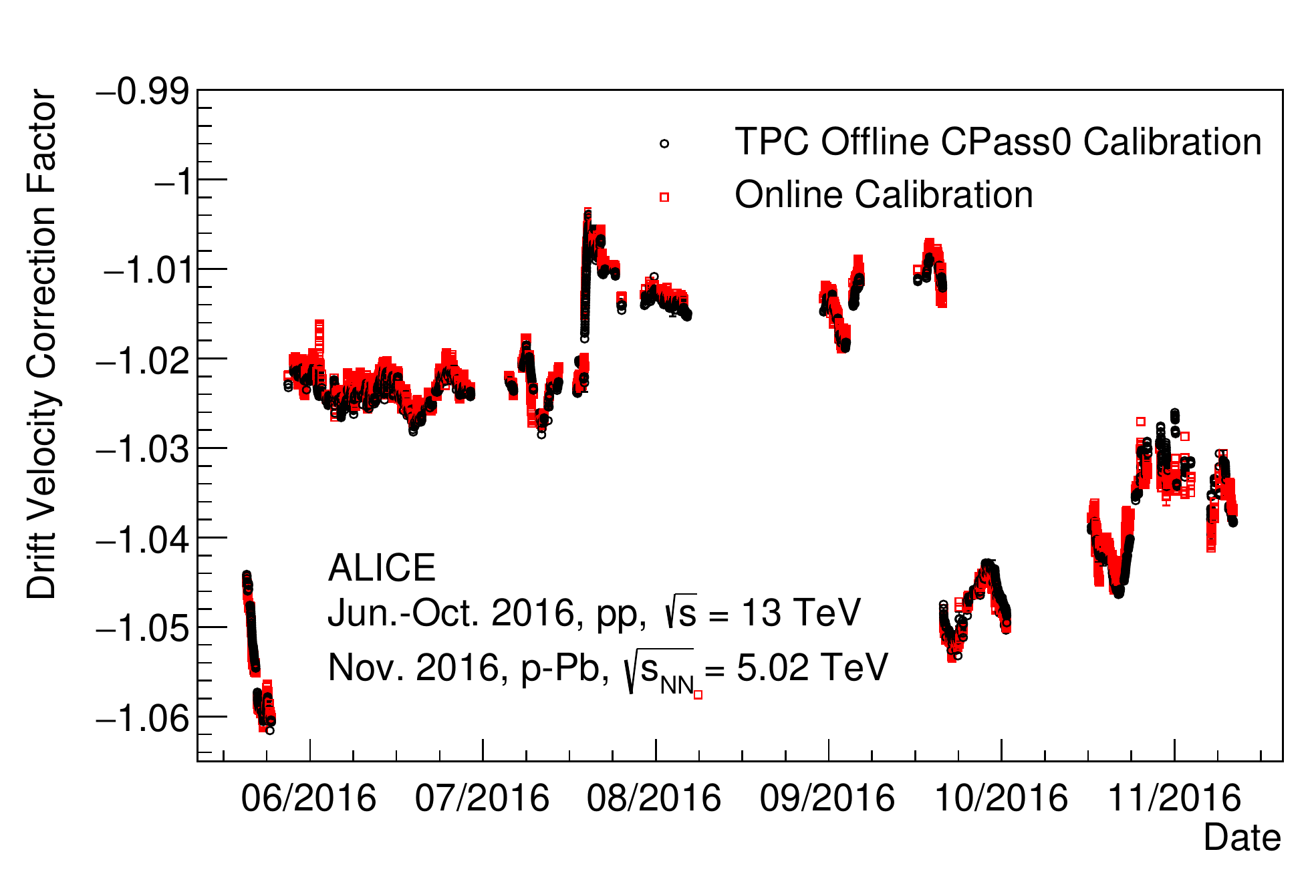}
 \subcaption{TPC drift-time correction factor for a period of 5 months measure during pp data taking at $\sqrt{s}$ = $13$\,TeV and p--Pb data taking at $\sqrt{s_{\rm{NN}}}$ = 5.02 TeV.}
 \label{fig:drifttime-long-term}
 \end{subfigure}

\caption{TPC drift-time correction factor obtained by the online and offline calibrations as a function of time for various periods of time.
 Gaps in the distributions correspond to periods without beam.
}
\label{fig:calib}
\end{figure}

One difference between online and offline calibration is the availability of real-time ambient pressure and temperature values.
Currently, the HLT only has access to the pressure value at the beginning of the run, and does not have access to the temperature at all.
In contrast, offline has the full pressure and temperature data over time.
This yields two effects shown in \fig{drifttime}, in which the drift-velocity correction factor is reported as a function of time.
First, the drift-velocity correction factor at the beginning of each run is shifted relative to the offline calibration, since the HLT calibration process uses an outdated temperature compared to offline.
Second, during the run the change of the pressure slightly affects the drift velocity.
In the offline case, this is accounted for, while HLT sticks to the pressure value at the beginning of the run.
For the spatial TPC cluster positions, it does not play a role whether the temperature change is accounted for by the base drift velocity estimation or by the correction factor.
\figur{drifttime-correct-temperature} shows a run in which the HLT uses the correct temperature at the beginning of the run, obtaining the exact same calibration as offline.
It should be noted that in the calibration procedure for \run{3}, the temperature value will be available a the beginning of each run.

\begin{figure}[htb]
  \begin{centering}
    \includegraphics[width=0.8\textwidth]{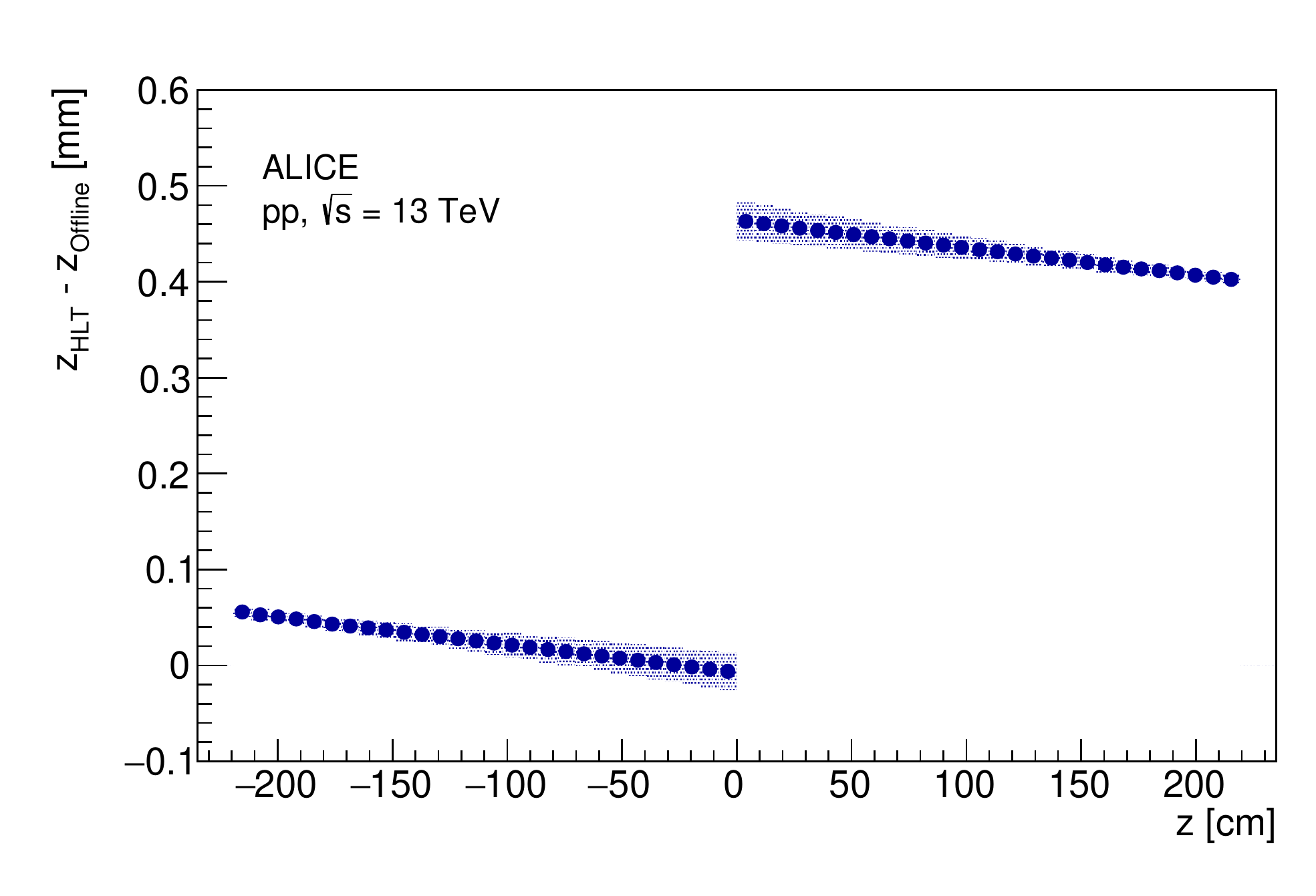}
\par\end{centering}
  \caption{Average differences of the TPC cluster position along $z$-axis calculated with drift-velocity correction factors from the online (HLT) and offline calibration.
    The differences are of the order of the intrinsic detector resolution.
    Calibration of the forward and backward halves of the TPC are computed independently.
    The error bands represent the statistical error, along with the $r$ and $\varphi$ dependent differences of online and offline calibration.}
  \label{fig:cluster-errors-calibrated}
\end{figure}

The drift-velocity calibration factors are in agreement for $90$\% of the runs.
The remaining $10$\% of the runs are primarily composed of short data taking runs, where there was not enough time to gather enough data for online calibration, or test runs in special conditions that prevented a TPC calibration.
Without online calibration, the TPC cluster position along the $z$-axis in the online reconstruction deviate by up to 3\,cm from the calibration position available offline.
The online calibration reduces this deviation down to 0.5\,mm, which is in the order of the intrinsic TPC space point resolution, see \fig{cluster-errors-calibrated}.
Online calibration objects can be used offline, but since the persistent data are not modified, calibration procedures can still run offline if needed.

\subsection{TPC data compression}

\label{sec:compression}

\begin{figure}[htb]
\begin{centering}
\includegraphics[width=0.8\textwidth]{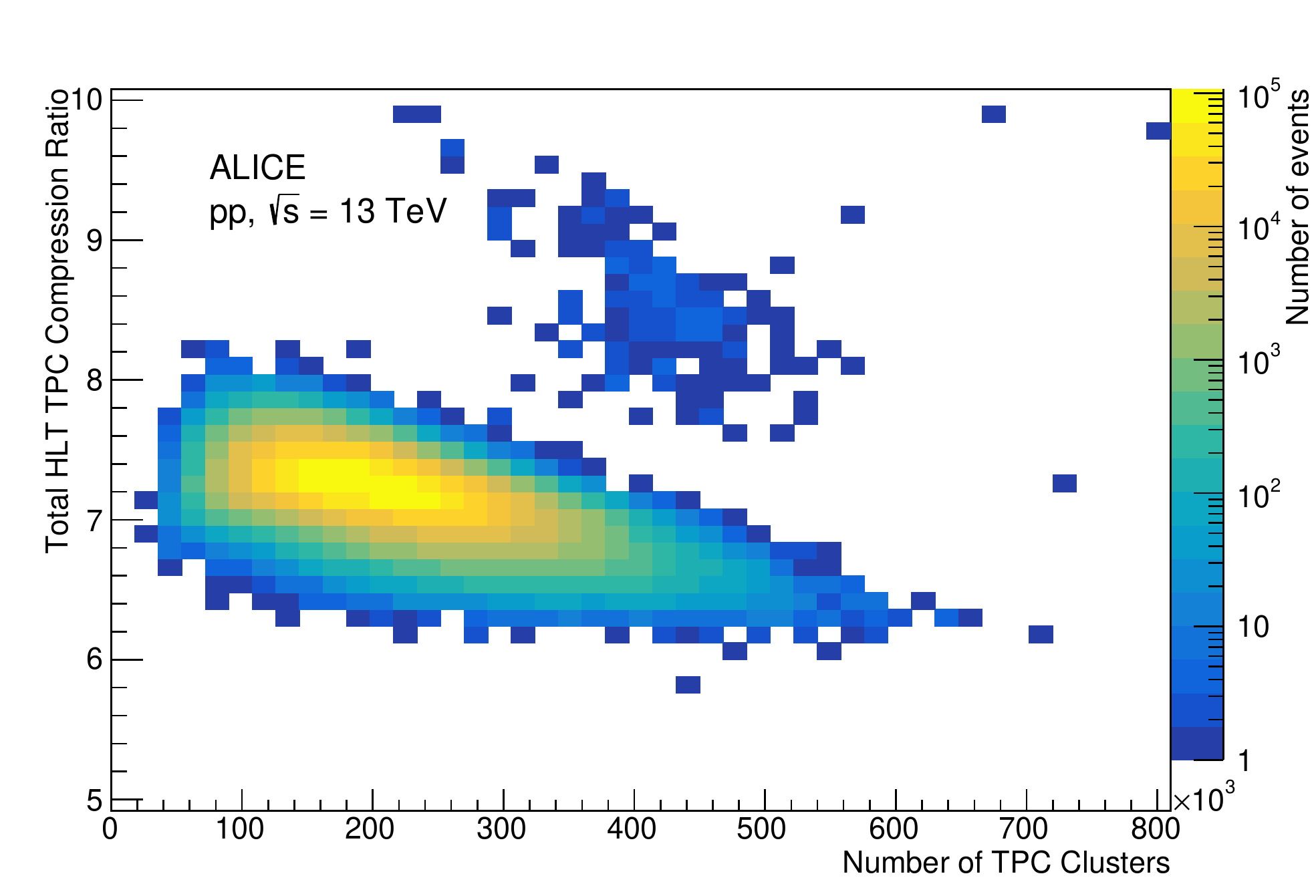}
\par\end{centering}
\caption{Total HLT TPC data compression ratio including improved TPC online cluster finder and Huffman compression in \run{2} on 2017 pp data as a function of the input data size expressed in terms of the number of TPC clusters.}
\label{fig:run1_huffman_compression}
\end{figure}

In parallel with the tracking and calibration, the data compression branch of the HLT chain compresses the TPC clusters and replaces the TPC raw data with these compressed clusters~\cite{compressionbase}.
The backbone of the data compression is Huffman entropy encoding~\cite{bib:huffman}.
Entropy encoding of the pure TPC ADC values achieves only a maximum compression factor of two, which is less than the compression achievable on the cluster level.
The data size is reduced in three consecutive steps.
It begins with the hardware cluster finder converting raw data into TPC clusters, calculating properties like total charge, width, and coordinates.
The second step converts the computed floating point properties into fixed point integers with the smallest unit equaling the detector resolution.
Finally, Huffman encoding compresses the fixed size properties.
During \run{1}, the average total compression factor was~$4.3$.
In preparation for \run{2} compression techniques were improved upon.
\figur{run1_huffman_compression} shows the compression ratio versus the input data size expressed in terms of the number of TPC clusters in~$2017$, when an average compression factor of~$7.3$ was achieved.

\begin{table*}[htb]
\begin{center}
\caption{Compression factors of the different processing steps for the TPC.}
\scalebox{0.68}{
\begin{tabular}{lr|rrrr|rr}
\hline
\textbf{Configuration}    & & & & & & & \\
Data taking period    & 2013      & 2015 Pb--Pb        & 2016 pp       & 2017 pp          & 2017 pp          & \textit{2016 pp}         & \textit{2015 Pb--Pb}      \\
TPC gas                   & neon      & argon           & argon             & neon             & neon             & \textit{argon}           & \textit{argon}           \\
RCU version               & 1         & 1              & 2                 & 2                & 2                & \textit{2}               & \textit{2}               \\
Cluster finder version              & run 1     & old             & old               & old              & improved         & \textit{improved}        & \textit{improved}        \\
Compression version       & run 1 / 2 & run 1 / 2       & run 1 / 2         & run 1 / 2        & run 1 / 2        & \textit{run 3 prototype} & \textit{run 3 prototype} \\
\hline
\textbf{Compression step} & & & & & & & \\
Cluster finder            & 1.20x       & 1.28x   & 1.50x                   & 1.42x            & 1.81x            & \textit{1.72x}           & \textit{1.70x}           \\
Branch merging            & 1.05x  & 1.05x    & -                           & -                & -                & \textit{-}               & \textit{-}               \\
Integer format            & 2.50x    & 2.50x      & 2.50x                   & 2.50x            & 2.40x            & \textit{2.40x}           & \textit{2.40x}           \\
(bits per cluster)        & 77 bits   & 77 bits         & 77 bits           & 77 bits          & 80 bits          & \textit{80 bits}         & \textit{80 bits}         \\
\textbf{Entropy reduction} & & & & & & & \\
\textit{(savings after entropy encoding)} & & & & & & & \\
Position differences      & -     & 16\% / -7.2 bits    & 2\% / -1.2 bits   & 2\% / -1.0 bits  & 2\% / -1.0 bits  & \textit{-1.0 bits}       & \textit{-4.5 bits}       \\
Track model               & -         & -               & -                 & -                & -                & \textit{-14.5 bits}      & \textit{-14.3 bits}      \\
Track model + differences & -         & -               & -                 & -                & -                & \textit{-8.0 bits}       & \textit{-8.41 bits}      \\
Logarithmic precision     & -         & -               & -                 & -                & 15\% / -6.6 bits & \textit{-7.3 bits}       & \textit{-7.3 bits}       \\
\textbf{Entropy encoding} & & & & & & & \\
Huffman coding            & 1.36x    & 1.75x    & 1.49x                     & 1.46x            & 1.68x            & \textit{2.08x}           & \textit{2.12x}           \\
Arithmetic coding         & -         & -               & -                 & -                & -                & \textit{2.18x}           & \textit{2.22x}           \\
\hline
Total compression         & 4.26x     & 5.89x     & 5.58x                   & 5.18x            & 7.28x            & \textit{9.00x}           & \textit{9.10x}           \\
(bits per cluster)        & 56.6 bits & 44.0 bits    & 51.7 bits                & 52.8 bits        & 47,7 bits        & \textit{36,7 bits}       & \textit{36,0 bits}       \\
\hline
\end{tabular}}
\label{tab:compression}
\end{center}
\end{table*}

\tabl{compression} gives an overview of the improvements of the HLT performance on the compression factors for different data-taking scenarios.
The baseline is the compression ratio of~$4.3$ achieved during \run{1}, shown in the leftmost column.
In this case, the cluster finding and merging of clusters at readout branch borders yielded a compression factor of 1.2.
Storing the cluster information in fixed point integer format reduced the size by a factor of~$2.5$, requiring~$77$~bits per cluster thereafter.
The entropy coding using Huffman compression reduced the average number of bits per cluster down to~$56.6$.

Several boundary conditions changed at the beginning of \run{2}.
The TPC gas was changed from neon to argon in 2015 and 2016 which led to a higher gain.
This increased the noise over the zero-suppression threshold, which led to a larger raw data size and an increase of the fake clusters.
The compression factor of the cluster finder itself increases, because the fraction of noise in the raw data that is rejected is larger than that in \run{1}.
In addition, the readout hardware was changed to the RCU2 and the \mbox{C-RORC} (see Section~\ref{sec:rorc}), allowing all incoming data of one TPC pad-row to be processed together.
Before, the pad-row was split into two branches which were processed independently and thus required a successive branch merging step to treat the clusters at the branch borders correctly.
This is now obsolete with the new hardware, leading to a better physics performance and higher compression during the cluster finding stage.

Additional processing steps can reduce the cluster entropy and improve the entropy encoding.
In particular, for high occupancy Pb--Pb events, the spatial distribution of the clusters is mostly uniform, but the distances between adjacent clusters are small.
Storing position differences instead of absolute positions reduces the entropy and yields a higher compression factor.
On average this saves~$7.2$~bits for Pb--Pb data, reducing the size by~$16$\%.
This is less efficient for pp data in which the occupancy is lower resulting in the position differences being much larger, leading to an average size reduction of only~$1.2$~bits.
Overall, this and other format optimizations have improved the compression factors to~$5.5$ for pp and to~$5.9$ for Pb--Pb for \run{2}.
For clusters associated to tracks, an alternative approach consists of storing the track properties and the residuals of the cluster-to-track position are stored~\cite{compressionbase,compressiontrackmodel}, listed as $"$Track model$"$ in \tab{compression}.
These residuals also have a small entropy and are ideally suited for Huffman compression.
This is particularly useful for pp data, where the position differences method does not perform well.

The two rightmost columns of \tab{compression} show compression factors obtained by a proof-of-concept prototype for the compression developed for \run{3}, using data from \run{2}.
The prototype includes an advanced version of the track model compression~\cite{Rohr:2017dlo} \comment{change citation once the proceedings are published}, which refits the track in a distorted coordinate system, yielding significantly smaller residuals than first track model compression during \run{1}~\cite{hlt-run1-2011}.
The track-model compression saves on average more than~$14$ bits per cluster, both for pp and Pb--Pb data.
In turn, it deteriorates the compression of the position differences method that is used for clusters not assigned to tracks because the occupancy of non-assigned clusters decreases, increasing the entropy of the differences.
The ``Track model + differences'' row of the table shows the total average savings for all clusters, calculated as the weighted average of the savings achieved by the track-model and the position-differences methods.
For Pb--Pb data, the result of~$8.41$ bits is only slightly better than the pure position differences method.
However, the compression factor of pp data reaches the one of Pb--Pb data.
There is an even more important benefit of track model compression.
It maintains the cluster to track association of HLT tracks for the offline analysis without requiring additional storage or a special data format.
In this way, the tracks that were found online by the HLT can be immediately used as seeds by the offline tracker.
Having access to the cluster association, the offline tracker can run the slower but more sophisticated routines on the tracks.
This approach saves memory and compute cycles during the offline track reconstruction and is currently being commissioned.
The \run{3} prototype also shows that, by using arithmetic compression instead of Huffman compression to obtain optimal entropy encoding, a savings of roughly~$5$\% is achieved.

The fixed point integer format is not ideal for all cluster properties.
For the cluster width and charge, only a certain relative precision but no absolute precision is needed.
Therefore, only a certain number of precision bits after the leading non-zero bit are allowed, and all following less significant bits are forced to zero, implementing proper rounding.
This practically emulates a floating point format, while the entropy compression already guarantees the best storage, optimizing away the invalid values with more non-zero bits.
By using only three non-significant bits for the cluster width and four for the charge, a savings of~$15$\% of the 2017 pp data volume was obtained.

With the argon gas being used in the TPC at the beginning of \run{2}, a significant overhead of fake clusters emerging from the increased noise was faced.
The cluster finder searches for charge peaks and merges them, creating fake clusters if the total adjacent noise exceeds a minimum threshold.
Therefore, the HLT cluster finder was improved for the 2017 data taking to reject this noise by an improved peak finding heuristic.
This improved hardware cluster finder (see Section~\ref{sec:hwcf}) reduced the amount of clusters reconstructed in the TPC in pp data collected in 2016 when argon was the TPC gas by~$32$\%.
The reduction was approximately~$21$\% for the pp data collected in 2016 when the TPC gas was neon.
It also sped up the tracking and yielded slightly better track parameters.
Note that the gain in compression after the Huffman encoding can differ between between the two data sets because noise clusters have different entropy.

Storage space is a limiting factor in data taking, even with the inclusion of HLT compression.
Currently \mbox{ALICE} uses almost the entire allocated capacity, which is roughly 10\,PB per year.
TPC data are by far the largest contributor taking up more than~$90$\% of the raw data volume.
The offline software employs built-in ROOT file compression on the raw data from the other detectors.
Their relative contribution increases significantly after the more than five-fold compression of the TPC data by the HLT in 2016.
Overall, the HLT compression increases the total number of events \mbox{ALICE} can record and store by more than a factor of~$4$ within the given storage budget.
In the case of Pb--Pb data taking, also the raw data bandwidth would exceed the available capacity necessitating the real-time compression in the HLT.
Aggregating all compression steps of the \run{3} prototype, a total compression factor of~$9$ was achieved for both pp and for Pb--Pb data.
In the future, additional compression steps are foreseen, like rejecting TPC clusters attached to tracks with transverse momenta below~$50$\,MeV/$c$, clusters attached to additional legs of looping tracks, and clusters attached to track segments with large inclination angles, which are not used in physics analyses.
Using this cluster rejection, an additional compression factor of~$2$ is expected, bringing the compression factor close to the foreseen factor of 20, which is necessary for the $\text{O}^2$ computing upgrade.

\subsection{Quality Assurance for TPC, EMCAL, and other detectors}

The HLT, in addition to online reconstruction and compression, also runs various types of QA and physics analysis components that allow for real-time monitoring of the physics performance of the \mbox{ALICE} apparatus.
These frameworks gather and process various types of information: from event, track and vertex properties to data compression parameters.
The HLT components executing these frameworks can be classified as fast, slow, and/or asynchronous.
The fast components (\eg EMCal and HLT's own QA) require the full data sample and therefore are considered prompt components, running in-chain.
Slow components that simply sample some of the reconstructed events are executed out-of-chain, subscribing to the main chain transiently on a dedicated monitoring node, processing events on a best effort basis.
Finally, some QA components run asynchronously on all nodes using the wrapper for the \mbox{ALICE} physics analysis task framework, which was developed for and is also used in online calibration \cite{bib:chep2016calibration} (compare Section~\ref{sec:calibration}).

In the asynchronous mode, the full statistics (or a subset proportional to the dedicated processing capacity) can be processed without disrupting the standard HLT operations.
Several tasks from the TPC team are now running within the HLT in this mode.
Another component that runs out-of-chain is the luminous region component, which provides information on the size and position of the region of the particle beams to the LHC team.
In this case, all event information of interest are processed synchronously, with the merging and fitting stages being performed out-of-chain.
The LHC is updated with these data in 30 second intervals.

In addition to running asynchronously the HLT also performs online monitoring syncronously.
This allows access to the full data sample, however the components must be very stable to not interfere with HLT operations.
With this infrastructure histograms can be created and modified on-the-fly to allow for \eg prompt studies of trigger selections per histogram.
This infrastructure also supports the correlation of arbitrary quantities like V0/T0/ZDC detector signals versus the number of ITS/TPC tracks.
The benefit is that all events can be processed by running syncronously at the full event rate.
The histograms produced by the in-chain components are continuously merged (asynchronously) and can be accessed at any time during the run.
The detector teams of TPC and EMCal have also implemented similar QA tasks, which run in an analogous manner.
The final monitoring histograms, run- and time-dependent, are published online for simple access.
Furthermore, the data for HLT monitoring are available on the data quality monitoring station utilized by the \mbox{ALICE} shift crew.

Data exchange between asynchronous components and a part of external communication (\eg related to QA) is handled by the ZeroMQ messaging library.
The processing components need to exchange a multitude of data types related to a single entity (\eg a triggered event).
Data originates from different sources, \eg shared memory holding raw or reconstructed data processed syncronously, buffers provided by serialization libraries and schema evolution data (\eg ROOT streamers).
Each data buffer needs to be uniquely identified to allow for correct decoding on the receiving end:
metadata that annotates the contents and serialization strategy of a data buffer are constructed separately and sent together with the data.
The association of metadata to data buffers is maintained at the transport layer level by ordering the header-payload pairs in a sequence.
The ZeroMQ multi-part functionality allows for the atomic transport of multiple buffers and buffer ordering preservation; it is wrapped by a thin abstraction layer to provide an easy to use vectored input/output-like interface.
In this scheme, many annotated data parts can efficiently be added to a single ZeroMQ message without the overhead being typically associated to message (de-)serialization \cite{bib:chep2016hlt}.

\section{Performance analysis of global HLT operation}

\label{sec:maxrate}

In addition to system stability, the HLT must ensure that during normal operations any throttling of the data taking of the experiment is avoided.
When one of the HLT processing components is too slow to process the incoming data, for example when the network can not manage the data rate or when the framework cannot schedule the events, the HLT internal buffers become full and this results in the HLT sending back-pressure to the experiment and pausing data taking until there is again buffer space to accept more data.
\\
\begin{table*}[htb]
\begin{center}
\caption{Maximum data rates and event rates in the HLT for different load scenarios in data replay.}
\scalebox{0.75}{
\begin{tabular}{llrrrl}
\hline
Scenario & Detectors & Input size & TPC rate & Total event rate & Limiting factor \\
\hline
Single Input Link & ZDC & $6$\,MB/s & 0 & $10$\,kHz & Framework \\
pp $5.02$\,TeV  & TPC, ITS, EMCAL, V0 & $8.3$\,GB/s & $4.5$\,kHz & $4.5$\,kHz & CPU load \\
pp $13$~TeV  & TPC, ITS, EMCAL, V0 & $48$\,GB/s & $2.4$\,kHz & $2.4$\,kHz & optical link bandwidth \\
Pb--Pb $5.02$~TeV & TPC, ITS, EMCAL, V0, ZDC & $48$\,GB/s & $950$\,Hz & $950$\,Hz & optical link bandwidth \\
Pb--Pb $5.02$~TeV & ITS, EMCAL, V0, ZDC & $3.5$\,GB/s & 0 & $6$\,kHz & Framework \\
\hline
pp $13$~TeV & All & $49$\,GB/s & $2.4$\,kHz & $6$\,kHz & optical link bandwidth / Framework \\
Pb--Pb $5.02$~TeV & All & $51$\,GB/s & $950$\,Hz & $3.75$\,kHz & optical link bandwidth / CPU \\
\hline
\end{tabular}}
\label{tab:hltrate}
\end{center}
\end{table*}

Data rate and event rate are, although related, two different factors.
For instance, small events at very high rate cause excessive load on the scheduling of related interprocess communication while the utilized network bandwidth can still be small.
On the other hand, a few large events can saturate the network.

In 2016, the HLT caused on average less than~$100$\,$\mu$s of back-pressure per run, an insignificant amount compared to the usual run duration of several hours.
Therefore, the HLT has a negligible effect on the data taking efficiency.
Besides observations during the operation, extensive data-replay based measurements were conducted to ensure that the HLT manages to process all data and event rates for all the foreseen data taking and trigger scenarios.

Data replay (see Section~\ref{sec:rorc}) allows for the evaluation of the HLT performance under a certain load scenario given the exact same conditions as in normal operation.
The maximum input data rate into the HLT is limited by the number and the link speed of the optical link fibers coming from the detectors.
The dominant contribution is from the TPC with~$216$~links, each running at~$3.125$\,GBit/s.
However, not all links can send data at full rate simultaneously.
This is due to the geometry of the TPC resulting in the number of channels sent per link not being constant.
Considering also the link protocol overhead, the maximum possible input rate from the TPC is~$48$\,GB/s.
Note that, during real operation, the TPC pauses the readout during sampling and that the TPC gating grid and detector busy time reduce the maximum rate.
Therefore, the real rate is below~$40$\,GB/s at less than~$2$\,kHz, which gives some additional margin.
Additionally~$10$\,GB/s can originate from the other detectors.
In the following, the data replay is analyzed using two data sets: pp events at high luminosity and maximum pile-up as well as minimum bias Pb--Pb events.
Considering the data size, the replay of the Pb--Pb data set was run at~$950$\,Hz and the pp data set at~$2.5$\,kHz, which correspond to a TPC input rate of~$48$\,GB/s in both cases.
Other detectors can operate at higher event rates than the TPC.
The \mbox{ALICE} trigger scenarios for \run{2} foresaw a rate below~$2$\,kHz for the central barrel detectors with an additional few hundred Hz from both the fast-interaction detectors and the muon detector.
This results in a total maximum aggregate event rate below~$3.5$\,kHz when all trigger clusters are at maximum rate at the same time.
A mixture of additional events without TPC contribution, to obtain a higher event rate for other detectors, was added to the replay data set.

\tabl{hltrate} gives an overview of the maximum rate handled by the HLT for various scenarios.
The HLT framework imposes an event-rate limit of~$10$\,kHz for a front-end node with a single input link, and a limit of~$6$\,kHz for event-merging of the twelve links of a fully connected \mbox{C-RORC}~\cite{bib:chep2016framework}.
Both limits do not apply in practice because the fastest foreseen trigger scenario peaks at an aggregate rate of~$3.5$\,kHz.
The table also shows that the CPU capacity will only become critical at event rates not supported by the detectors.
The current GPU-based tracking achieves a peak TPC processing rate of~$2.4$\,kHz for Pb--Pb data with an~$8$\,kHz interaction rate, if it runs locally and standalone using all compute nodes.
Currently, this leaves a~$50$\% margin on the GPU capacity, which can be used for the implementation of additional compute-intensive online reconstruction steps.

The experience during \run{1} demonstrates that the performance of the event merging task is critical for the maximum achievable rate.
Consequently, both the interprocess communication in the framework (see Section~\ref{sec:framework}) as well as the HLT configuration was improved by having a more balanced layout, thereby reducing the load on the event fragment merger.
These changes improved the maximum rates from~$3$ to~$6$\,kHz for pp and from~$500$ to~$950$\,Hz for Pb--Pb collisions.

In addition to the input and processing capacity, the HLT must have sufficient bandwidth of the internal network and of the output links to DAQ.
The above scenarios lead to a maximum outgoing network bandwidth of~$1.38$\,GB/s per input node, and for the current TPC data compression factor, a maximum of~$1.53$\,GB/s received per output node.
In total, $10.7$\,GB/s are sent to DAQ.
Using the current HLT chain without processing, the framework has been tested up to input and output rates of~$2.4$\,GB/s per node.
This leaves already a margin of more than~$50$\%, and close to~$6$\,GB/s of network bandwidth per node accessible through the use of multiple transport streams.
The optical link rates to DAQ and data storage systems have been tested to a maximum aggregate transfer speed of~$12$\,GB/s, which is above the upper bound for the output rate of~$10.7$\,GB/s.
For the current HLT chain, the limiting factor is the actual link speed of the~$28$ fibers to DAQ, which are running at the highest possible speed of~$5.3$\,GBit.
The output rate could be increased by using more physical fibers, for which the HLT already has spare ports available.
As presented in Section~\ref{sec:rorc}, the HLT C-RORC and the PCI Express bus in the FEP nodes are also able to handle any incoming data from the up to twelve optical links per node.

Overall, the current HLT farm handles all foreseen workloads for \run{2} without imposing backpressure.
Looking ahead to \run{3},  the available resources will be used to test and prototype the many new features planned for the $\text{O}^2$ system as early as possible under real conditions in the HLT.

\subsection{HLT operation stability}

\label{sec:stability}

The HLT is an integral part of the \mbox{ALICE} data taking chain and its operational stability is critical because a failure would interrupt the data taking.
Moreover, without the HLT compression a maximum readout rate is no longer possible due to the bottlenecks described above.
In addition, storage space becomes a problem due to the fact that uncompressed raw data quadruples storage requirements.

\begin{figure}[htb]
\begin{centering}
\includegraphics[width=0.8\textwidth]{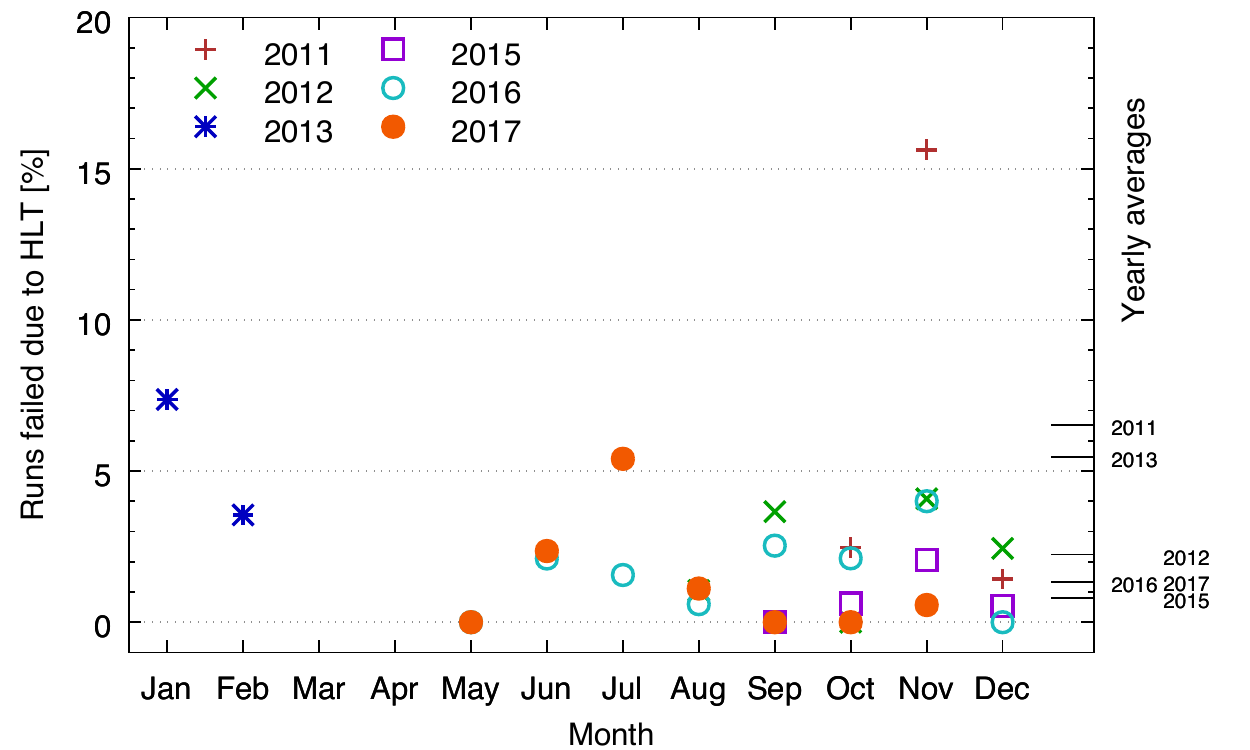}
\par\end{centering}
\caption{Number of data taking runs terminated due to failure in the HLT during \run{1} and \run{2} since 2011, when the TPC data compression in the HLT was introduced.
  Missing months correspond to long shutdowns, end-of-year shutdowns, commissioning phases for the data compression and recommissioning for updated TPC readout.
  The yearly averages are shown as long tick marks along the right-side y-axis.
}
\label{fig:stability}
\end{figure}

\figur{stability} shows the stability of the HLT over the period from 2011 to 2017.
The figure includes only runs in which \mbox{ALICE} was collecting physics data and with the HLT performing TPC data compression.
Overall, only a small percentage of data taking runs ended due to HLT problems.
Since October 2011, there were only three months during which the HLT caused more than~$5$\% of the data taking runs to fail.
The largest percentage of failures occurred in November 2011.
At that time the TPC compression was still in the commissioning phase and for the first time Pb--Pb data were collected at a higher interaction rate with respect to \run{1}, which in turn demanded additional fine tuning.
On average the end-of-run reasons associated with HLT failures were less than~$2$\%.
Compared to the beginning of data taking during \run{1}~\cite{hlt-run1-2012}, the fraction of runs failing due to HLT issues was reduced by roughly a factor of~$2$.
The absolute rate, which was above~$100$ failures per year in \run{1} decreased considerably.
In total, only~$18$ physics runs failed due to an HLT issue in 2016.
The causes were:
GPU driver problems causing reboots solved by driver update (2 runs),
GPU stuck due to driver problems (4 runs),
malfunctioning GPU (2 runs),
malfunctioning CPU (2 runs),
unexpected node reboot (1 run),
uncorrectable machine check exception (4 runs), and
network communication problems (3 runs).

A significant fraction of failures are due to GPU driver problems, which are still not fully resolved by the vendor.
A workaround was implemented that outsources the GPU reconstruction to a different operating system process.
If the GPU or the driver get stuck, the HLT chain continues normal operation and skips the track reconstruction for the few events scheduled for that process.
This reduces the statistics for online QA and calibration only negligibly.

Despite several failures that happened during the debugging of a reoccurring hardware problem at the beginning of July, the HLT had a low failure rate in 2017.
In the future, additional preemptive measures will be deployed for network, hard disk, and machine check failures, to reduce the failure rate even further.

For a better estimate of how much data taking time is lost due to HLT failures, the total time between the occurrence of the problem and the moment in which data are recorded is again calculated.
This includes possible unsuccessful measures to continue data taking without a full restart as well as stop and startup time.
This is a more realistic measurement of the dead time, compared to an estimate based on the time between end-of-run and start of the next data taking run.
However, this is an upper bound, as it also includes time for unrelated actions performed in between.
In this metric, the~$18$ failures in 2016 interrupted the data taking for~$11233$~seconds in total out of~$1409$~hours of data taking with HLT in its full configuration.
This means that the HLT failures amounted to less than~$0.22$\% of the available data taking time.

\section{Outlook}

The \mbox{ALICE} HLT has been operational since November~2009 with the first pp collisions at LHC at $\sqrt{s}$ = $0.9$\,TeV and has since then processed all subsequent data.
Operated with a combination of a fast FPGA hardware cluster finder and GPU tracker the \mbox{ALICE} HLT pioneered the use of hardware accelerator technologies in real-time computing at the LHC.

During the LHC \run{3} \mbox{ALICE} will collect 100 times more data with respect to what was recorded during \run{1} and \run{2}.
The increase in statistics will be made possible by a tenfold increase of the LHC luminosity as well as the change of the detector readout mode from triggered to continuous, allowing the readout of the full Pb--Pb interaction rate of up to 50 kHz.
The data stream has to be compressed by a factor of 20 in order to be transported to the storage element for permanent storage.
Achieving this compression ratio requires a paradigm shift in processing: all data will be reconstructed and calibrated online synchronous to the data taking.
In addition to compression schemes already discussed in Section~\ref{sec:compression},
parts of the data, \eg clusters positively identified to be disposable, will be discarded making the overall compression scheme lossy.
The quality of online reconstruction and calibration will therefore be paramount.

Concepts and technologies which are developed as part of the HLT (described in this paper and in Sections~\ref{sec:framework}, \ref{sec:tracking}, and \ref{sec:compression}) are being studied, prototyped and tested already now in a production environment, also being adapted and further developed in the software framework of $\text{O}^2$.

\newenvironment{acknowledgement}{\relax}{\relax}
\begin{acknowledgement}
\section*{Acknowledgements}
\input{fa_2018-11-23.tex}
We thank AMD and ASUS for their support in commissioning the compute farm.
\end{acknowledgement}

\bibliographystyle{utphys.bst}   
\bibliography{citations.bib}

\newpage
\appendix
\section{The ALICE Collaboration}
\label{app:collab}
\input{2018-11-24-Alice_Authorlist_2018-Nov-24.tex}  
\end{document}

%% file: fa_2018-11-23.tex

The ALICE Collaboration would like to thank all its engineers and technicians for their invaluable contributions to the construction of the experiment and the CERN accelerator teams for the outstanding performance of the LHC complex.
The ALICE Collaboration gratefully acknowledges the resources and support provided by all Grid centres and the Worldwide LHC Computing Grid (WLCG) collaboration.
The ALICE Collaboration acknowledges the following funding agencies for their support in building and running the ALICE detector:
A. I. Alikhanyan National Science Laboratory (Yerevan Physics Institute) Foundation (ANSL), State Committee of Science and World Federation of Scientists (WFS), Armenia;
Austrian Academy of Sciences and Nationalstiftung f\"{u}r Forschung, Technologie und Entwicklung, Austria;
Ministry of Communications and High Technologies, National Nuclear Research Center, Azerbaijan;
Conselho Nacional de Desenvolvimento Cient\'{\i}fico e Tecnol\'{o}gico (CNPq), Universidade Federal do Rio Grande do Sul (UFRGS), Financiadora de Estudos e Projetos (Finep) and Funda\c{c}\~{a}o de Amparo \`{a} Pesquisa do Estado de S\~{a}o Paulo (FAPESP), Brazil;
Ministry of Science \& Technology of China (MSTC), National Natural Science Foundation of China (NSFC) and Ministry of Education of China (MOEC) , China;
Croatian Science Foundation and Ministry of Science and Education, Croatia;
Centro de Aplicaciones Tecnol\'{o}gicas y Desarrollo Nuclear (CEADEN), Cubaenerg\'{\i}a, Cuba;
Ministry of Education, Youth and Sports of the Czech Republic, Czech Republic;
The Danish Council for Independent Research | Natural Sciences, the Carlsberg Foundation and Danish National Research Foundation (DNRF), Denmark;
Helsinki Institute of Physics (HIP), Finland;
Commissariat \`{a} l'Energie Atomique (CEA) and Institut National de Physique Nucl\'{e}aire et de Physique des Particules (IN2P3) and Centre National de la Recherche Scientifique (CNRS), France;
Bundesministerium f\"{u}r Bildung, Wissenschaft, Forschung und Technologie (BMBF) and GSI Helmholtzzentrum f\"{u}r Schwerionenforschung GmbH, Germany;
General Secretariat for Research and Technology, Ministry of Education, Research and Religions, Greece;
National Research, Development and Innovation Office, Hungary;
Department of Atomic Energy Government of India (DAE), Department of Science and Technology, Government of India (DST), University Grants Commission, Government of India (UGC) and Council of Scientific and Industrial Research (CSIR), India;
Indonesian Institute of Science, Indonesia;
Centro Fermi - Museo Storico della Fisica e Centro Studi e Ricerche Enrico Fermi and Istituto Nazionale di Fisica Nucleare (INFN), Italy;
Institute for Innovative Science and Technology , Nagasaki Institute of Applied Science (IIST), Japan Society for the Promotion of Science (JSPS) KAKENHI and Japanese Ministry of Education, Culture, Sports, Science and Technology (MEXT), Japan;
Consejo Nacional de Ciencia (CONACYT) y Tecnolog\'{i}a, through Fondo de Cooperaci\'{o}n Internacional en Ciencia y Tecnolog\'{i}a (FONCICYT) and Direcci\'{o}n General de Asuntos del Personal Academico (DGAPA), Mexico;
Nederlandse Organisatie voor Wetenschappelijk Onderzoek (NWO), Netherlands;
The Research Council of Norway, Norway;
Commission on Science and Technology for Sustainable Development in the South (COMSATS), Pakistan;
Pontificia Universidad Cat\'{o}lica del Per\'{u}, Peru;
Ministry of Science and Higher Education and National Science Centre, Poland;
Korea Institute of Science and Technology Information and National Research Foundation of Korea (NRF), Republic of Korea;
Ministry of Education and Scientific Research, Institute of Atomic Physics and Romanian National Agency for Science, Technology and Innovation, Romania;
Joint Institute for Nuclear Research (JINR), Ministry of Education and Science of the Russian Federation, National Research Centre Kurchatov Institute, Russian Science Foundation and Russian Foundation for Basic Research, Russia;
Ministry of Education, Science, Research and Sport of the Slovak Republic, Slovakia;
National Research Foundation of South Africa, South Africa;
Swedish Research Council (VR) and Knut \& Alice Wallenberg Foundation (KAW), Sweden;
European Organization for Nuclear Research, Switzerland;
National Science and Technology Development Agency (NSDTA), Suranaree University of Technology (SUT) and Office of the Higher Education Commission under NRU project of Thailand, Thailand;
Turkish Atomic Energy Agency (TAEK), Turkey;
National Academy of  Sciences of Ukraine, Ukraine;
Science and Technology Facilities Council (STFC), United Kingdom;
National Science Foundation of the United States of America (NSF) and United States Department of Energy, Office of Nuclear Physics (DOE NP), United States of America.

%% file: 2018-11-24-Alice_Authorlist_2018-Nov-24.tex

\begingroup
\small
\begin{flushleft}
S.~Acharya\Irefn{org140}\And 
F.T.-.~Acosta\Irefn{org20}\And 
D.~Adamov\'{a}\Irefn{org93}\And 
S.P.~Adhya\Irefn{org140}\And 
A.~Adler\Irefn{org74}\And 
J.~Adolfsson\Irefn{org80}\And 
M.M.~Aggarwal\Irefn{org98}\And 
G.~Aglieri Rinella\Irefn{org34}\And 
M.~Agnello\Irefn{org31}\And 
Z.~Ahammed\Irefn{org140}\And 
S.~Ahmad\Irefn{org17}
% ...
\And
\renewcommand\labelenumi{\textsuperscript{\theenumi}~}

\section*{Affiliation notes}
\renewcommand\theenumi{\roman{enumi}}
\begin{Authlist}
\item \Adef{org*}Deceased
\item \Adef{orgI}Dipartimento DET del Politecnico di Torino, Turin, Italy
\item \Adef{orgII}M.V. Lomonosov Moscow State University, D.V. Skobeltsyn Institute of Nuclear, Physics, Moscow, Russia
\item \Adef{orgIII}Department of Applied Physics, Aligarh Muslim University, Aligarh, India
\item \Adef{orgIV}Institute of Theoretical Physics, University of Wroclaw, Poland
\end{Authlist}

\section*{Collaboration Institutes}
\renewcommand\theenumi{\arabic{enumi}~}
\begin{Authlist}
\item \Idef{org1}A.I. Alikhanyan National Science Laboratory (Yerevan Physics Institute) Foundation, Yerevan, Armenia
\item \Idef{org2}Bogolyubov Institute for Theoretical Physics, National Academy of Sciences of Ukraine, Kiev, Ukraine
\item \Idef{org3}Bose Institute, Department of Physics  and Centre for Astroparticle Physics and Space Science (CAPSS), Kolkata, India
\item \Idef{org4}Budker Institute for Nuclear Physics, Novosibirsk, Russia
\item \Idef{org5}California Polytechnic State University, San Luis Obispo, California, United States
\item \Idef{org6}Central China Normal University, Wuhan, China
\item \Idef{org7}Centre de Calcul de l'IN2P3, Villeurbanne, Lyon, France
\item \Idef{org8}Centro de Aplicaciones Tecnol\'{o}gicas y Desarrollo Nuclear (CEADEN), Havana, Cuba
\item \Idef{org9}Centro de Investigaci\'{o}n y de Estudios Avanzados (CINVESTAV), Mexico City and M\'{e}rida, Mexico
\item \Idef{org10}Centro Fermi - Museo Storico della Fisica e Centro Studi e Ricerche ``Enrico Fermi', Rome, Italy
\item \Idef{org11}Chicago State University, Chicago, Illinois, United States
\item \Idef{org12}China Institute of Atomic Energy, Beijing, China
\item \Idef{org13}Chonbuk National University, Jeonju, Republic of Korea
\item \Idef{org14}Comenius University Bratislava, Faculty of Mathematics, Physics and Informatics, Bratislava, Slovakia
\item \Idef{org15}COMSATS Institute of Information Technology (CIIT), Islamabad, Pakistan
\item \Idef{org16}Creighton University, Omaha, Nebraska, United States
\item \Idef{org17}Department of Physics, Aligarh Muslim University, Aligarh, India
\item \Idef{org18}Department of Physics, Pusan National University, Pusan, Republic of Korea
\item \Idef{org19}Department of Physics, Sejong University, Seoul, Republic of Korea
\item \Idef{org20}Department of Physics, University of California, Berkeley, California, United States
\item \Idef{org21}Department of Physics, University of Oslo, Oslo, Norway
\item \Idef{org22}Department of Physics and Technology, University of Bergen, Bergen, Norway
\item \Idef{org23}Dipartimento di Fisica dell'Universit\`{a} 'La Sapienza' and Sezione INFN, Rome, Italy
\item \Idef{org24}Dipartimento di Fisica dell'Universit\`{a} and Sezione INFN, Cagliari, Italy
\item \Idef{org25}Dipartimento di Fisica dell'Universit\`{a} and Sezione INFN, Trieste, Italy
\item \Idef{org26}Dipartimento di Fisica dell'Universit\`{a} and Sezione INFN, Turin, Italy
\item \Idef{org27}Dipartimento di Fisica e Astronomia dell'Universit\`{a} and Sezione INFN, Bologna, Italy
\item \Idef{org28}Dipartimento di Fisica e Astronomia dell'Universit\`{a} and Sezione INFN, Catania, Italy
\item \Idef{org29}Dipartimento di Fisica e Astronomia dell'Universit\`{a} and Sezione INFN, Padova, Italy
\item \Idef{org30}Dipartimento di Fisica `E.R.~Caianiello' dell'Universit\`{a} and Gruppo Collegato INFN, Salerno, Italy
\item \Idef{org31}Dipartimento DISAT del Politecnico and Sezione INFN, Turin, Italy
\item \Idef{org32}Dipartimento di Scienze e Innovazione Tecnologica dell'Universit\`{a} del Piemonte Orientale and INFN Sezione di Torino, Alessandria, Italy
\item \Idef{org33}Dipartimento Interateneo di Fisica `M.~Merlin' and Sezione INFN, Bari, Italy
\item \Idef{org34}European Organization for Nuclear Research (CERN), Geneva, Switzerland
\item \Idef{org35}Faculty of Electrical Engineering, Mechanical Engineering and Naval Architecture, University of Split, Split, Croatia
\item \Idef{org36}Faculty of Engineering and Science, Western Norway University of Applied Sciences, Bergen, Norway
\item \Idef{org37}Faculty of Nuclear Sciences and Physical Engineering, Czech Technical University in Prague, Prague, Czech Republic
\item \Idef{org38}Faculty of Science, P.J.~\v{S}af\'{a}rik University, Ko\v{s}ice, Slovakia
\item \Idef{org39}Frankfurt Institute for Advanced Studies, Johann Wolfgang Goethe-Universit\"{a}t Frankfurt, Frankfurt, Germany
\item \Idef{org40}Gangneung-Wonju National University, Gangneung, Republic of Korea
\item \Idef{org41}Gauhati University, Department of Physics, Guwahati, India
\item \Idef{org42}Helmholtz-Institut f\"{u}r Strahlen- und Kernphysik, Rheinische Friedrich-Wilhelms-Universit\"{a}t Bonn, Bonn, Germany
\item \Idef{org43}Helsinki Institute of Physics (HIP), Helsinki, Finland
\item \Idef{org44}High Energy Physics Group,  Universidad Aut\'{o}noma de Puebla, Puebla, Mexico
\item \Idef{org45}Hiroshima University, Hiroshima, Japan
\item \Idef{org46}Hochschule Worms, Zentrum  f\"{u}r Technologietransfer und Telekommunikation (ZTT), Worms, Germany
\item \Idef{org47}Horia Hulubei National Institute of Physics and Nuclear Engineering, Bucharest, Romania
\item \Idef{org48}Indian Institute of Technology Bombay (IIT), Mumbai, India
\item \Idef{org49}Indian Institute of Technology Indore, Indore, India
\item \Idef{org50}Indonesian Institute of Sciences, Jakarta, Indonesia
\item \Idef{org51}INFN, Laboratori Nazionali di Frascati, Frascati, Italy
\item \Idef{org52}INFN, Sezione di Bari, Bari, Italy
\item \Idef{org53}INFN, Sezione di Bologna, Bologna, Italy
\item \Idef{org54}INFN, Sezione di Cagliari, Cagliari, Italy
\item \Idef{org55}INFN, Sezione di Catania, Catania, Italy
\item \Idef{org56}INFN, Sezione di Padova, Padova, Italy
\item \Idef{org57}INFN, Sezione di Roma, Rome, Italy
\item \Idef{org58}INFN, Sezione di Torino, Turin, Italy
\item \Idef{org59}INFN, Sezione di Trieste, Trieste, Italy
\item \Idef{org60}Inha University, Incheon, Republic of Korea
\item \Idef{org61}Institut de Physique Nucl\'{e}aire d'Orsay (IPNO), Institut National de Physique Nucl\'{e}aire et de Physique des Particules (IN2P3/CNRS), Universit\'{e} de Paris-Sud, Universit\'{e} Paris-Saclay, Orsay, France
\item \Idef{org62}Institute for Nuclear Research, Academy of Sciences, Moscow, Russia
\item \Idef{org63}Institute for Subatomic Physics, Utrecht University/Nikhef, Utrecht, Netherlands
\item \Idef{org64}Institute for Theoretical and Experimental Physics, Moscow, Russia
\item \Idef{org65}Institute of Experimental Physics, Slovak Academy of Sciences, Ko\v{s}ice, Slovakia
\item \Idef{org66}Institute of Physics, Homi Bhabha National Institute, Bhubaneswar, India
\item \Idef{org67}Institute of Physics of the Czech Academy of Sciences, Prague, Czech Republic
\item \Idef{org68}Institute of Space Science (ISS), Bucharest, Romania
\item \Idef{org69}Institut f\"{u}r Kernphysik, Johann Wolfgang Goethe-Universit\"{a}t Frankfurt, Frankfurt, Germany
\item \Idef{org70}Instituto de Ciencias Nucleares, Universidad Nacional Aut\'{o}noma de M\'{e}xico, Mexico City, Mexico
\item \Idef{org71}Instituto de F\'{i}sica, Universidade Federal do Rio Grande do Sul (UFRGS), Porto Alegre, Brazil
\item \Idef{org72}Instituto de F\'{\i}sica, Universidad Nacional Aut\'{o}noma de M\'{e}xico, Mexico City, Mexico
\item \Idef{org73}iThemba LABS, National Research Foundation, Somerset West, South Africa
\item \Idef{org74}Johann-Wolfgang-Goethe Universit\"{a}t Frankfurt Institut f\"{u}r Informatik, Fachbereich Informatik und Mathematik, Frankfurt, Germany
\item \Idef{org75}Joint Institute for Nuclear Research (JINR), Dubna, Russia
\item \Idef{org76}Korea Institute of Science and Technology Information, Daejeon, Republic of Korea
\item \Idef{org77}KTO Karatay University, Konya, Turkey
\item \Idef{org78}Laboratoire de Physique Subatomique et de Cosmologie, Universit\'{e} Grenoble-Alpes, CNRS-IN2P3, Grenoble, France
\item \Idef{org79}Lawrence Berkeley National Laboratory, Berkeley, California, United States
\item \Idef{org80}Lund University Department of Physics, Division of Particle Physics, Lund, Sweden
\item \Idef{org81}Nagasaki Institute of Applied Science, Nagasaki, Japan
\item \Idef{org82}Nara Women{'}s University (NWU), Nara, Japan
\item \Idef{org83}National and Kapodistrian University of Athens, School of Science, Department of Physics , Athens, Greece
\item \Idef{org84}National Centre for Nuclear Research, Warsaw, Poland
\item \Idef{org85}National Institute of Science Education and Research, Homi Bhabha National Institute, Jatni, India
\item \Idef{org86}National Nuclear Research Center, Baku, Azerbaijan
\item \Idef{org87}National Research Centre Kurchatov Institute, Moscow, Russia
\item \Idef{org88}Niels Bohr Institute, University of Copenhagen, Copenhagen, Denmark
\item \Idef{org89}Nikhef, National institute for subatomic physics, Amsterdam, Netherlands
\item \Idef{org90}NRC Kurchatov Institute IHEP, Protvino, Russia
\item \Idef{org91}NRNU Moscow Engineering Physics Institute, Moscow, Russia
\item \Idef{org92}Nuclear Physics Group, STFC Daresbury Laboratory, Daresbury, United Kingdom
\item \Idef{org93}Nuclear Physics Institute of the Czech Academy of Sciences, \v{R}e\v{z} u Prahy, Czech Republic
\item \Idef{org94}Oak Ridge National Laboratory, Oak Ridge, Tennessee, United States
\item \Idef{org95}Ohio State University, Columbus, Ohio, United States
\item \Idef{org96}Petersburg Nuclear Physics Institute, Gatchina, Russia
\item \Idef{org97}Physics department, Faculty of science, University of Zagreb, Zagreb, Croatia
\item \Idef{org98}Physics Department, Panjab University, Chandigarh, India
\item \Idef{org99}Physics Department, University of Jammu, Jammu, India
\item \Idef{org100}Physics Department, University of Rajasthan, Jaipur, India
\item \Idef{org101}Physikalisches Institut, Eberhard-Karls-Universit\"{a}t T\"{u}bingen, T\"{u}bingen, Germany
\item \Idef{org102}Physikalisches Institut, Ruprecht-Karls-Universit\"{a}t Heidelberg, Heidelberg, Germany
\item \Idef{org103}Physik Department, Technische Universit\"{a}t M\"{u}nchen, Munich, Germany
\item \Idef{org104}Research Division and ExtreMe Matter Institute EMMI, GSI Helmholtzzentrum f\"ur Schwerionenforschung GmbH, Darmstadt, Germany
\item \Idef{org105}Rudjer Bo\v{s}kovi\'{c} Institute, Zagreb, Croatia
\item \Idef{org106}Russian Federal Nuclear Center (VNIIEF), Sarov, Russia
\item \Idef{org107}Saha Institute of Nuclear Physics, Homi Bhabha National Institute, Kolkata, India
\item \Idef{org108}School of Physics and Astronomy, University of Birmingham, Birmingham, United Kingdom
\item \Idef{org109}Secci\'{o}n F\'{\i}sica, Departamento de Ciencias, Pontificia Universidad Cat\'{o}lica del Per\'{u}, Lima, Peru
\item \Idef{org110}Shanghai Institute of Applied Physics, Shanghai, China
\item \Idef{org111}St. Petersburg State University, St. Petersburg, Russia
\item \Idef{org112}Stefan Meyer Institut f\"{u}r Subatomare Physik (SMI), Vienna, Austria
\item \Idef{org113}SUBATECH, IMT Atlantique, Universit\'{e} de Nantes, CNRS-IN2P3, Nantes, France
\item \Idef{org114}Suranaree University of Technology, Nakhon Ratchasima, Thailand
\item \Idef{org115}Technical University of Ko\v{s}ice, Ko\v{s}ice, Slovakia
\item \Idef{org116}Technische Universit\"{a}t M\"{u}nchen, Excellence Cluster 'Universe', Munich, Germany
\item \Idef{org117}The Henryk Niewodniczanski Institute of Nuclear Physics, Polish Academy of Sciences, Cracow, Poland
\item \Idef{org118}The University of Texas at Austin, Austin, Texas, United States
\item \Idef{org119}Universidad Aut\'{o}noma de Sinaloa, Culiac\'{a}n, Mexico
\item \Idef{org120}Universidade de S\~{a}o Paulo (USP), S\~{a}o Paulo, Brazil
\item \Idef{org121}Universidade Estadual de Campinas (UNICAMP), Campinas, Brazil
\item \Idef{org122}Universidade Federal do ABC, Santo Andre, Brazil
\item \Idef{org123}University College of Southeast Norway, Tonsberg, Norway
\item \Idef{org124}University of Cape Town, Cape Town, South Africa
\item \Idef{org125}University of Houston, Houston, Texas, United States
\item \Idef{org126}University of Jyv\"{a}skyl\"{a}, Jyv\"{a}skyl\"{a}, Finland
\item \Idef{org127}University of Liverpool, Liverpool, United Kingdom
\item \Idef{org128}University of Science and Techonology of China, Hefei, China
\item \Idef{org129}University of Tennessee, Knoxville, Tennessee, United States
\item \Idef{org130}University of the Witwatersrand, Johannesburg, South Africa
\item \Idef{org131}University of Tokyo, Tokyo, Japan
\item \Idef{org132}University of Tsukuba, Tsukuba, Japan
\item \Idef{org133}Universit\'{e} Clermont Auvergne, CNRS/IN2P3, LPC, Clermont-Ferrand, France
\item \Idef{org134}Universit\'{e} de Lyon, Universit\'{e} Lyon 1, CNRS/IN2P3, IPN-Lyon, Villeurbanne, Lyon, France
\item \Idef{org135}Universit\'{e} de Strasbourg, CNRS, IPHC UMR 7178, F-67000 Strasbourg, France, Strasbourg, France
\item \Idef{org136} Universit\'{e} Paris-Saclay Centre d¿\'Etudes de Saclay (CEA), IRFU, Department de Physique Nucl\'{e}aire (DPhN), Saclay, France
\item \Idef{org137}Universit\`{a} degli Studi di Foggia, Foggia, Italy
\item \Idef{org138}Universit\`{a} degli Studi di Pavia, Pavia, Italy
\item \Idef{org139}Universit\`{a} di Brescia, Brescia, Italy
\item \Idef{org140}Variable Energy Cyclotron Centre, Homi Bhabha National Institute, Kolkata, India
\item \Idef{org141}Warsaw University of Technology, Warsaw, Poland
\item \Idef{org142}Wayne State University, Detroit, Michigan, United States
\item \Idef{org143}Westf\"{a}lische Wilhelms-Universit\"{a}t M\"{u}nster, Institut f\"{u}r Kernphysik, M\"{u}nster, Germany
\item \Idef{org144}Wigner Research Centre for Physics, Hungarian Academy of Sciences, Budapest, Hungary
\item \Idef{org145}Yale University, New Haven, Connecticut, United States
\item \Idef{org146}Yonsei University, Seoul, Republic of Korea
\end{Authlist}
\endgroup

%% file: cpc-hlt.bbl
\providecommand{\href}[2]{#2}\begingroup\raggedright\begin{thebibliography}{10}

\bibitem{bib:alice_citation}
{\bfseries ALICE} Collaboration, K.~Aamodt {\em et~al.}, ``{The ALICE
  experiment at the CERN LHC},'' {\em Journal of Instrumentation} {\bfseries 3}
  no.~08, (2008) S08002.

\bibitem{bib:lhc}
L.~Evans and P.~Bryant, ``{LHC Machine},''
\href{http://dx.doi.org/10.1088/1748-0221/3/08/S08001}{{\em JINST} {\bfseries
  3} (2008) S08001}.

\bibitem{MultPbPb5}
{\bfseries ALICE} Collaboration, J.~Adam {\em et~al.}, ``{Centrality dependence
  of the charged-particle multiplicity density at midrapidity in Pb-Pb
  collisions at $\sqrt{s_{\rm NN}}$ = 5.02 TeV},''
  \href{http://dx.doi.org/10.1103/PhysRevLett.116.222302}{{\em Phys. Rev.
  Lett.} {\bfseries 116} no.~22, (2016) 222302},
\href{http://arxiv.org/abs/1512.06104}{{\ttfamily arXiv:1512.06104 [nucl-ex]}}.

\bibitem{tpc}
{\bfseries ALICE} Collaboration, J.~Alme {\em et~al.}, ``The {ALICE} {TPC}, a
  large 3-dimensional tracking device with fast readout for ultra-high
  multiplicity events,''
  \href{http://dx.doi.org/10.1016/j.nima.2010.04.042}{{\em Nuclear Instruments
  and Methods in Physics Research Section A: Accelerators, Spectrometers,
  Detectors and Associated Equipment} {\bfseries 622} no.~1, (2010) 316--367}.

\bibitem{aliceits}
{\bfseries ALICE} Collaboration, G.~Dellacasa {\em et~al.}, ``{ALICE Inner
  Tracking System (ITS): Technical Design Report},'' Tech. Rep.
  CERN-LHCC-99-12, Geneva, 1999.

\bibitem{Acharya:2017lco}
{\bfseries ALICE} Collaboration, S.~Acharya {\em et~al.}, ``{The ALICE
  Transition Radiation Detector: construction, operation, and performance},''
  \href{http://dx.doi.org/10.1016/j.nima.2017.09.028}{{\em Nucl. Instrum.
  Meth.} {\bfseries A881} (2018) 88--127},
\href{http://arxiv.org/abs/1709.02743}{{\ttfamily arXiv:1709.02743
  [physics.ins-det]}}.

\bibitem{Cortese:2002kf}
{\bfseries ALICE} Collaboration, P.~Cortese {\em et~al.}, ``{ALICE: Addendum to
  the Technical Design Report of the Time Of Flight System (TOF)},'' Tech. Rep.
  CERN-LHCC-2002-016, Geneva, 2002.

\bibitem{phos}
{\bfseries ALICE} Collaboration, G.~Dellacasa {\em et~al.}, ``{ALICE Technical
  Design Report of the Photon Spectrometer (PHOS)},'' Tech. Rep.
  CERN-LHCC-99-04, Geneva, 1999.

\bibitem{emcal}
{\bfseries ALICE} Collaboration, P.~Cortese {\em et~al.}, ``{ALICE
  Electromagnetic Calorimeter Technical Design report},'' Tech. Rep.
  CERN-LHCC-2008-014, CERN-ALICE-TDR-014, Geneva, 2008.

\bibitem{bib:muon}
{\bfseries ALICE} Collaboration,
``{ALICE technical design report of the dimuon forward spectrometer},''.

\bibitem{forward}
{\bfseries ALICE} Collaboration, P.~Cortese {\em et~al.}, ``{ALICE forward
  detectors: FMD, TO and VO: Technical Design Report},'' Tech. Rep.
  CERN-LHCC-2004-025, Geneva, 2004.

\bibitem{zdc}
{\bfseries ALICE} Collaboration, G.~Dellacasa {\em et~al.}, ``{ALICE
  Zero-Degree Calorimeter (ZDC): Technical Design Report},'' Tech. Rep.
  CERN-LHCC-99-05, Geneva, 1999.

\bibitem{trigger1}
C.~Adler {\em et~al.}, ``From the big bang to massive data flow: Parallel
  computing in high energy physics experiments,'' in {\em Proceedings of the
  5th International Workshop on Applied Parallel Computing, New Paradigms for
  HPC in Industry and Academia}, PARA '00, pp.~333--341.
\newblock Springer-Verlag, London, UK, UK, 2001.

\bibitem{trigger2}
V.~{Lindenstruth} and I.~{Kisel}, ``{Overview of trigger systems},''
  \href{http://dx.doi.org/10.1016/j.nima.2004.07.267}{{\em Nuclear Instruments
  and Methods in Physics Research A} {\bfseries 535} (Dec., 2004) 48--56}.

\bibitem{trigger3}
V.~Lindenstruth, ``An extreme processor for an extreme experiment,''
  \href{http://dx.doi.org/10.1109/MM.2006.29}{{\em IEEE Micro} {\bfseries 26}
  no.~2, (Mar., 2006) 48--57}.

\bibitem{bib:rcu2}
J.~Alme {\em et~al.}, ``{RCU2 The ALICE TPC readout electronics consolidation
  for Run2},'' {\em Journal of Instrumentation} {\bfseries 8} no.~12, (2013)
  C12032. \url{http://stacks.iop.org/1748-0221/8/i=12/a=C12032}.

\bibitem{hlt-run1-2011}
T.~Kollegger, \href{http://dx.doi.org/10.1109/RTC.2012.6418366}{``{The ALICE
  High Level Trigger: The 2011 run experience},''} in {\em 2012 18th IEEE-NPSS
  Real Time Conference}, pp.~1--4.
\newblock June, 2012.

\bibitem{charmcard}
R.~E. Panse, {\em CHARM-Card: Hardware Based Cluster Control And Management
  System}.
\newblock PhD thesis, University of Heidelberg, 2009.

\bibitem{ipmi}
``Intelligent platform management interface.''
\newblock
  \url{https://www.intel.com/content/www/us/en/servers/ipmi/ipmi-second-gen-interface-spec-v2-rev1-1.html}.

\bibitem{bib:sanampaper}
D.~Rohr, S.~Kalcher, M.~Bach, A.~Alaqeeli, H.~Alzaid, D.~Eschweiler,
  V.~Lindenstruth, A.~Sakhar, A.~Alharthi, A.~Almubarak, I.~Alqwaiz, and
  R.~{Bin Suliman}, \href{http://dx.doi.org/10.1109/HPCC.2014.14}{``{An
  Energy-Efficient Multi-GPU Supercomputer},''} in {\em Proceedings of the 16th
  IEEE International Conference on High Performance Computing and
  Communications, HPCC 2014, Paris, France. IEEE}, pp.~42--45.
\newblock IEEE, 2014.

\bibitem{bib:o2}
P.~Buncic, M.~Krzewicki, and P.~{Vande Vyvre}, ``{Technical Design Report for
  the Upgrade of the Online-Offline Computing System},'' Tech. Rep.
  CERN-LHCC-2015-006. ALICE-TDR-019, CERN, Geneva, Apr, 2015.
\newblock \url{https://cds.cern.ch/record/2011297}.

\bibitem{bib:torsten_diss}
T.~Alt, {\em {An FPGA based pre-processor for the ALICE High Level-Trigger}}.
\newblock PhD thesis, Goethe-University Frankfurt, 2017.

\bibitem{c-rorc}
A.~Borga {\em et~al.}, ``{The C-RORC PCIe card and its application in the ALICE
  and ATLAS experiments},'' {\em Journal of Instrumentation} {\bfseries 10}
  no.~02, (2015) C02022.
  \url{http://stacks.iop.org/1748-0221/10/i=02/a=C02022}.

\bibitem{eschweiler:2014:pda}
D.~Eschweiler and V.~Lindenstruth, ``{The Portable Driver Architecture},'' in
  {\em {Proceedings of the 16th Real-Time Linux Workshop}}.
\newblock Open Source Automation Development Lab (OSADL), Duesseldorf, Germany,
  October, 2014.

\bibitem{bib:twepp2015}
H.~Engel, T.~Alt, T.~Breitner, A.~G. Ramirez, T.~Kollegger, M.~Krzewicki,
  J.~Lehrbach, D.~Rohr, and U.~Kebschull, ``{The ALICE High-Level Trigger
  Read-Out Upgrade for LHC Run 2},'' {\em JINST} {\bfseries 11} no.~01, (2016)
  C01041. \url{http://stacks.iop.org/1748-0221/11/i=01/a=C01041}.

\bibitem{Foreman}
``Foreman lifecycle management tool.''
\newblock \url{https://theforeman.org}. Accessed: 2018-12-03.

\bibitem{Puppet}
``Puppet software configuration management tool.''
\newblock \url{https://puppet.com/}. Accessed: 2018-12-03.

\bibitem{zabbix}
``{Zabbix LLC}.''
\newblock \url{http://www.zabbix.com/}. Accessed: 2018-12-03.

\bibitem{Bird:2014ctt}
{\bfseries ALICE} Collaboration, I.~Bird {\em et~al.}, ``{Update of the
  Computing Models of the WLCG and the LHC Experiments},'' Tech. Rep.
  CERN-LHCC-2014-014, LCG-TDR-002, Geneva, 2014.
\newblock \url{https://cds.cern.ch/record/1695401?ln=en}.

\bibitem{openstack}
T.~Rosado and J.~Bernardino,
  \href{http://dx.doi.org/10.1145/2628194.2628195}{``An overview of openstack
  architecture,''} in {\em Proceedings of the 18th International Database
  Engineering}, IDEAS '14, pp.~366--367.
\newblock 2014.
\newblock \url{http://doi.acm.org/10.1145/2628194.2628195}.

\bibitem{docker}
D.~Merkel, ``Docker: Lightweight linux containers for consistent development
  and deployment,'' {\em Linux J.} {\bfseries 2014} no.~239, (Mar., 2014) .
  \url{http://dl.acm.org/citation.cfm?id=2600239.2600241}.

\bibitem{bib:root}
R.~Brun and F.~Rademakers, ``{ROOT: An object oriented data analysis
  framework},''
\href{http://dx.doi.org/10.1016/S0168-9002(97)00048-X}{{\em Nucl. Instrum.
  Meth.} {\bfseries A389} (1997) 81--86}.

\bibitem{bib:hlt1}
M.~Richter, T.~Alt, S.~Bablok, C.~Cheshkov, P.~T. Hille, V.~Lindenstruth,
  G.~Ovrebekk, M.~Ploskon, S.~Popescu, D.~Rohrich, T.~M. Steinbeck, and J.~M.
  Thader, ``{High Level Trigger Applications for the ALICE Experiment},''
  \href{http://dx.doi.org/10.1109/TNS.2007.913469}{{\em IEEE Transactions on
  Nuclear Science} {\bfseries 55} no.~1, (2007) 133--138}.

\bibitem{framework1}
T.~M. Steinbeck, V.~Lindenstruth, and M.~W. Schulz, ``An object-oriented
  network-transparent data transportation framework,''
  \href{http://dx.doi.org/10.1109/TNS.2002.1003773}{{\em IEEE Transactions on
  Nuclear Science} {\bfseries 49} no.~2, (Apr, 2002) 455--459}.

\bibitem{framework2}
T.~M. Steinbeck, V.~Lindenstruth, D.~R{\"o}hrich, A.~S. Vestbo, and
  A.~Wiebalck, {\em A Framework for Building Distributed Data Flow Chains in
  Clusters}, \href{http://dx.doi.org/10.1007/3-540-48051-X\_45}{pp.~454--464}.
\newblock Springer Berlin Heidelberg, Berlin, Heidelberg, 2002.

\bibitem{bib:chep2016hlt}
M.~Krzewicki, V.~Lindenstruth, and A.~Collaboration, ``{ALICE HLT Run 2
  performance overview.},'' {\em Journal of Physics: Conference Series}
  {\bfseries 898} no.~3, (2017) 032056.
  \url{http://stacks.iop.org/1742-6596/898/i=3/a=032056}.

\bibitem{hltglobal}
B.~Becker, S.~Chattopadhyay, C.~Cicalo, J.~Cleymans, G.~de~Vaux, R.~.~W.
  Fearick, V.~Lindenstruth, M.~Richter, D.~Rohrich, F.~Staley, T.~M. Steinbeck,
  A.~Szostak, H.~Tilsner, R.~Weis, and Z.~Z. Vilakazi, ``Real time global tests
  of the {ALICE High Level Trigger} data transport framework,''
  \href{http://dx.doi.org/10.1109/TNS.2008.918521}{{\em IEEE Transactions on
  Nuclear Science} {\bfseries 55} no.~2, (April, 2008) 703--709}.

\bibitem{moore}
G.~E. Moore, ``{Cramming More Components onto Integrated Circuits},''
  \href{http://dx.doi.org/10.1109/jproc.1998.658762}{{\em Electronics}
  {\bfseries 38} no.~8, (Apr., 1965) 114--117}.

\bibitem{fpga2}
G.~{Grastveit}, H.~{Helstrup}, V.~{Lindenstruth}, C.~{Loizides}, D.~{Roehrich},
  B.~{Skaali}, T.~{Steinbeck}, R.~{Stock}, H.~{Tilsner}, K.~{Ullaland},
  A.~{Vestbo}, and T.~{Vik}, ``{FPGA Co-processor for the ALICE High Level
  Trigger},'' {\em ArXiv Physics e-prints} (June, 2003) ,
  \href{http://arxiv.org/abs/physics/0306017}{{\ttfamily physics/0306017}}.

\bibitem{bib:simdkalman}
S.~Gorbunov, U.~Kebschull, I.~Kisel, V.~Lindenstruth, and W.~F.~J.
  M{\"{u}}ller, ``{Fast SIMDized Kalman filter based track fit},'' {\em
  Computer Physics Communications} {\bfseries 178} (2008) 374--383.

\bibitem{catracking}
I.~Kisel, ``Event reconstruction in the {CBM} experiment,''
  \href{http://dx.doi.org/10.1016/j.nima.2006.05.040}{{\em Nuclear Instruments
  and Methods in Physics Research Section A: Accelerators, Spectrometers,
  Detectors and Associated Equipment} {\bfseries 566} no.~1, (2006) 85 -- 88}.
  Proceedings of the 1st Workshop on Tracking in High Multiplicity
  Environments1st Workshop on Tracking in High Multiplicity Environments.

\bibitem{bib:kalman}
R.~E. Kalman, ``{A new approach to linear filtering and prediction problems},''
  \href{http://dx.doi.org/10.1109/ICASSP.1982.1171734}{{\em Journal Of Basic
  Engineering} {\bfseries 82} (1960) 35--45}.

\bibitem{chep2016gpu}
D.~Rohr, S.~Gorbunov, V.~Lindenstruth, and A.~Collaboration, ``Gpu-accelerated
  track reconstruction in the alice high level trigger,'' {\em Journal of
  Physics: Conference Series} {\bfseries 898} no.~3, (2017) 032030.
  \url{http://stacks.iop.org/1742-6596/898/i=3/a=032030}.

\bibitem{bib:chep}
D.~Rohr, S.~Gorbunov, A.~Szostak, M.~Kretz, T.~Kollegger, T.~Breitner, and
  T.~Alt, ``{ALICE HLT TPC Tracking of Pb-Pb Events on GPUs},'' {\em Journal of
  Physics: Conference Series, Proceedings of 19th International Conference on
  Computing in High Energy and Nuclear Physics} {\bfseries 396} no.~1, (2012)
  012044. \url{http://stacks.iop.org/1742-6596/396/i=1/a=012044}.

\bibitem{vc}
M.~Kretz and V.~Lindenstruth, ``{Vc}: A {C++} library for explicit
  vectorization,'' \href{http://dx.doi.org/10.1002/spe.1149}{{\em Softw. Pract.
  Exper.} {\bfseries 42} no.~11, (Nov., 2012) 1409--1430}.
  \url{http://dx.doi.org/10.1002/spe.1149}.

\bibitem{bib:parallelism2ts}
``{Technical Specification for C++ Extensions for Parallelism Version 2},''
  {\em ISO/IEC JTC1 SC22 WG21 N4744} (2018) .
  \url{http://www.open-std.org/jtc1/sc22/wg21/docs/papers/2018/n4742.html}.

\bibitem{bib:chep2015}
{\bfseries ALICE} Collaboration, D.~Rohr, S.~Gorbunov, M.~Krzewicki,
  T.~Breitner, M.~Kretz, and V.~Lindenstruth, ``{Fast TPC Online Tracking on
  GPUs and Asynchronous Data Processing in the ALICE HLT to facilitate Online
  Calibration},'' \href{http://dx.doi.org/10.1088/1742-6596/664/8/082047}{{\em
  J. Phys. Conf. Ser.} {\bfseries 664} no.~8, (2015) 082047},
\href{http://arxiv.org/abs/1712.09416}{{\ttfamily arXiv:1712.09416
  [physics.ins-det]}}.

\bibitem{bib:fair}
{H.H. Gutbrod et. al. (Eds.)}, {\em {FAIR - Baseline Technical Report}}.
\newblock 2006.
\newblock {ISBN: } 3-9811298-0-6.

\bibitem{Gyulassy:1994ew}
M.~Gyulassy and X.-N. Wang, ``{HIJING 1.0: A Monte Carlo program for parton and
  particle production in high-energy hadronic and nuclear collisions},''
  \href{http://dx.doi.org/10.1016/0010-4655(94)90057-4}{{\em Comput. Phys.
  Commun.} {\bfseries 83} (1994) 307},
\href{http://arxiv.org/abs/nucl-th/9502021}{{\ttfamily arXiv:nucl-th/9502021
  [nucl-th]}}.

\bibitem{Krzewicki:2013dkt}
M.~Krzewicki, {\em {Anisotropic flow of identified hadrons in heavy-ion
  collisions at the LHC}}.
\newblock PhD thesis, Utrecht University,
2013.
\newblock

\bibitem{compressionbase}
J.~Berger, U.~Frankenfeld, V.~Lindenstruth, P.~Plamper, D.~R{\"o}hrich,
  E.~Sch{\"a}fer, M.~W. Schulz, T.~M. Steinbeck, R.~Stock, K.~Sulimma,
  A.~Vestb{\o}, and A.~Wiebalck, ``{TPC} data compression,''
  \href{http://dx.doi.org/http://dx.doi.org/10.1016/S0168-9002(02)00792-1}{{\em
  Nuclear Instruments and Methods in Physics Research Section A: Accelerators,
  Spectrometers, Detectors and Associated Equipment} {\bfseries 489} no.~13,
  (2002) 406 -- 421}.

\bibitem{bib:huffman}
D.~A. Huffman, ``A method for the construction of minimum-redundancy codes,''
  \href{http://dx.doi.org/10.1109/JRPROC.1952.273898}{{\em Proceedings of the
  IRE} {\bfseries 40} no.~9, (Sept, 1952) 1098--1101}.

\bibitem{compressiontrackmodel}
D.~R{\"o}hrich and A.~Vestb{\o}, ``Efficient {TPC} data compression by track
  and cluster modeling,''
  \href{http://dx.doi.org/http://dx.doi.org/10.1016/j.nima.2006.06.056}{{\em
  Nuclear Instruments and Methods in Physics Research Section A: Accelerators,
  Spectrometers, Detectors and Associated Equipment} {\bfseries 566} no.~2,
  (2006) 668 -- 674}.

\bibitem{Rohr:2017dlo}
{\bfseries ALICE} Collaboration, D.~Rohr, ``{Tracking performance in high
  multiplicities environment at ALICE},'' in {\em {5th Large Hadron Collider
  Physics Conference (LHCP 2017) Shanghai, China, May 15-20, 2017}}.
\newblock 2017.
\newblock
\href{http://arxiv.org/abs/1709.00618}{{\ttfamily arXiv:1709.00618
  [physics.ins-det]}}.
\newblock

\bibitem{bib:chep2016calibration}
M.~Krzewicki, D.~Rohr, C.~Zampolli, J.~Wiechula, S.~Gorbunov, A.~Chauvin,
  I.~Vorobyev, S.~Weber, K.~Schweda, R.~Shahoyan, V.~Lindenstruth, and
  A.~Collaboration, ``Support for online calibration in the alice hlt
  framework,'' {\em Journal of Physics: Conference Series} {\bfseries 898}
  no.~3, (2017) 032055. \url{http://stacks.iop.org/1742-6596/898/i=3/a=032055}.

\bibitem{bib:chep2016framework}
D.~Rohr, M.~Krzwicki, H.~Engel, J.~Lehrbach, and V.~Lindenstruth,
  ``{Improvements of the ALICE HLT data transport framework for LHC Run 2},''
  {\em Journal of Physics: Conference Series} {\bfseries 898} no.~3, (2017)
  032031. \url{http://stacks.iop.org/1742-6596/898/i=3/a=032031}.

\bibitem{hlt-run1-2012}
A.~Szostak, ``Operational experience with the {ALICE High Level Trigger},''
  {\em Journal of Physics: Conference Series} {\bfseries 396} no.~1, (2012)
  012048. \url{http://stacks.iop.org/1742-6596/396/i=1/a=012048}.

\end{thebibliography}\endgroup
